\documentclass{aa}
\usepackage{subfigure}
\bibpunct{(}{)}{;}{a}{}{,} 
\usepackage{graphicx}
\usepackage{txfonts}
\def\tm{$\theta_\mathrm{max}$ }

\begin{document} 
        

   \title{Testing the cosmological principle with the Pantheon+ sample and the region-fitting method}
   \titlerunning{Testing the cosmological principle}

    \author{J. P. Hu\inst{1} 
        \and Y. Y. Wang\inst{2} 
        \and J. Hu\inst{3}
        \and F. Y. Wang\inst{1,4} 
   }

   \institute{School of Astronomy and Space Science, Nanjing University, Nanjing 210093, China\\
             \email{fayinwang@nju.edu.cn}
         \and
             Anton Pannekoek Institute for Astronomy, University of Amsterdam, Science Park 904, 1098 XH Amsterdam, Netherlands  \\
         \and
             Institute of Astronomy and Information, Dali University, Dali 671003, China \\
         \and 
            Key Laboratory of Modern Astronomy and Astrophysics (Nanjing University), Ministry of Education, Nanjing 210093, China
             }

   \date{Received date; accepted date}

 
  \abstract
   {The cosmological principle is fundamental to the standard cosmological model. It assumes that the Universe is homogeneous and isotropic on very large scales. As the basic assumption, it must stand the test of various observations. In this work, we investigated the properties of the Pantheon+ sample, including redshift distribution and position distribution, and we give its constraint on the flat $\Lambda$CDM model: $\Omega_{m}$ = 0.36$\pm$0.02 and $H_{0}$ = 72.83$\pm$0.23 km s$^{-1}$ Mpc$^{-1}$. Then, using the region fitting (RF) method, we mapped the all-sky distribution of cosmological parameters ($\Omega_{m}$ and $H_{0}$) and find that the distribution significantly deviates from isotropy. A local matter underdensity region exists toward (${308.4^{\circ}}$$_{-48.7}^{+47.6}$, ${-18.2^{\circ}}$$_{-28.8}^{+21.1}$) as well as a preferred direction of the cosmic anisotropy (${313.4^{\circ}}$$_{-18.2}^{+19.6}$, ${-16.8^{\circ}}$$_{-10.7}^{+11.1}$) in galactic coordinates. Similar directions may imply that local matter density might be responsible for the anisotropy of the accelerated expansion of the Universe. Results of statistical isotropy analyses including Isotropy and Isotropy with real-data positions (RP) show high confidence levels. For the local matter underdensity, the statistical significances are 2.78$\sigma$ (isotropy) and 2.34$\sigma$ (isotropy RP). For the cosmic anisotropy, the statistical significances are 3.96$\sigma$ (isotropy) and 3.15$\sigma$ (isotropy RP). The comparison of these two kinds of statistical isotropy analyses suggests that inhomogeneous spatial distribution of real sample can increase the deviation from isotropy. The similar results and findings are also found from reanalyses of the low-redshift sample (lp+) and the lower screening angle (\tm = 60$\degr$), but with a slight decrease in statistical significance. Overall, our results provide clear indications for a possible cosmic anisotropy. This possibility must be taken seriously. Further testing is needed to better understand this signal. }

\maketitle

%

\section{Introduction} \label{sec:intro} 
The $\Lambda$CDM model is generally accepted as the standard cosmological model, which is consistent with most astronomical observations \citep{2018ApJ...859..101S,2019PhRvL.122q1301A,2020MNRAS.492.4456K,2022ApJ...938..110B,2022MNRAS.513.5686C,2022MNRAS.514.1828D,2022MNRAS.516.2575J,2022ApJ...931...50L,2022PhRvD.106j3530P,2022ApJ...924...97W,2023PhRvD.107f3522D,2023MNRAS.522.1247K,2023MNRAS.521.4406L}. It is based on the fundamental assumption of the cosmological principle, namely that the Universe is statistically isotropic and homogeneous on sufficiently large scales. Despite its many successes, there have been also several analyses of observations that indicate that the Universe may be inhomogeneous and anisotropic; for instance, the fine-structure constant \citep{2011PhRvL.107s1101W,2012MNRAS.422.3370K,2017ChPhC..41f5102L,2022arXiv221202458M}, the direct measurement of the Hubble parameter \citep{2006PhRvL..96s1302B,2021PhRvL.126w1101K}, the cosmic microwave background (CMB)
\citep{2003ApJS..148....1B,2003PhRvD..68l3523T,2004MNRAS.355.1283B,2011ApJS..192...17B,2011MNRAS.411.1445G,2020MNRAS.492.3994G,2020A&A...641A...7P}, the anisotropic dark energy \citep{2008JCAP...06..018K,2012PhRvD..86h3517M,2020JCAP...10..019B,2021PDU....3200806M}, the large dipole of radio source counts \citep{2014A&A...565A.111R,2017MNRAS.471.1045C,2018MNRAS.477.1772R,2019PhRvD.100f3501S,2023MNRAS.524.3636S}, quasar dipoles \citep{2005AA...441..915H,2016AA...590A..53P,2019AA...622A.113T,2020A&A...643A..93H,2021ApJ...908L..51S,2021EPJC...81..948Z,2021EPJC...81..694Z,2023MNRAS.525..231D}, the anisotropic Hubble constant \citep{2022PhRvD.105j3510L,2023arXiv230402718M} and the type Ia supernovae (SNe Ia) dipole \citep{2011MNRAS.414..264C,2014MNRAS.437.1840Y,2015ApJ...810...47J,2018MNRAS.478.5153S,2018MNRAS.474.3516W,Tang_2023}, and local matter underdensity \citep{2020PhRvD.102b3520K}. These analyses hint that the Universe may have a local void and a preferred expanding direction. In addition, it is assumed that the $\Lambda$CDM model also triggers a serious Hubble constant discrepancy between the \emph{Planck} CMB \citep{2020A&A...641A...6P} and the local distance ladder \citep{2019ApJ...876...85R,2022ApJ...934L...7R}. This is known as the Hubble tension, and its statistical significance has reached 5.0 $\sigma$. Such a high confidence level could not be explained by systematic uncertainty alone, and might imply new physics beyond the $\Lambda$CDM model. There has been intense discussion focused on this issue, and a lot of theoretical explanations have been proposed. We refer the readers to some review articles \citep{2021CQGra..38o3001D,2021A&ARv..29....9S,2022JHEAp..34...49A,2022NewAR..9501659P,2023Univ....9...94H,2023CQGra..40i4001K,2023arXiv230911552K,2023arXiv230810954R,2023Univ....9..393V} for more detailed information about the cosmological anomalies and tension.

It is worth noting that some researchers claim that considering a void model can successfully explain the cosmic dipole and the Hubble tension \citep{2020A&A...633A..19B,2020MNRAS.499.2845H,2020MNRAS.491.2075L,2022CQGra..39r4001C,2022arXiv221106857C,2023MNRAS.525.3274M}. \citet{2020MNRAS.499.2845H} showed that the KBC void \citep{2013ApJ...775...62K} could naturally resolve the Hubble tension in Milgromian dynamics for the first time. By computing the Hubble constant in an inhomogeneous universe and adopting model selection via both the Bayes factor and the Akaike information criterion, \citet{2022CQGra..39r4001C} found that the lambda Lemaître-Tolman-Bondi ($\Lambda$LTB) model is favored with respect to the $\Lambda$CDM model at low-redshift range (0.023 $< z <$ 0.15), and this can be used to explain the Hubble tension. After that, \citet{2022arXiv221106857C} proposed that a gigaparsec-scale void can reconcile the CMB and quasar dipolar tension. If considering a large and thick void, their setup can also ease the Hubble tension. At the same time, there are some findings that could be explained by a local void model. For example, inspired by the H0LiCOW results \citep{2020A&A...639A.101M,2020MNRAS.498.1420W}, \citet{2022MNRAS.517..576H} reported a late-time transition of $H_{0;}$ that is, $H_{0}$ changes from being consistent with the CMB result to being consistent with the distance ladder one from an early-to-late cosmic time, which can be explicated by the local void. The late-time transition of $H_{0}$ was found from the observational Hubble parameter $H(z)$ data combining the Gaussian process (GP) method \citep{2011JMLR...12.2825P} and can be used to effectively relieve the Hubble tension (a mitigation level of around 70 \%). In addition, a similar $H_{0}$ descending behavior has also been discovered by utilizing various observations (such as SNe Ia, $H(z)$, baryon acoustic oscillations and megamasers) and their combinations \citep{2020PhRvD.102j3525K,2021PhRvD.103j3509K,2022Galax..10...24D,2022PhRvD.106d1301O,2022A&A...668A..34H,2022arXiv220611447C,2023A&A...674A..45J,2023arXiv230112725M}. Of course, there are also some opposing voices believing that the void model alone cannot solve the Hubble tension \citep{2019ApJ...875..145K,2021PhRvD.103l3539C,2022JCAP...07..003C}. Therefore, further research on this controversial topic is necessary and would certainly be worthwhile. So far, there has been no research using the Pantheon+ sample to simultaneously map matter-density ($\Omega_{m}$) distribution and the Hubble expansion ($H_{0}$) distribution to test the cosmological principle.
 
In this work, we tested the cosmological principle by the region fitting (RF) method with the latest Pantheon+ sample. Compared to the previous work \citep{2021CQGra..38r4001K,2022PhRvD.105f3514K,2023arXiv230402718M}, we improved our research methodology and carried out the necessary statistical analyses. Usually, considering the simple flat $\Lambda$CDM model, $\Omega_{m}$ and $H_{0}$ are considered to be negatively correlated. Therefore, for the convenience of analysis, one parameter is usually fixed. For example, some recent works fix $\Omega_{m}$ to 0.30 and regard $H_{0}$ as a free parameter \citep{2021CQGra..38r4001K,2022PhRvD.105f3514K,2023arXiv230402718M}. In our fitting calculation process, all cosmological parameters ($\Omega_{m}$ and $H_{0}$) are free to investigate the local properties of our Universe. We would like to plot the all-sky distributions of $\Omega_{m}$ and $H_{0}$ to find out the local matter underdensity region and the preferred direction of expansion (cosmic anisotropy), respectively. The influence of redshift and the screening angle \tm on the final results is also considered. We found the suitable angle of the RF method for the Pantheon+ sample. Then, we analysed a combination of the local void \citep{2008GReGr..40..451E,2008JCAP...04..003G,2013ApJ...775...62K,2013MNRAS.432.3025W,2016ApJ...818..173H,2022A&A...658A..20H,2023ApJ...952...59S}, cosmic anisotropy \citep{2019EPJC...79..783S,2020PhRvD.102l4059A,2020A&A...636A..15M,2021AA...649A.151M,2022JPhCS2191a2001A,2022A&A...668A..34H,2022MNRAS.514..139R,2023PDU....3901162A,2023MNRAS.519.4841D,2023arXiv230516177E}, and Hubble tension \citep{2019NatAs...3..384C,2019ApJ...876...85R,2019NatAs...3..891V,2020A&A...641A...6P,2020NatRP...2...10R,2022ApJ...934L...7R,2023PDU....4001201C} in detail. Finally, we compared our results with those of previous similar research \citep{2010JCAP...12..012A,2012JCAP...02..004C,2013AA...553A..56K,2014MNRAS.443.1680W,2014MNRAS.437.1840Y,2015MNRAS.446.2952C,2016MNRAS.456.1881L,2019MNRAS.486.1658C,2020A&A...643A..93H,2022PhRvD.105j3510L,2023arXiv230402718M} and other observations, including the CMB dipole \citep{2016AA...594A...1P,2020AA...641A...1P}, dark flow \citep{2022JHEAp..34...49A}, bulk flow \citep{2012MNRAS.420..447T,2017MNRAS.468.1420F,2023MNRAS.524.1885W}, and galaxy cluster \citep{2021AA...649A.151M}.

The outline of this paper is as follows. In Sect. 2, we give a detailed description of the Pantheon+ sample, including redshift distribution, location distribution, and corresponding density contour, and compare it to the Pantheon sample. Section 3 briefly introduces the RF method, which we used to map the all-sky distribution of cosmological parameters. In Sect. 4, we present the results from the whole and low-redshift Pantheon+ sample ($z < 0.30$, lp+) and discuss the impact of RF method with different screening angles \tm on the final results from the Pantheon+ sample. The corresponding investigation and discussion are given in Sect. 5. Finally, conclusions and perspectives are presented in Sect. 6.

\section{Pantheon+ sample}\label{sec:data} 
\begin{figure}[hpt]
        \centering
        \includegraphics[width=0.35\textwidth]{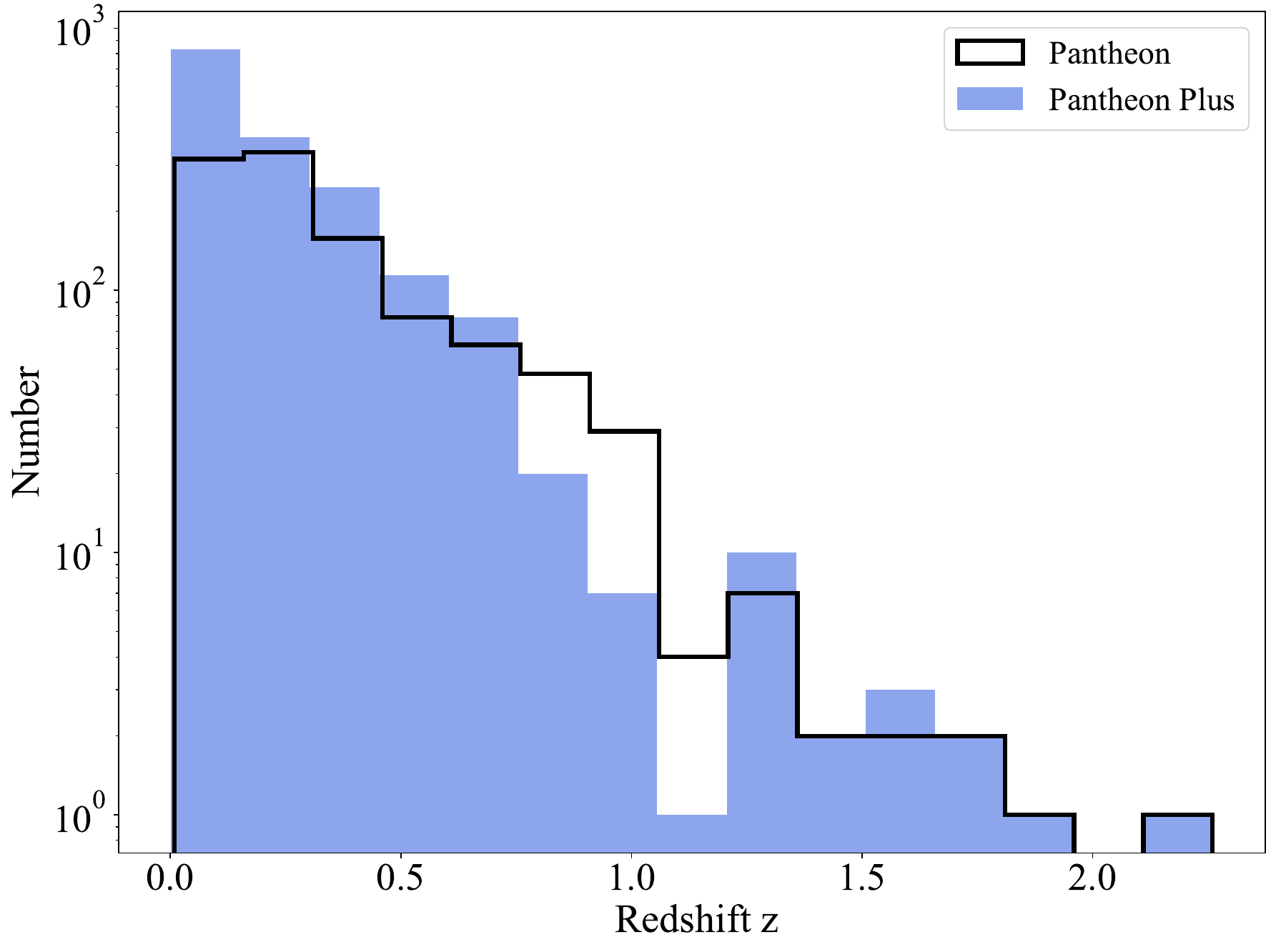}
        \caption{Redshift distribution of the Pantheon and Pantheon+ samples.}
        \label{F1}       
\end{figure}

Pantheon+, as the latest sample of SNe Ia, consists of 1701 SNe Ia light curves observed from 1550 distinct SNe and covers redshift range from 0.001 to 2.26 \citep{2022ApJ...938..110B,2022ApJ...938..113S}. 
The redshift distributions of the Pantheon \citep{2018ApJ...859..101S} and Pantheon+ samples are shown in Fig. \ref{F1}. From the redshift distribution of these two samples, it is not difficult to find that there are two main differences between the new sample and the Pantheon sample. One is that the SN number has increased significantly at low redshift. There are more than 700 additional SNe in the range of $z <$ 0.8, of which more than 500 SNe originated from $z <$ 0.08. This is mainly because the new sample adds five large samples, including the Foundation Supernova Survey \citep[Foundation;][]{2018MNRAS.475..193F}, the Swift Optical/Ultraviolet Supernova Archive \citep[SOUSA;][]{2014Ap&SS.354...89B}, the Lick Observatory Supernova Search \citep[LOSS1;][]{2010ApJS..190..418G}, the second sample from LOSS \citep[LOSS2;][]{2019MNRAS.490.3882S}, and the Dark Energy Survey Year 3  \citep[DES;][]{2019ApJ...874..106B,2020AJ....160..267S}, all of which are low-redshift surveys, except DES. The other difference is that there is a significant deletion in Pantheon+ statistics between $0.8 < z < 1.0$. The reason is that \citet{2022ApJ...938..113S} did not use SNe from the Supernova Legacy Survey (SNLS) at $z >$ 0.8 considering sensitivity to the $U$ band in model training (56 SNe in total). 


\begin{figure*}[htp]
        \centering
        \includegraphics[width=0.45\textwidth]{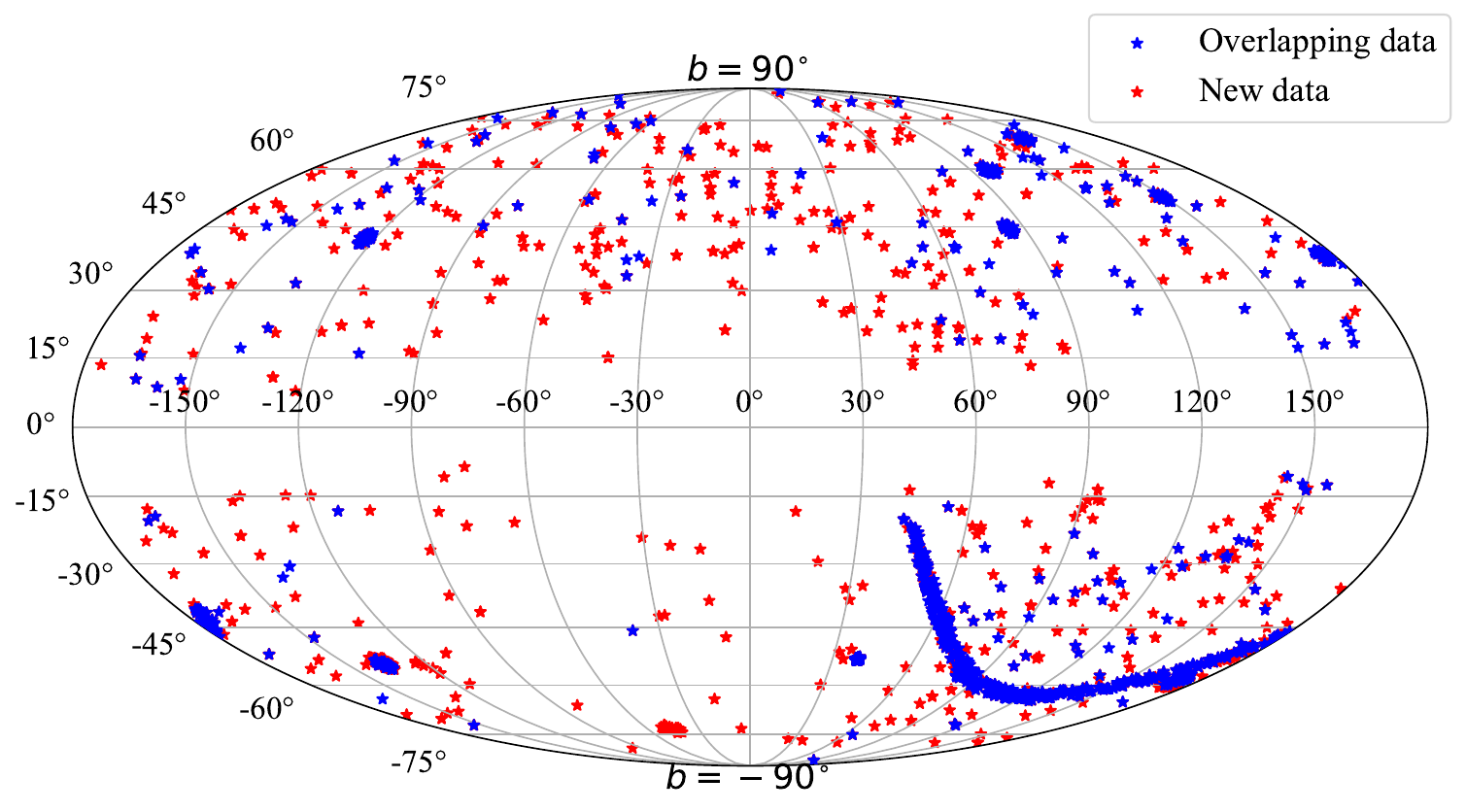}
        \includegraphics[width=0.45\textwidth]{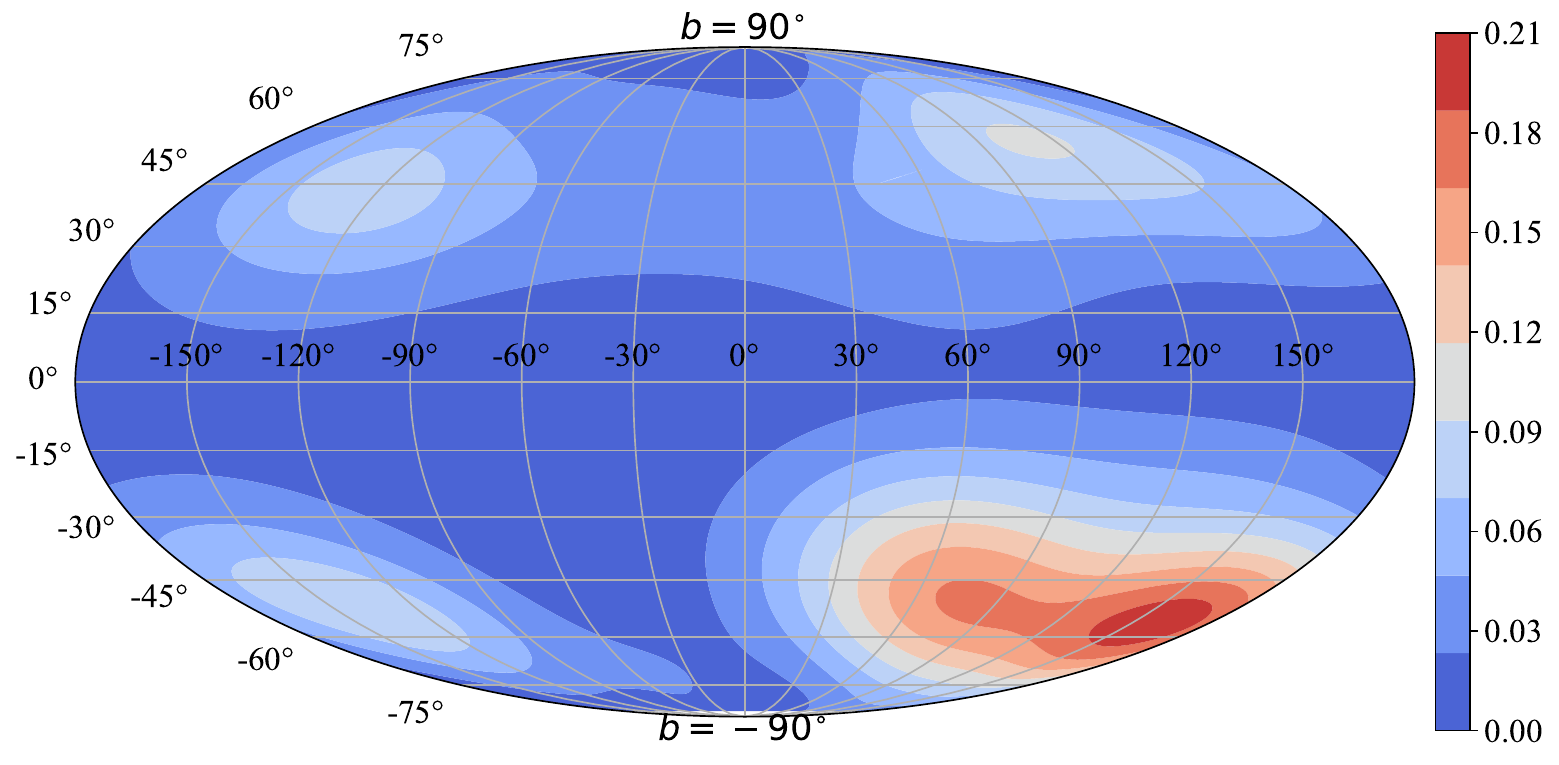}
        \caption{Distribution and corresponding density contour in the galactic coordinate system. Left panel shows the SNe distribution of the Pantheon sample. Red and blue points represent the new added SNe and the SNe where the Pantheon sample and the Pan sample overlap, respectively. Right panel shows the corresponding density contour.}
        \label{F2}       
\end{figure*}

Figure \ref{F2} shows the distribution (left panel) and corresponding density (right panel) of the Pantheon+ sample in the sky of the galactic coordinate system. Some previous work has pointed out that the Pantheon SNe Ia are not uniformly distributed in the sky; half of them are located in the galactic southeast. Some SNe are very concentrated, forming a belt-like structure that is the SDSS sample \citep{2019MNRAS.486.5679Z}. At the same time, \citet{2019MNRAS.486.5679Z} also investigated the effect of the inhomogeneous distribution of the Pantheon sample on the cosmic anisotropy and found that the belt-like structure plays the most important role in the Pantheon sample. In the left panel, to highlight the difference between these two samples, we mark the newly added SNe Ia in red. From the distribution of the Pantheon+ sample, we find that the distribution of the new data is still uneven across the sky. There are only fewer observations near (l, b)$\sim$(300$^{\circ}$, -30$^{\circ}$). Its incomplete sky coverage is primarily due to the fact that the Pantheon+ sample consists of different subsamples, following multiple measurement strategies. In order to make it easier to comprehend this focus, we also plot the density distribution of the Pantheon+ sample utilizing the $plt.contour()$ function\footnote{https://matplotlib.org/stable/gallery/images$_{-}$contours$_{-}$and$_{-}$fields/ir regulardatagrid.html}, as shown in the right panel of Fig. \ref{F2}. Color-coded values represent sample fraction per unit area. From the density distribution, we can more intuitively feel the inhomogeneity of sample distribution. As can be seen, the belt-like part of the Pantheon+ sample still plays the most important role, as in the case of the Pantheon sample, while the maximum density value changes from 0.24 to 0.21 \citep{2020A&A...643A..93H}. This means that the Pantheon+ sample is more uniform than the Pantheon sample. The newly added SNe Ia effectively weaken the dominance of the belt structure in the distribution. The Pantheon+ sample that is relatively uniform and rich in low redshift is very suitable for analyzing the local property of the Universe \citep{2018ApJ...865..119A,2022PhRvD.105j3510L,2023PhRvD.107b3507K}. Here, we used the RF method combined with the Pantheon+ sample to describe the all-sky distribution of cosmological parameters to study the local structure of the Universe, and the detailed analysis process is shown in the next section.

\section{Region fitting method}
The hemisphere comparison (HC) method proposed by \citet{2007AA...474..717S} has been widely used in the investigation of the cosmic anisotropy, such as the anisotropy of cosmic expansion \citep{2018EPJC...78..755D,2022MNRAS.511.5661Z}, the acceleration scale of modified Newtonian dynamics \citep{2017ApJ...847...86Z,2018ChPhC..42k5103C,2019MNRAS.486.1658C}, and the temperature anisotropy of the CMB \citep{2004MNRAS.354..641H,2013ApJS..208...20B,2016JCAP...01..046G,2021PhRvD.104f3503F}. The RF method we used is similar to it. Here, we provide a detailed introduction to this method. Its goal is to map the all-sky distribution of the cosmological parameters. The most important step is to generate random directions $\hat{D}$ $(l$, $b)$ to pick out SNe data located in specific regions to construct subdatasets, where $l\in(0^{\circ}$, $360^{\circ})$ and $b\in(-90^{\circ}$, $90^{\circ})$ are the longitude and latitude in the galactic coordinate system, respectively. The specific region is given by condition ($\theta < \theta_\mathrm{max}$). $\theta$ is the angle between the random direction $\hat{D}$ $(l$, $b)$ and the SN position. Here we refer to $\theta_\mathrm{max}$ as the screening angle used to obtain regions of different sizes, and the value range is (0$^{\circ}$, 180$^{\circ}$). The remaining steps can be divided into three parts, which we present in the following subsections.

\subsection{Fitting parameters}
According to subdatasets obtained in the previous step, the corresponding best fits of the cosmological parameters are obtained by minimizing value of $\chi^{2}$,
\begin{equation}
        \chi^{2} = \Delta \mu \, \mathbf{C}^{-1}_\mathrm{stat+syst} \, \Delta \mu ^\mathrm{T} ,
        \label{chi}
\end{equation}
where $\Delta \mu$ is the difference between the observational distance modulus $\mu_\mathrm{obs}$ and the theoretical distance modulus $\mu_\mathrm{th}$:
\begin{equation}
        \Delta \mu = \mu_\mathrm{obs}(z_{i}) - \mu_\mathrm{th}(\Omega_{m}, H_{0},z_{i}) . 
        \label{Dmu}
\end{equation}
For the flat $\Lambda$CDM model, the corresponding form of $\mu_\mathrm{th}$ can be written as 
\begin{equation}
        \mu_\mathrm{th}(\Omega_{m}, H_{0},z_{i}) = m - M = 5 \log_{10} \frac{d_{L}(\Omega_{m}, H_{0},z_{i})}{\textnormal{Mpc}} + 25;
        \label{mu}
\end{equation}
here, $z_{i}$ is the peculiar-velocity-corrected CMB-frame redshift of each SN \citep{2022PASA...39...46C}, $m$ is the apparent magnitude of the source, $M$ is the absolute magnitude, and $d_{L}$ is the luminosity distance expressed in megaparsec, defined in the following equation: 
\begin{equation}
        d_{L} = \frac{c(1+z)}{H_{0}} \int_{0}^{z} 
        \frac{\mathrm{d} z'}{\sqrt{\Omega_{m} (1+z')^{3} + (1-\Omega_{m})}},
        \label{dl}
\end{equation}
where $c$ is the speed of light. 

The statistical ($\mathbf{C}_\mathrm{stat}$) and systematic covariance matrices ($\mathbf{C}_\mathrm{sys}$) are combined and adopted to constrain the cosmological parameters: 
\begin{equation}
    \mathbf{C}_\mathrm{stat+sys} = \mathbf{C}_\mathrm{stat} +  \mathbf{C}_\mathrm{sys}.
        \label{C}
\end{equation}
The datasets we used, $\mu_\mathrm{obs,}$ and $\mathbf{C}_\mathrm{stat+sys,}$ are provided by \citet{2022ApJ...938..110B} and can be obtained online\footnote{https://github.com/PantheonPlusSH0ES/DataRelease}. $\mathbf{C}_\mathrm{stat+sys}$ includes all the covariance between SNe (and also Cepheid host covariance) due to systematic uncertainties \citep{2022ApJ...938..110B}. $\mathbf{C}_\mathrm{stat}$ mainly includes the full distance error and measurement noise. $\mathbf{C}_\mathrm{sys}$ can manifest in three key places in the analysis: (1) from changing aspects affecting the light-curve fitting; (2) from changing redshifts that propagate to changes in distance modului relative to a cosmological model; and (3) from changes in the simulations used for bias corrections \citep{2022ApJ...938..110B}. More detailed information about the covariance matrices can be found in Sect. 2.2 of \citet{2022ApJ...938..110B}. Unlike the previous analyses \citep{2014A&A...568A..22B,2019A&A...631L..13C}, we did not introduce an intrinsic scatter. So, in the fitting process, there are only two free parameters ($\Omega_{m}$ and $H_{0}$), which makes them strongly correlated. For the different subdatasets, $\Omega_{m}$ and $H_{0}$ are fitted simultaneously. The $\mathbf{C}_\mathrm{stat+sys}$ we used were obtained by cropping the total covariance matrix according to the SNe Ia subsample. In this work, the minimization was performed employing a Bayesian Markov chain Monte Carlo \citep[MCMC;][]{2013PASP..125..306F} method with the $emcee$ package\footnote{https://emcee.readthedocs.io/en/stable/}. All the fittings in this paper were obtained adopting this python package. The MCMC samples were plotted utilizing the $getdist$ package \citep{2019arXiv191013970L}.

\subsection{All-sky distribution}
During the calculation, we repeated 5000 random directions $\hat{D}$ $(l$, $b)$, that is, 5000 sets of best fitting results ($\Omega_{m}$ and $H_{0}$). Based on the fitting results, the all-sky distributions of $\Omega_{m}$ and $H_{0}$ are mapped, respectively. From the distribution of $\Omega_{m}$ and $H_{0}$, we can obtain information concerning the local matter underdensity and cosmic anisotropy, respectively. In order to describe the degree of deviation from the cosmological principle, we define a parameter $D_\mathrm{max}(\sigma)$ whose form is as follows:
\begin{equation}
        D_\mathrm{max}(\sigma) = \frac{P_\mathrm{max} - P_\mathrm{min}}{\sqrt{\sigma_{\rm P_\mathrm{max}}^{2}+\sigma_{P_\mathrm{min}}^{2}}}.
        \label{CS}
\end{equation}
Here, $P_\mathrm{min}$ and $P_\mathrm{max}$ are the minimum value and the maximum value of the best fitting results, respectively. $\sigma_{P_\mathrm{min}}$ and $\sigma_{P_\mathrm{max}}$ are the corresponding 1$\sigma$ error. The local matter underdensity direction and the preferred direction of cosmic anisotropy are marked by the corresponding location of the lowest $\Omega_\mathrm{m}$ and the largest $H_{0}$. The corresponding 1$\sigma$ regions are plotted in terms of the values of $P_{0,}$ which is calculated from
\begin{equation}
        1.00(\sigma) \geqslant \frac{P_{0} - P_\mathrm{fit}}{\sqrt{\sigma_{ P_{0}}^{2}+\sigma_{ P_\mathrm{fit}}^{2}}},
        \label{DA}
\end{equation}
where $P_\mathrm{fit}$, representing the lowest $\Omega_{m}$ constraint or the largest $H_{0}$ constraint, is used to find the 1$\sigma$ range of the local matter underdensity direction or the preferred direction. $\sigma_{P_\mathrm{fit}}$ represents the corresponding error. $P_{0}$ represents the constraints that are consistent with $P_\mathrm{fit}$ within 1$\sigma$ error, and $\sigma_{P_{0}}$ are the corresponding 1$\sigma$ values. We note that $P_{0}$ and $\sigma_{P_{0}}$ are filtered from the total $\Omega_{m}$/$H_{0}$ fitting results depending on Eq. \ref{DA}. Here, $P_{0}$ and $P_\mathrm{fit}$ are completely independent and were obtained using different SNe subsamples.

\subsection{Statistical analyses} \label{stat}
In order to examine whether the discrepancy degree of the cosmological parameters from the Pantheon+ sample is consistent with statistical isotropy, we plan to carry out statistical isotropic analyses. To achieve this, we spread the original data set evenly across the sky. After that, we were able to obtain the $D_{max}$ for the isotropic dataset. Meanwhile, an additional isotropic analysis was also considered. We preserved the spatial inhomogeneity of real sample and then randomly distributed the real dataset, which randomly redistributed the distance moduli and redshift combination to real-data positions (RP) only. Given the limitations of computing time, we repeated it 500 times; this gave acceptable statistics. For convenience, we refer to these two approaches as isotropy analysis and isotropy RP analysis.

\section{Results}
We first give the best fitting results in the flat $\Lambda$CDM model employing the full Pantheon+ sample, $\Omega_{m}$ = 0.36$\pm$0.02, and $H_{0}$ = 72.83 $\pm$ 0.23 km s$^{-1}$ Mpc$^{-1}$. The results are in line with the previous research \citep{2022ApJ...938..110B}, except that the 1$\sigma$ error of $H_{0}$ is reduced. The main reason is that we utilized the standardized distance modulus \citep[$\mu_\mathrm{obs}$;][]{1998A&A...331..815T}, where fiducial $M$ has been determined from SH0ES 2021 Cepheid host distances \citep{2022ApJ...934L...7R}.


\begin{figure}[h]
        \centering
        \includegraphics[width=0.42\textwidth]{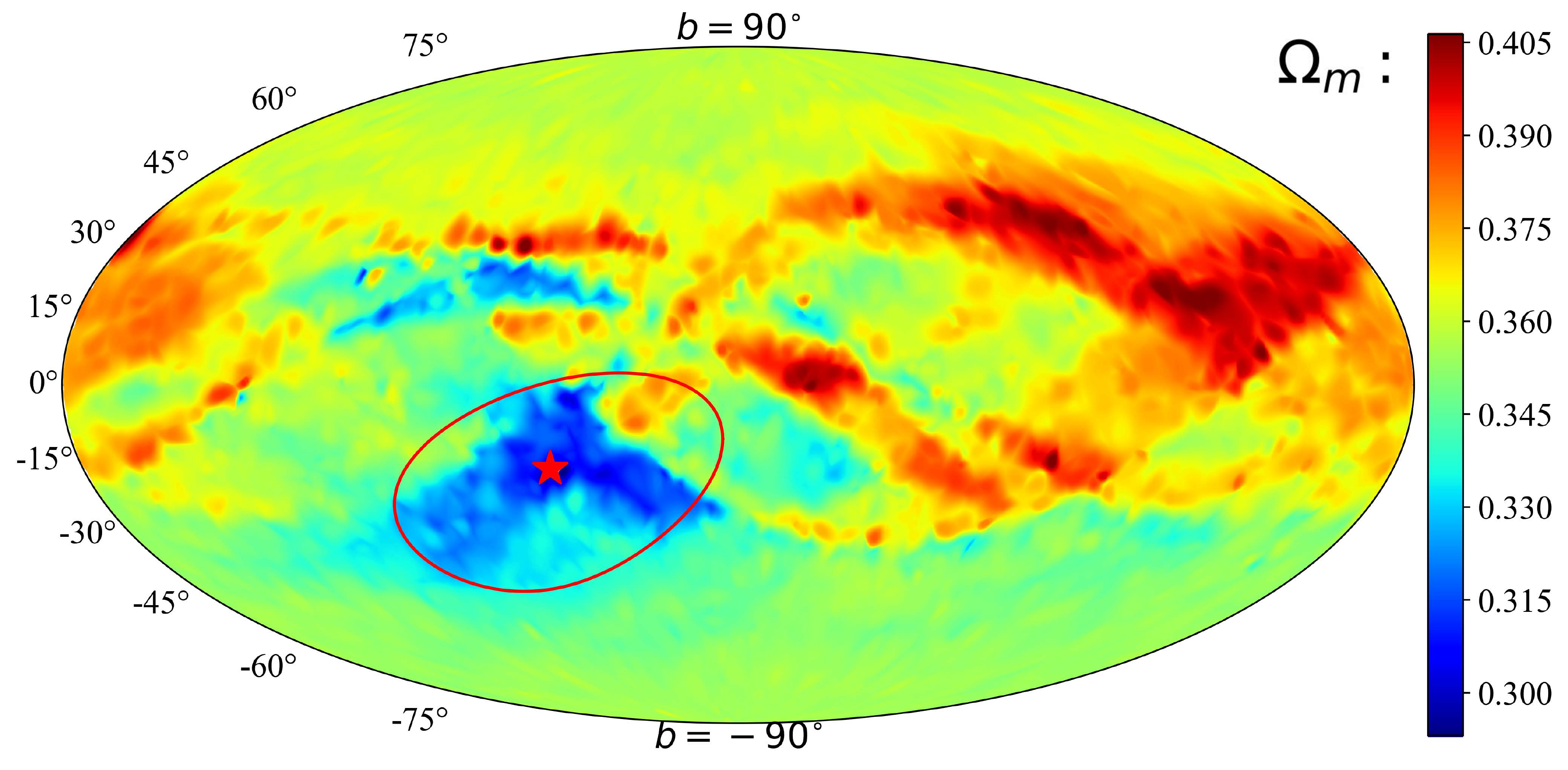}
        \includegraphics[width=0.42\textwidth]{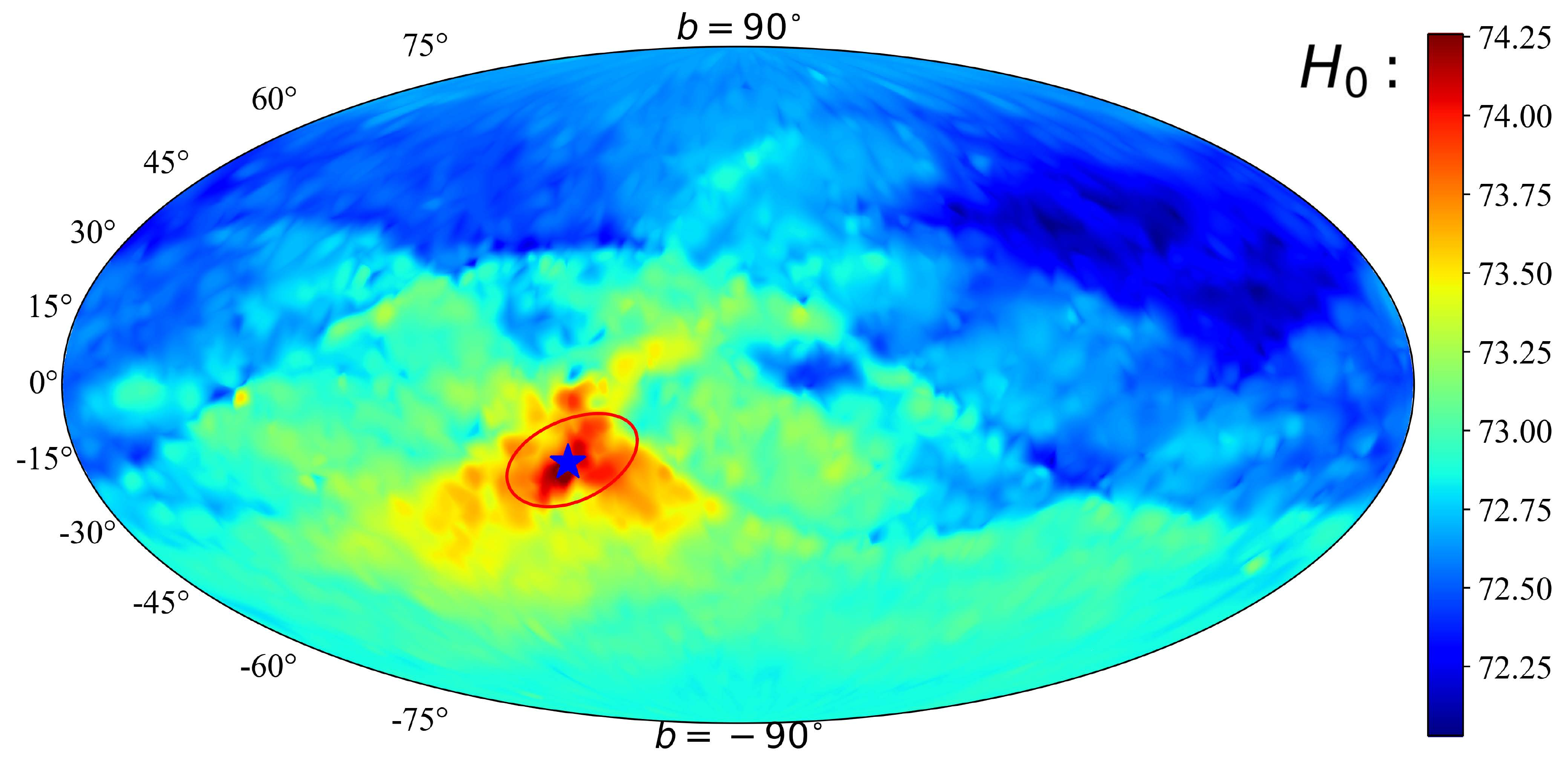}
        \caption{All-sky distribution of cosmological parameters utilizing the Pantheon+ sample combined with the 90$^{\circ}$ RF method. The upper and the lower panels show the results of $\Omega_{m}$ and $H_{0}$, respectively. The corresponding values of $D_\mathrm{max}(\sigma)$ are 3.29$\sigma$ and 4.48$\sigma$, respectively. The star marks the directions of the lowest $\Omega_{m}$ (upper panel) and the largest $H_{0}$ (lower panel), and the red circle outlines the corresponding 1$\sigma$ areas. The directions and 1$\sigma$ areas are parameterized as $\Omega_{m, \mathrm{min}}$ (${308.4^{\circ}}_{-48.7}^{+47.6}$, ${-18.2^{\circ}}_{-28.8}^{+21.1}$) and $H_{0, \mathrm{max}}$ (${313.4^{\circ}}_{-18.2}^{+19.6}$, ${-16.8^{\circ}}_{-10.7}^{+11.1}$). }
        \label{F4}       
\end{figure}

\begin{figure}[h]
        \centering
        \includegraphics[width=0.4\textwidth]{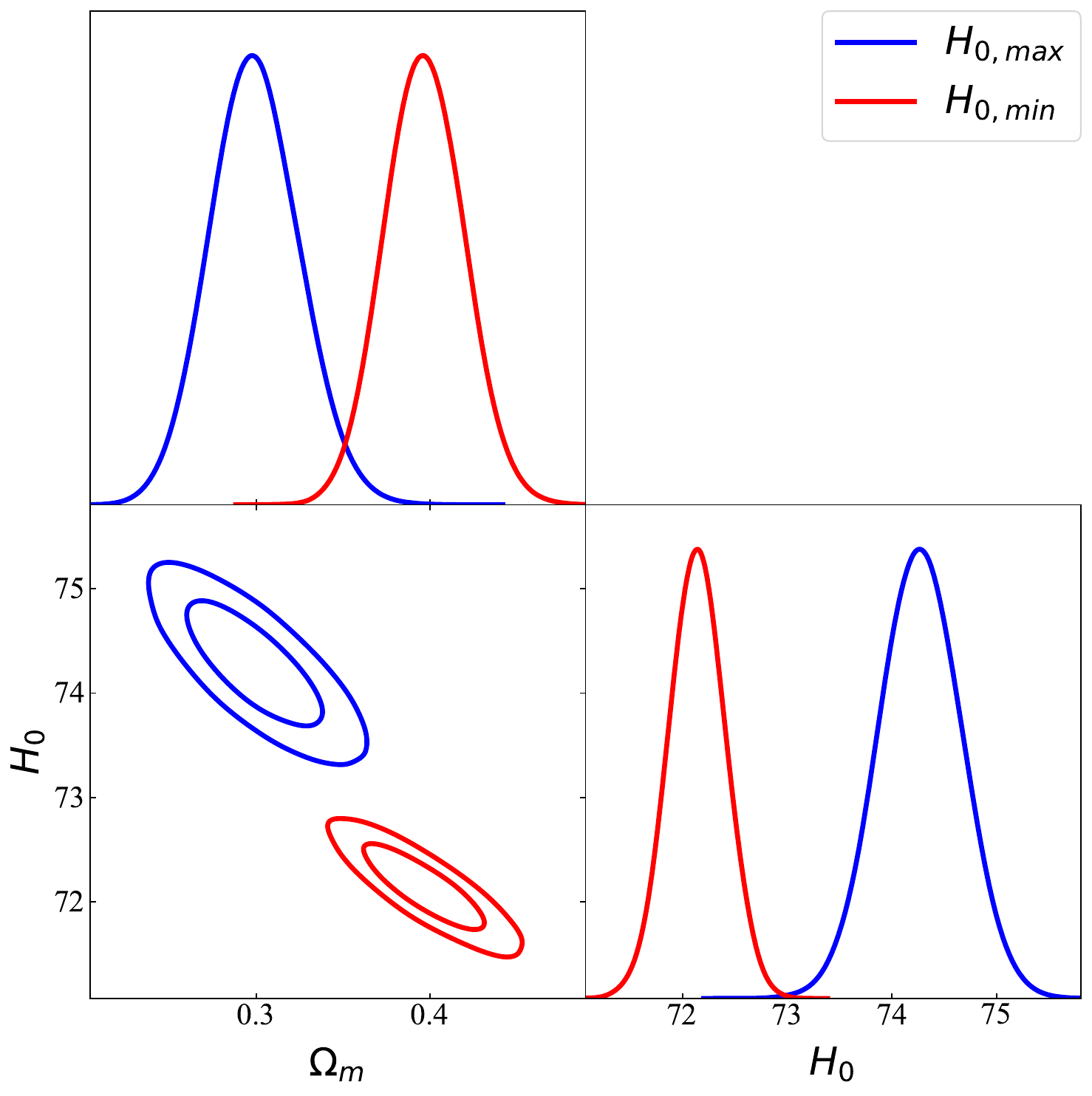}
        \caption{Confidence contours ($1\sigma$ and $2\sigma$) and marginalized likelihood distributions for the parameter space ($\Omega_{m}$ and $H_{0}$) in the spatially flat $\rm \Lambda$CDM model from the SNe Ia subsamples, which corresponds to $H_{0, \mathrm{min}}$ (red line) and $H_{0, \mathrm{max}}$ (blue line). The best fitting results of preferred direction are $H_{0, \mathrm{max}}$ = 74.26$^{+0.40}_{-0.39}$ km s$^{-1}$ Mpc$^{-1}$, $\Omega_{m}$ = 0.30$^{+0.03}_{-0.03}$ (blue line). The best fitting results of opposite directions are $H_{0, \mathrm{min}}$ = 72.14$^{+0.27}_{-0.27}$ km s$^{-1}$ Mpc$^{-1}$ and $\Omega_{m}$ = 0.40$^{+0.02}_{-0.02}$ (red line).}
        \label{total}       
\end{figure}

After that, using the RF method with a screening angle $\theta_\mathrm{max} = 90^{\circ}$, we mapped the all-sky distribution of $\Omega_{m}$ and $H_{0}$ and draw the 1$\sigma$ regions of local matter underdensity and cosmic anisotropy, as shown in the upper panel and lower panel of Fig. \ref{F4}, respectively. In Fig. \ref{F4}, the range of $\Omega_{m}$ and $H_{0}$ are (0.29, 0.41) and (72.03, 74.26), respectively. The corresponding differences are $\Delta \Omega_{m}$ = 0.12 and $\Delta H_{0}$ = 2.23 km s$^{-1}$ Mpc$^{-1}$. The values of $D_\mathrm{max}$ to $\Omega_{m}$ and $H_{0}$ are $D_{\mathrm{max},\Omega_{m}}$ = 3.29$\sigma$ and $D_{\mathrm{max}, H_{0}}$ = 4.48$\sigma$, respectively. In the upper panel of Fig. \ref{F4}, the minimum constraint of $\Omega_{m}$ is $\Omega_{m, \mathrm{min}}$ = 0.29$^{+0.03}_{-0.02}$ (1$\sigma$) and the corresponding constraint for $H_{0}$ is 74.11$^{+0.40}_{-0.40}$ km s$^{-1}$ Mpc$^{-1}$. The direction and corresponding 1$\sigma$ range that can be adopted to describe the local matter underdensity are 
\begin{equation}
        (l, b) = ({308.4^{\circ}}_{-48.7}^{+47.6}, {-18.2^{\circ}}_{-28.8}^{+21.1}).
        \label{lbF4}
\end{equation}
The lower panel of Fig. \ref{F4} shows the corresponding all-sky distribution of the Hubble expansion. The maximum constraint of $H_{0}$ is $H_{0,\mathrm{max}}$ = 74.26$\pm$ 0.39 km s$^{-1}$ Mpc$^{-1}$, and the corresponding constraint of $\Omega_{m}$ is 0.30$^{+0.03}_{-0.03}$. Its confidence contours are shown in Fig. \ref{total}, marked with blue lines. The preferred direction of cosmic anisotropy and corresponding 1$\sigma$ range are
\begin{equation}
        (l, b) = ({313.4^{\circ}}_{-18.2}^{+19.6}, {-16.8^{\circ}}_{-10.7}^{+11.1}).
        \label{lbF4H}
\end{equation} 
In Fig. \ref{total}, we also give the best fitting results of opposite directions; that is, $H_{0, \mathrm{min}}$ = 72.14$^{+0.27}_{-0.27}$ km s$^{-1}$ Mpc$^{-1}$ and $\Omega_{m}$ = 0.40$^{+0.02}_{-0.02}$ , which are marked with red lines.

\begin{figure}
        \centering
        \includegraphics[width=0.23\textwidth]{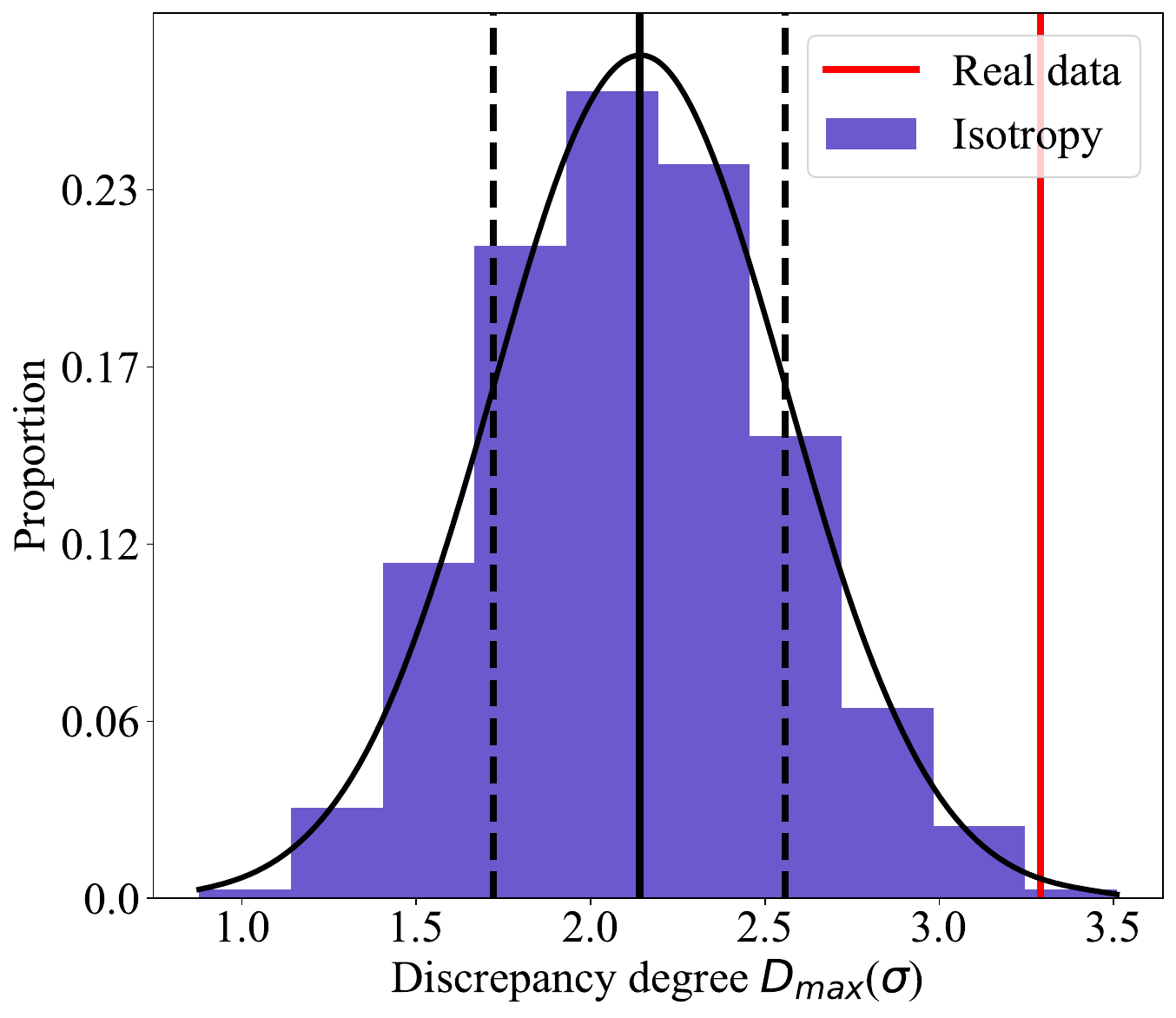}
        \includegraphics[width=0.23\textwidth]{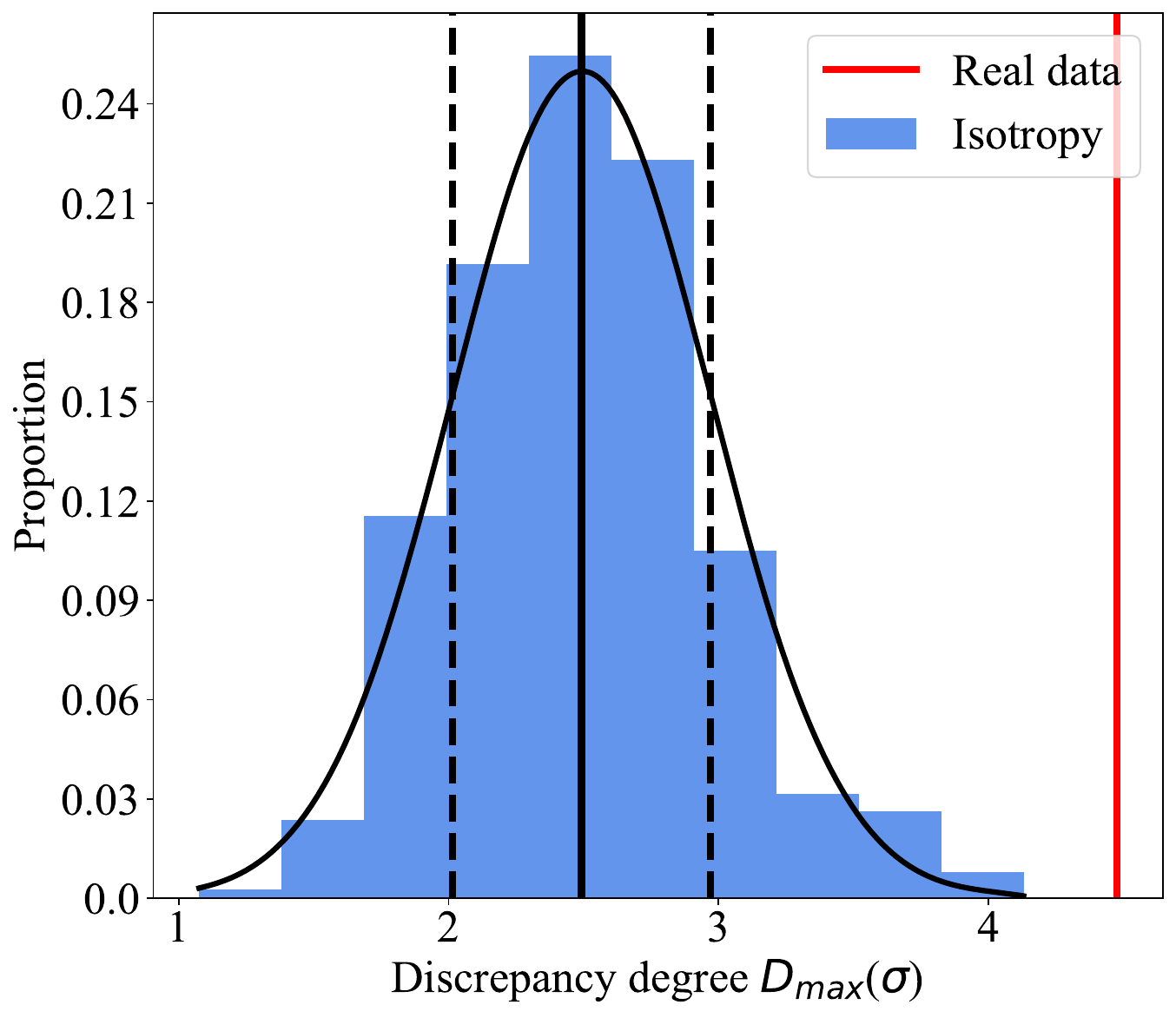}
        \includegraphics[width=0.23\textwidth]{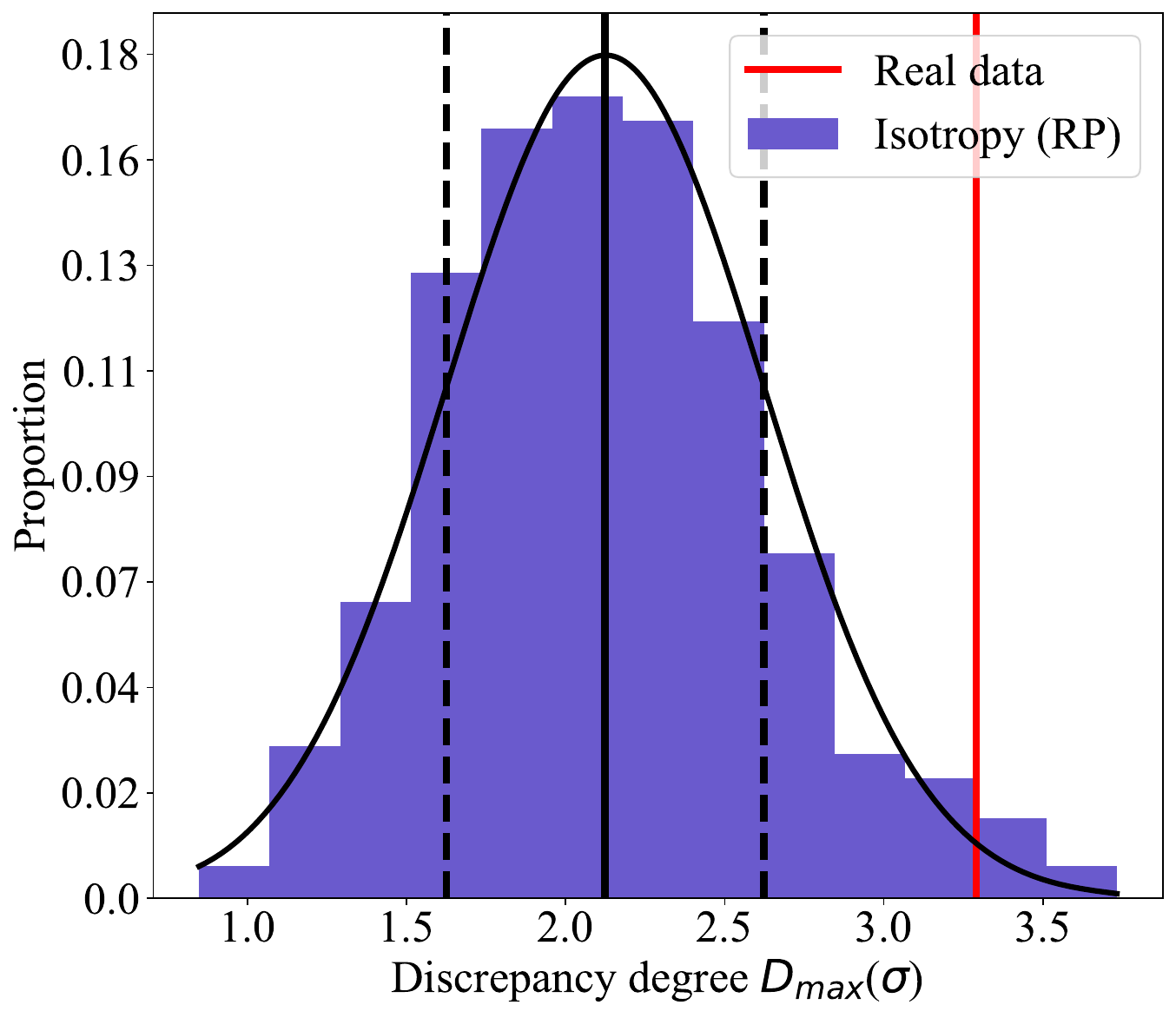}
        \includegraphics[width=0.23\textwidth]{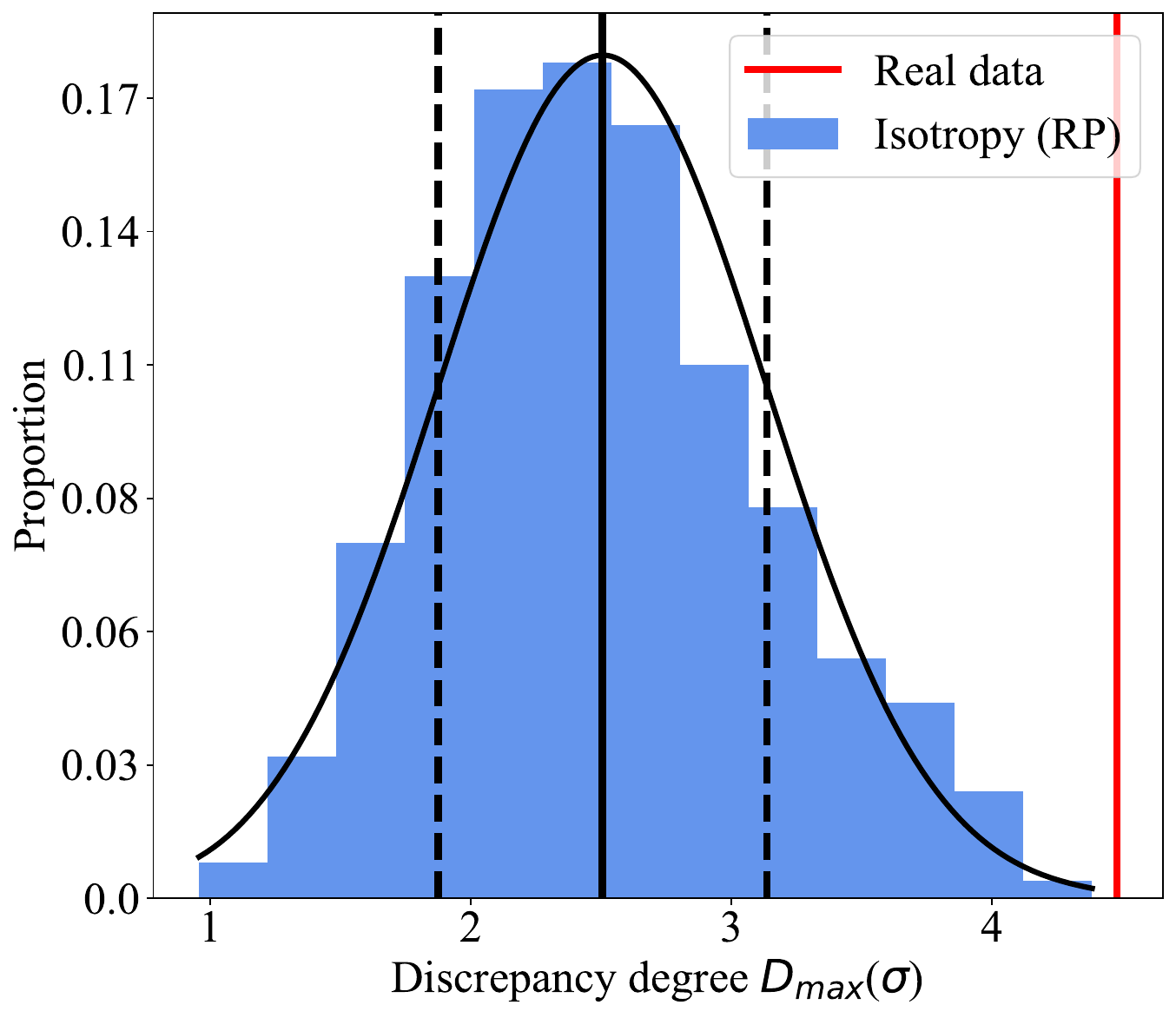}
        \caption{Distribution of discrepancy degree $D_\mathrm{max}$ in 500 simulated isotropic datasets. The upper two panels show the results of statistical isotropic analyses (isotropy). The lower two panels show the results of statistical isotropic analyses that preserve the spatial inhomogeneity of real data (isotropy RP). Purple and blue represent the statistical results of $\Omega_{m}$ and $H_{0}$. The black curve is the best fit to the Gaussian function. The solid black and vertical dashed lines are commensurate with the mean and the standard deviation, respectively. The red lines represent the discrepancy degree from the real data. For the isotropy analyses, the statistical significance of the real data are 2.78$\sigma$ for $\Omega_{m}$ anisotropy and 3.96$\sigma$ for $H_{0}$ anisotropy. For the isotropy RP analyses, the statistical significance of the real data are 2.34$\sigma$ and 3.15$\sigma$ for $\Omega_{m}$ and $H_{0}$ anisotropy, respectively. }
        \label{F5}       
\end{figure}

The statistical isotropic results shown in Fig. \ref{F5} can be well described by Gaussian functions. For the isotropy analysis, the statistical significances ($\alpha$) of the real data are 2.78$\sigma$ for $\Omega_{m}$ anisotropy (upper, purple panel) and 3.96$\sigma$ for $H_{0}$ anisotropy (upper, blue panel). The statistical significance ($\beta$) of the real data given by the isotropy RP analysis are 2.34$\sigma$ for $\Omega_{m}$ anisotropy (lower, purple panel) and 3.15$\sigma$ for $H_{0}$ anisotropy (lower, blue panel).

\subsection{Reanalyses at low redshift}
Recently, \citet{2022MNRAS.517..576H} reported a late-time transition of $H_{0}$; that is, $H_{0}$ changes from a low value to a high one from early to late cosmic time by combining the GP method and $H(z)$ data. The $H_{0}$ transition occurs at $z \sim$ 0.49. From other observations, a similar descending behavior of $H_{0}$ has been found adopting other methods (see \citet{2023Univ....9...94H} for a review of $H_{0}$ descending trend). Around $z \sim$ 0.40, $H_{0}$ starts to decrease \citep{2023A&A...674A..45J}. Both the transition behavior and the descending trend of $H_{0}$ can effectively alleviate the Hubble tension. \citet{2023Sci...380.1322K} gave new $H_{0}$ measurements from the gravitationally lensed SNe Ia Refsdal \citep{1964MNRAS.128..295R}. The redshifts of the lens and the source are 0.54 and 1.49 \citep{2015Sci...347.1123K}, respectively. Utilizing eight cluster lens models, they inferred $H_{0}$ = 64.80$_{-4.30}^{+4.40}$ km s$^{-1}$ Mpc$^{-1}$. Then, using the two models most consistent with observations, they found $H_{0}$ = 66.60$_{-3.30}^{+4.10}$ km s$^{-1}$ Mpc$^{-1}$. Marking the values of lensing redshift and $H_{0}$ on Fig. 3 in \citet{2022MNRAS.517..576H} and Fig. 4 in \citet{2023A&A...674A..45J}, we find that these results are in good agreement with the evolutionary behaviors of $H_{0}$. Motivated by the $H_{0}$ special behaviors in the late-time universe, we plan to construct a low-redshift subsample based on the Pantheon+ sample to make a reanalyses. According to the previous research \citep{2022MNRAS.517..576H,2023A&A...674A..45J}, and considering the need for a sufficient sample size, we decided to select 1218 SNe with redshift less than 0.30 in the Pantheon+ sample to construct sub-sample (named the lp+ sample). Detailed steps and results are as follows. 

\begin{figure}[h]
        \centering
        \includegraphics[width=0.40\textwidth]{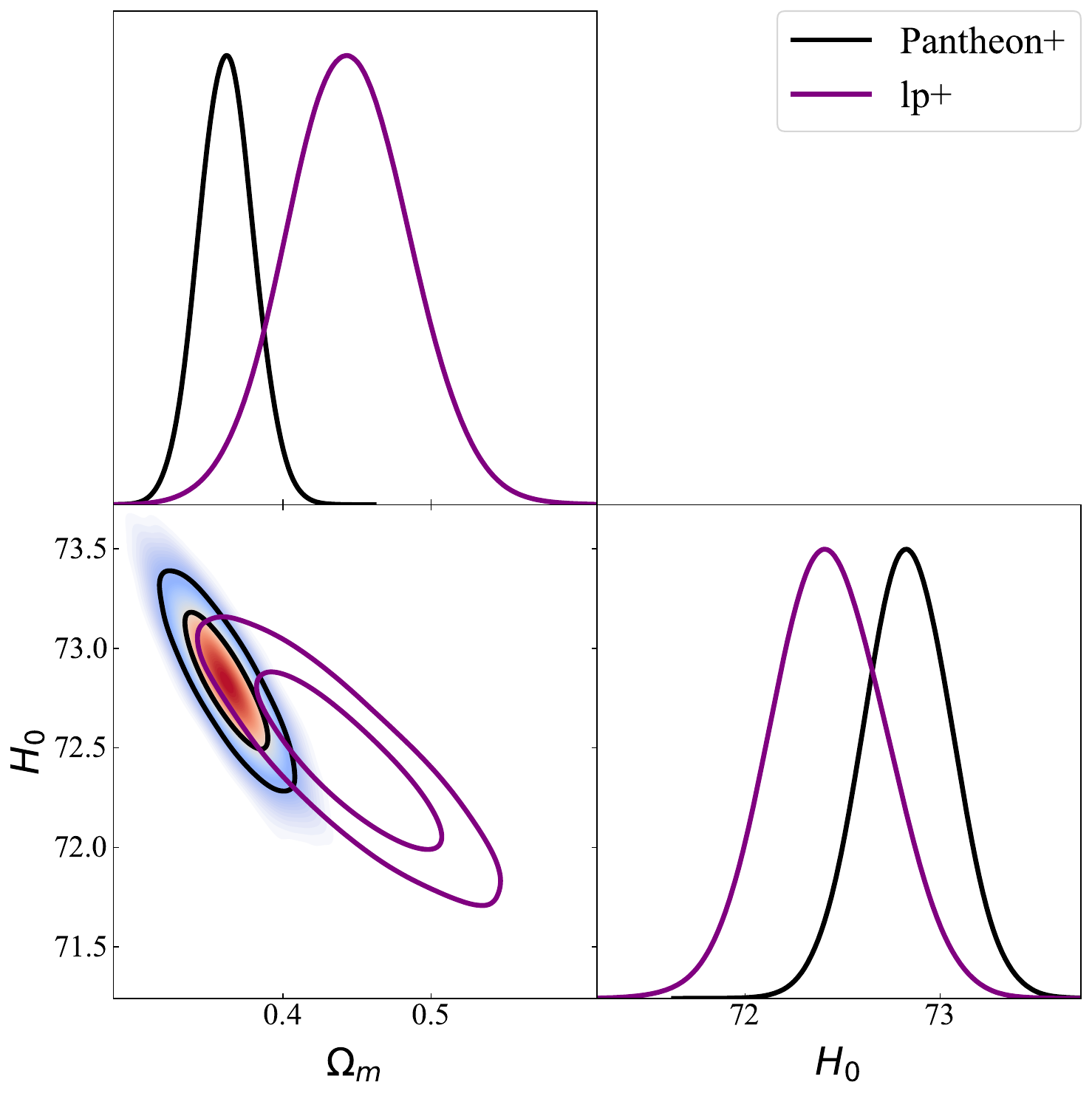}
        \caption{Confidence contours ($1\sigma$ and $2\sigma$) and marginalized likelihood distributions for parameters space ($\Omega_{m}$ and $H_{0}$) in the spatially flat $\rm \Lambda$CDM model from the Pantheon+ sample and the lp+ sample. For the former, the best fits are $\Omega_{m}$ = 0.36$\pm$0.02 and $H_{0}$ = 72.83 $\pm$ 0.23 km s$^{-1}$ Mpc$^{-1}$ (black line). For the latter, the best fits are $\Omega_{m}$ = 0.44$\pm$0.04 and $H_{0}$ = 72.43$\pm$0.30 km s$^{-1}$ Mpc$^{-1}$ (purple line).}
        \label{low1}       
\end{figure}
\begin{figure}[h]
        \centering
        \includegraphics[width=0.42\textwidth]{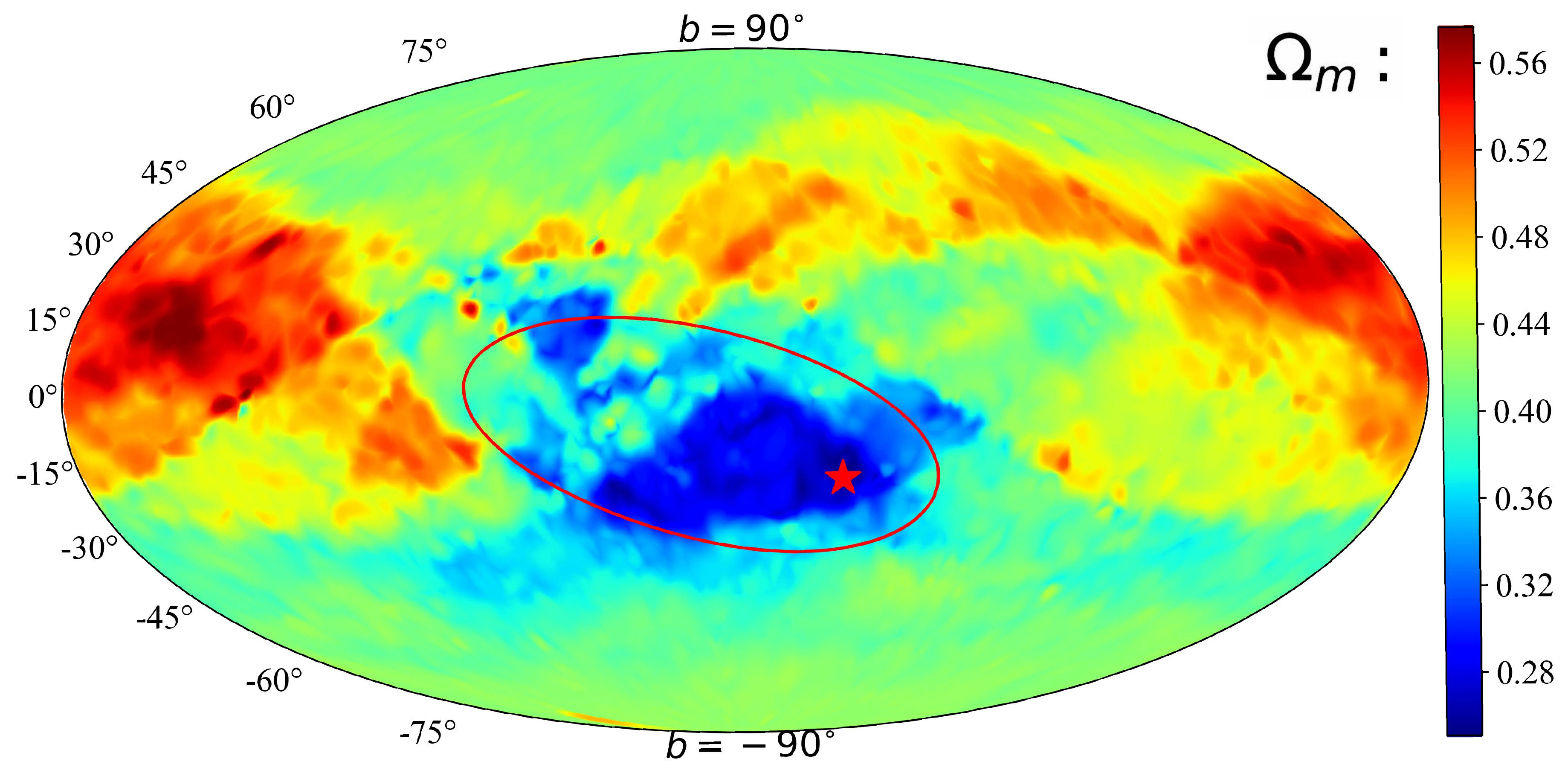}
        \includegraphics[width=0.42\textwidth]{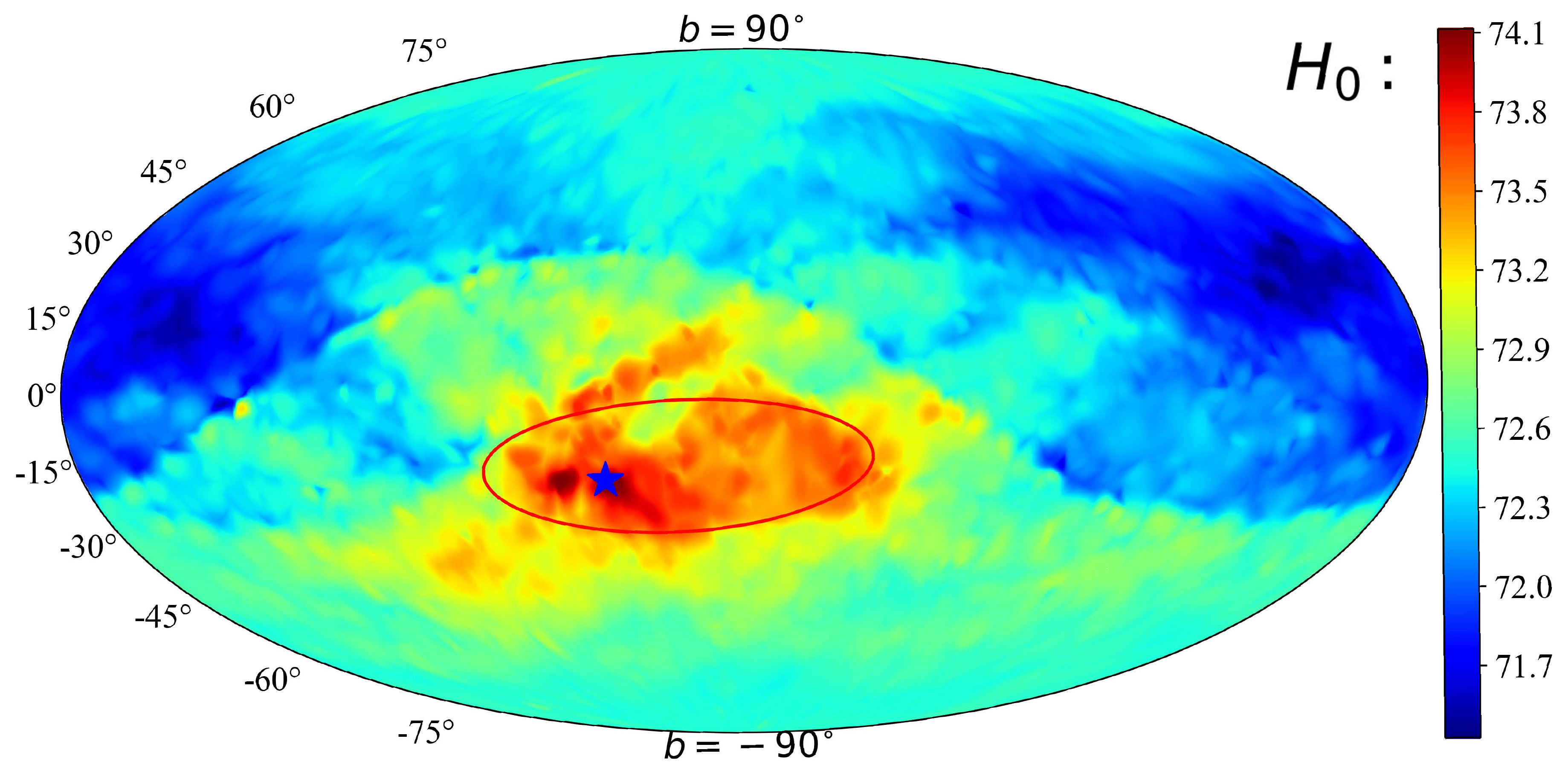}
        \caption{ All-sky distribution of cosmological parameters utilizing the z $<$ 0.30 part of Pantheon+ sample (lp+ sample) combined with the 90$^{\circ}$ RF method. The upper and the lower panel show the results of $\Omega_{m}$ and $H_{0}$, respectively. The corresponding values of $D_\mathrm{max}(\sigma)$ are 4.17$\sigma$ and 4.49$\sigma$, respectively. The star marks the directions of the lowest $\Omega_{m}$ (upper panel) and the largest $H_{0}$ (lower panel), and the red circle outlines the corresponding 1$\sigma$ areas. The directions and 1$\sigma$ areas are parameterized as $\Omega_{m}$ (${26.4^{\circ}}_{-100.9}^{+26.3}$, ${-19.3^{\circ}}_{-17.0}^{+35.4}$), $H_{0}$ (${321.9^{\circ}}_{-33.5}^{+72.5}$, ${-18.9^{\circ}}_{-11.5}^{+16.6}$). }
        \label{F6}       
\end{figure}

\begin{figure}[h]
        \centering
        \includegraphics[width=0.4\textwidth]{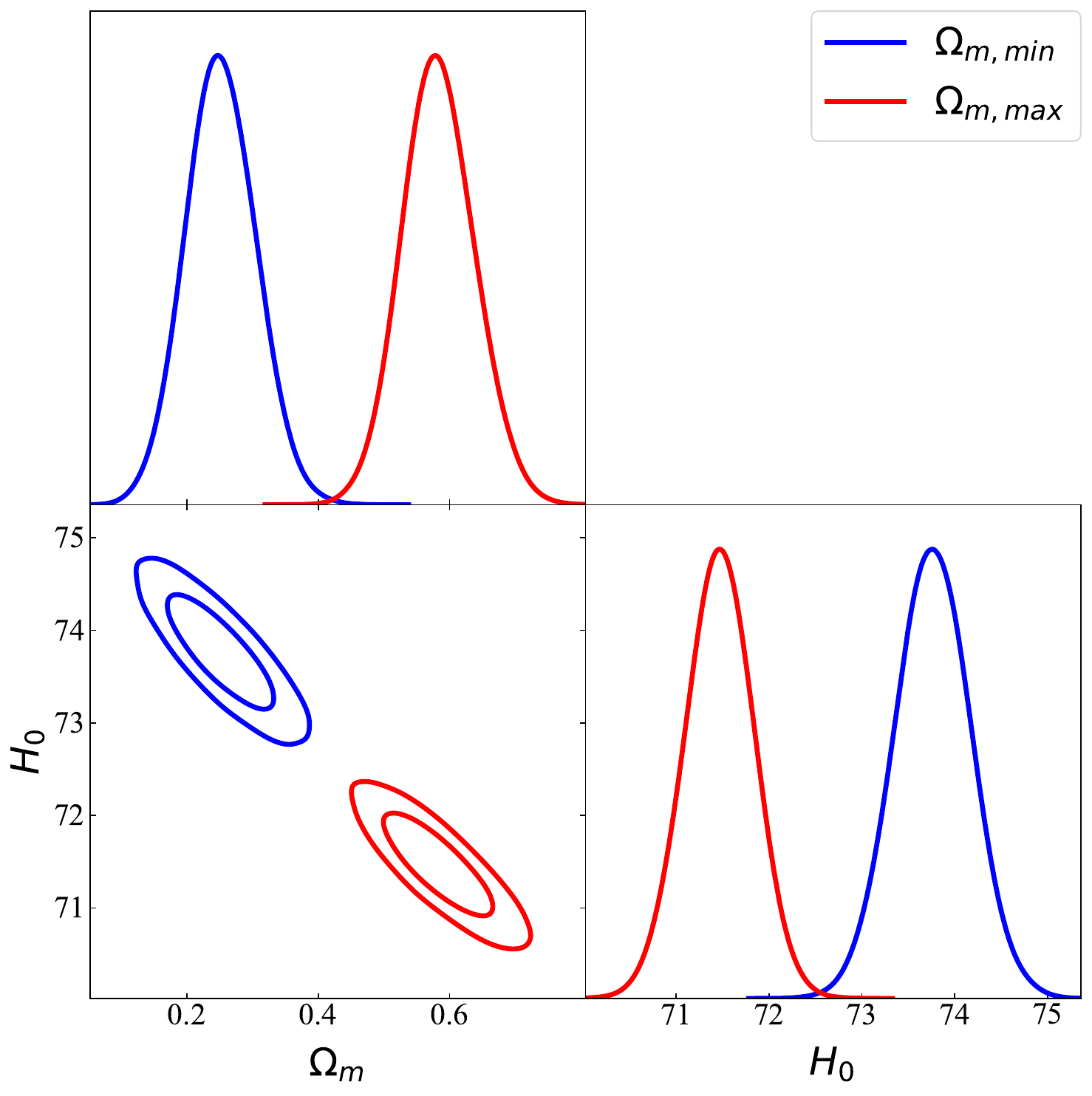}
        \caption{Confidence contours ($1\sigma$ and $2\sigma$) and marginalized likelihood distributions for the parameters space ($\Omega_{m}$ and $H_{0}$) in the spatially flat $\rm \Lambda$CDM model from the SNe Ia subsamples, which corresponds to $\Omega_{m, \mathrm{min}}$ (blue line) and $\Omega_{m, \mathrm{max}}$ (red line). The best fitting results of preferred direction are $\Omega_{m, \mathrm{min}}$ = 0.25$^{+0.05}_{-0.05}$ and $H_{0}$ = 73.76$^{+0.41}_{+0.41}$ km s$^{-1}$ Mpc$^{-1}$ (blue line). The best fitting results of opposite direction are $\Omega_{m, \mathrm{max}}$ = 0.58$^{+0.06}_{-0.05}$ and $H_{0}$ = 71.46$^{+0.37}_{+0.37}$ km s$^{-1}$ Mpc$^{-1}$ (red line).}           
        \label{lowr}       
\end{figure}

\begin{figure}[h]
        \centering
        \includegraphics[width=0.23\textwidth]{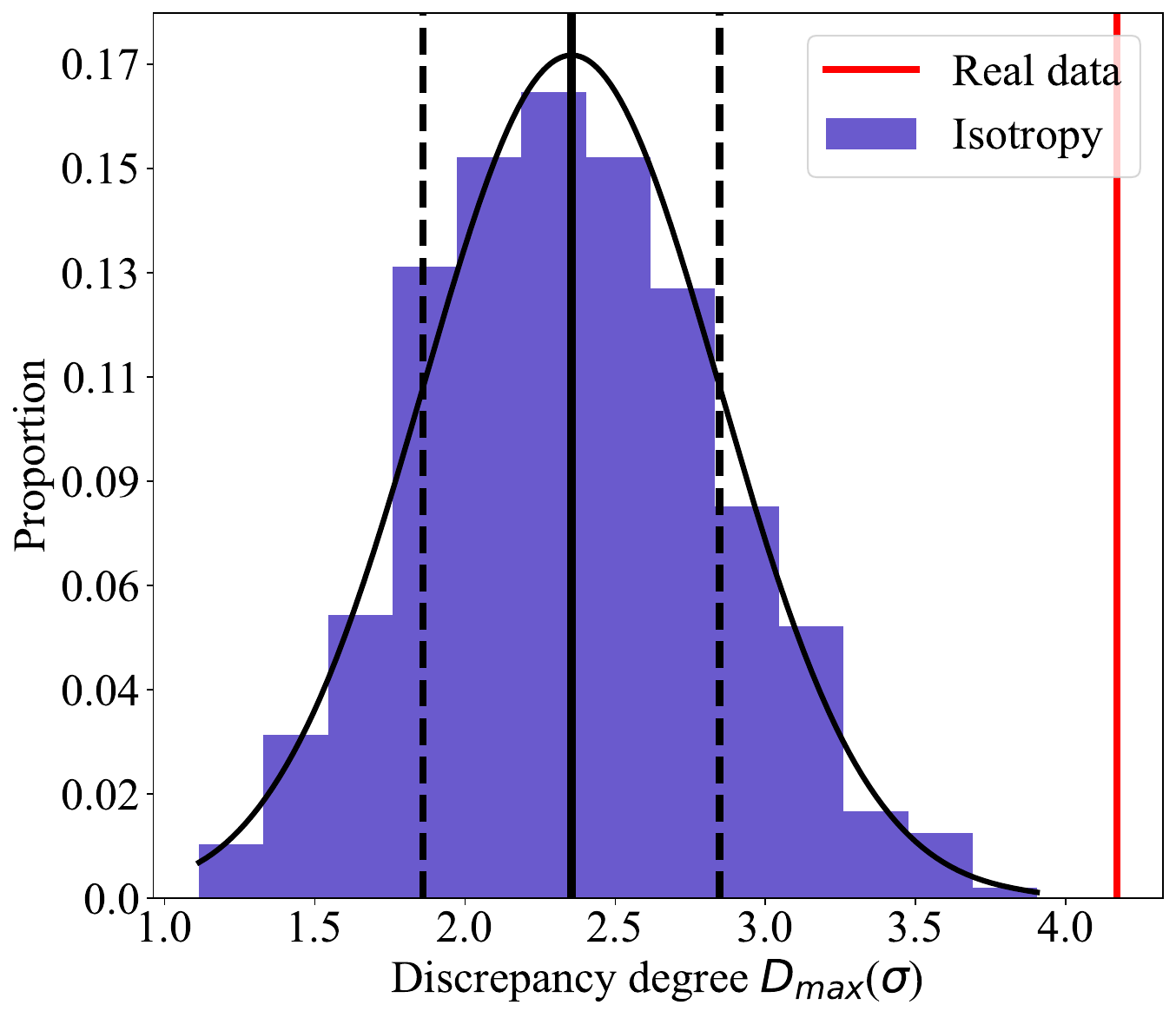}
        \includegraphics[width=0.23\textwidth]{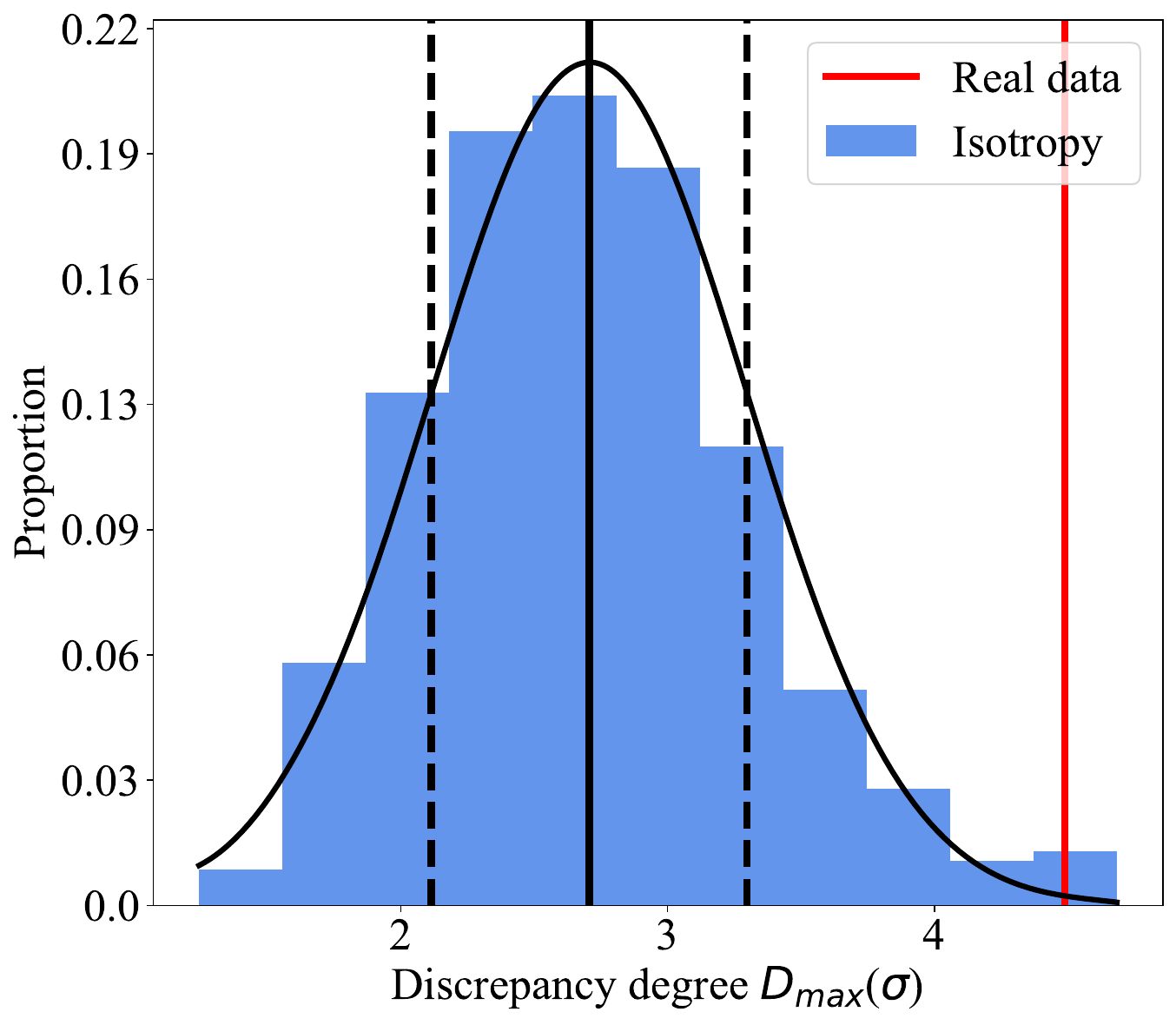}
        \includegraphics[width=0.23\textwidth]{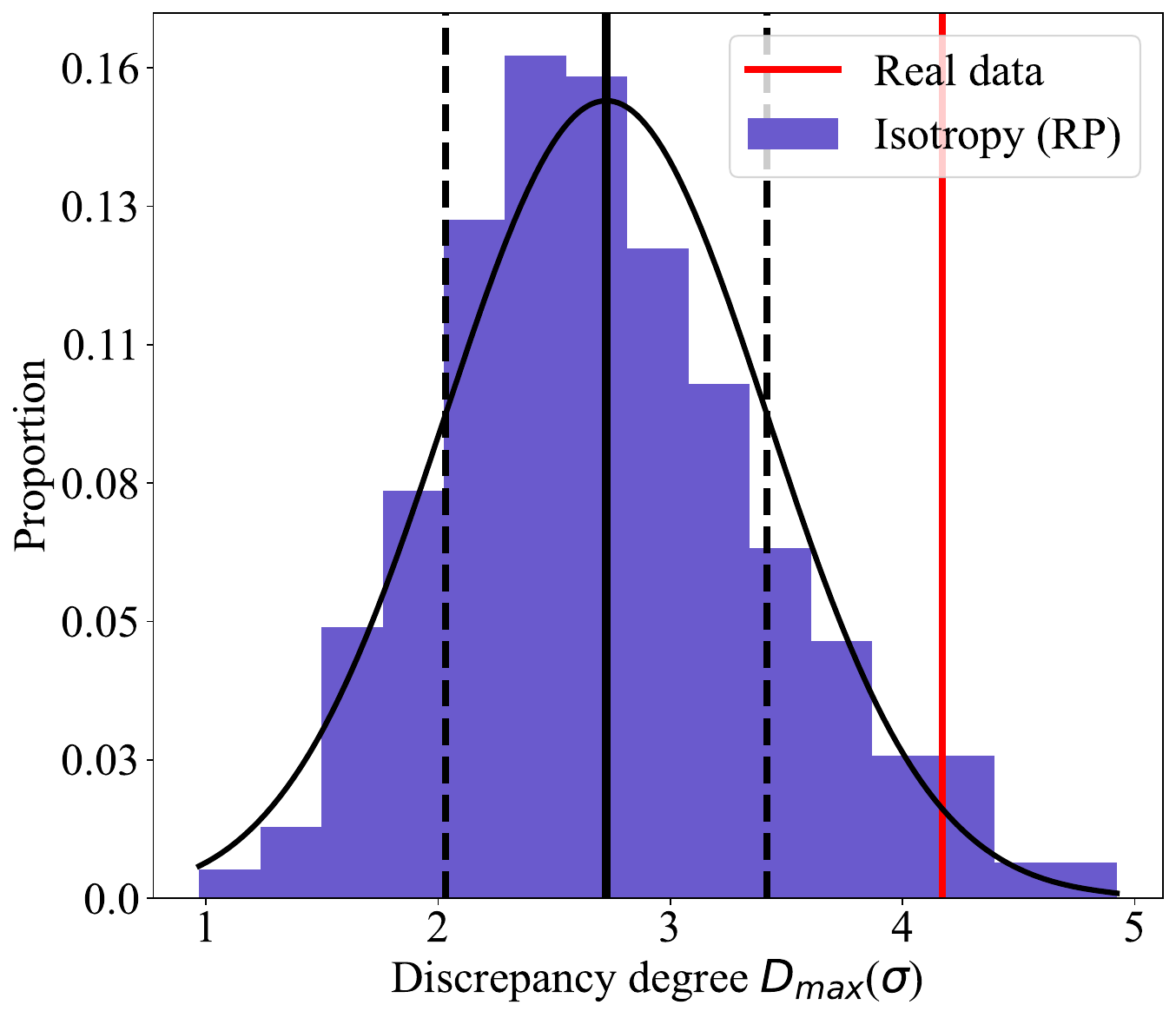}
        \includegraphics[width=0.23\textwidth]{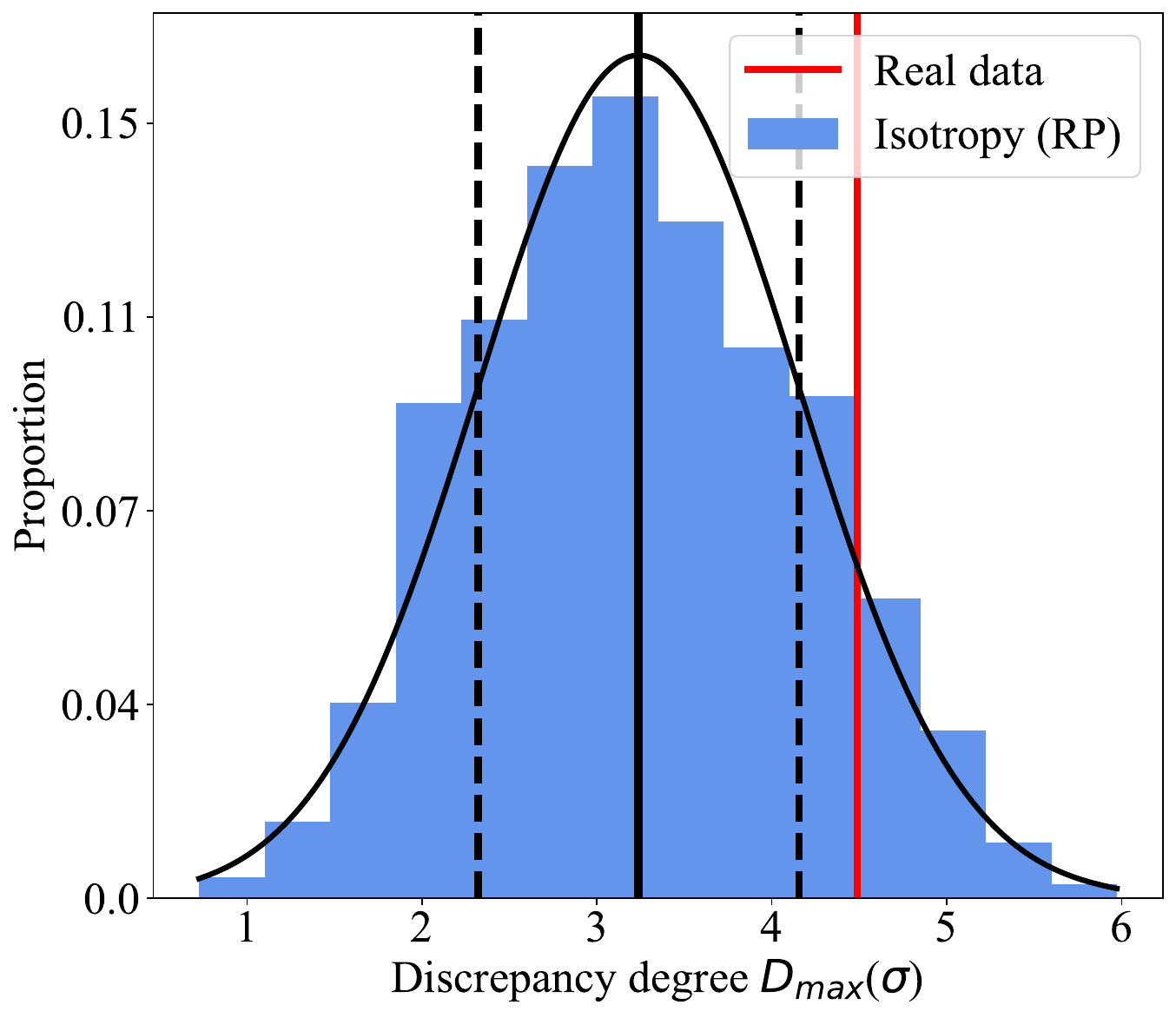}
        \caption{Distribution of discrepancy degree $D_\mathrm{max}$ in 500 simulated isotropic datasets. The upper two panels show the results of statistical isotropic analyses (isotropy). The lower two panels show the results of statistical isotropic analyses that preserve the spatial inhomogeneity of real data (isotropy RP). The purple and blue represent the statistical results of $\Omega_{m}$ and $H_{0}$. The black curve is the best fit to the Gaussian function. The solid black and vertical dashed lines are commensurate with the mean and the standard deviation, respectively. The red lines represent the discrepancy degree from the real data. For the isotropy analyses, the statistical significance of the real data are 3.68$\sigma$ for $\Omega_{m}$ anisotropy and 3.01$\sigma$ for $H_{0}$ anisotropy. For the isotropy RP analyses, the statistical significance of the real data are 2.09$\sigma$ and 1.35$\sigma$ for $\Omega_{m}$ and $H_{0}$ anisotropy, respectively.}
        \label{F7}       
\end{figure}

At first, we give the cosmological constraint in the flat $\Lambda$CDM model using the lp+ sample; that is, $\Omega_{m}$ = 0.44$\pm$0.04 and $H_{0}$ = 72.43$\pm$0.30 km s$^{-1}$ Mpc$^{-1}$, which is consistent with the results from the full Pantheon+ sample and from Fig. 16 in \citet{2022ApJ...938..110B}. For ease of comparison, we give the constraint results of the Pantheon+ and lp+ samples in Fig. \ref{low1}. Then, we reproduced the previous analyses. The all-sky distribution, the cosmological constraints, and the isotropic statistic results are shown in Figs. \ref{F6}, \ref{lowr}, and \ref{F7}, respectively. From Fig. \ref{F6}, it is easy to obtain the ranges of constraints: $\Omega_{m}$ (0.25, 0.58) and $H_{0}$ (71.43, 74.11) km s$^{-1}$ Mpc$^{-1}$. The corresponding differences are $\Delta \Omega_{m}$ = 0.33 and $\Delta{H_{0}}$ = 2.68 km s$^{-1}$ Mpc$^{-1}$. By calculation, we obtain $D_{\mathrm{max},\Omega_{m}}$ = 4.17$\sigma$ and $D_{\mathrm{max}, H_{0}}$ = 4.49$\sigma$. The directions and 1$\sigma$ regions of the local matter underdensity and cosmic anisotropy given by the reanalyses using the lp+ sample are 
\begin{equation}
        (l, b) = ({26.4^{\circ}}_{-100.9}^{+26.3}, {-19.3^{\circ}}_{-17.0}^{+35.4})
        \label{lbF6}
\end{equation}
and 
\begin{equation}
        (l, b) = ({321.9^{\circ}}_{-33.5}^{+72.5}, {-18.9^{\circ}}_{-11.5}^{+16.6}).
        \label{lbF6H}
\end{equation} 


The constraints corresponding to preferred directions of the local matter underdensity and cosmic anisotropy are $\Omega_{m, \mathrm{min}}$ = 0.25$\pm$0.05 and $H_{0}$ = 73.76$\pm$0.41 km s$^{-1}$ Mpc$^{-1}$, and $H_{0, \mathrm{max}}$ = 74.11$^{+0.47}_{-0.48}$ km s$^{-1}$ Mpc$^{-1}$ and $\Omega_{m}$ = 0.27$\pm$0.06. In Fig. \ref{lowr}, we show the confidence contours in the preferred and opposite directions of the local matter underdensity. The isotropic statistical results from the lp+ sample are shown in Fig. \ref{F7}. For the isotropy analysis, the statistical significance of the real data are 3.68$\sigma$ for $\Omega_{m}$ anisotropy (upper, purple panel) and 3.01$\sigma$ for $H_{0}$ anisotropy (upper, blue panel). The isotropy RP analysis gives the statistical significance of the real data, which is 2.09$\sigma$ for $\Omega_{m}$ anisotropy (lower, purple panel) and 1.35$\sigma$ for $H_{0}$ anisotropy (lower, blue panel).

\subsection{Different screening angles \tm }\label{theta}
\begin{figure}[h]
        \centering
        \includegraphics[width=0.43\textwidth]{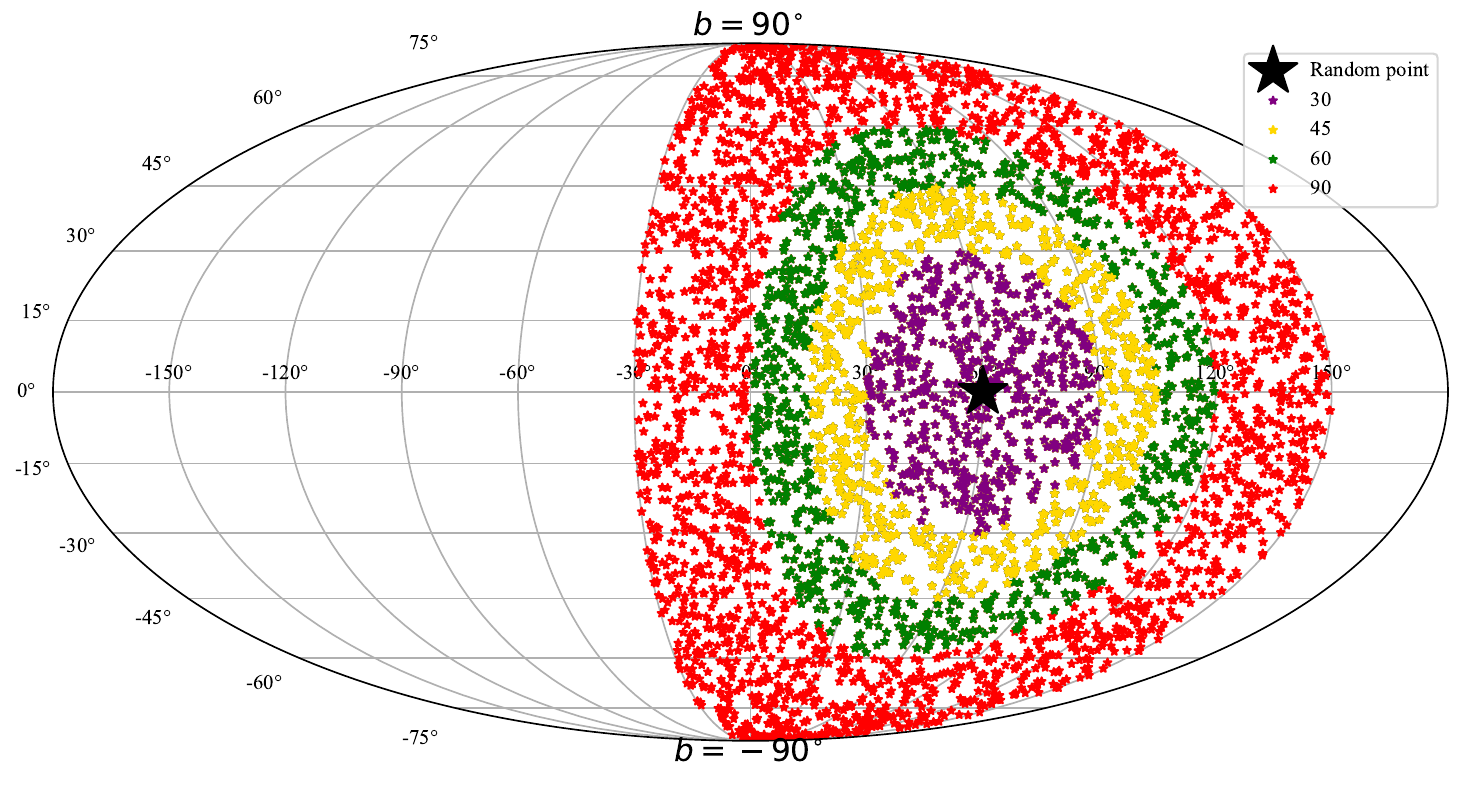}
        \caption{Schematic diagram of the RF method with different screening angles including 30$\degr$, 45$\degr$, 60$\degr,$ and 90$\degr$. Black star represents the random direction $\hat{D}$ (60$\degr$, 0$\degr$).}
        \label{F8}       
\end{figure}

Theoretically, if there are enough SNe data distributed at different redshifts in a certain direction, the constraints of the cosmological parameters in this direction can be obtained \citep{2016MNRAS.460..617L,2018PhRvD..97l3515D,2023CQGra..40i4001K}. Due to the limitation of the number of observational data, it is currently impossible to directly give the result of the limitation of cosmological parameters in a certain direction. Therefore, the 90$\degr$ RF method is usually used to derive the cosmological constraint of the single direction, that is the HC method \citep{2007AA...474..717S}. In this subsection, we try to find out the limit value of $\theta_\mathrm{max}$ that is suitable for the Pantheon+ sample by mapping the all-sky distributions of the cosmological parameters $\Omega_{m}$ and $H_{0}$ with different $\theta_\mathrm{max}$. Considering the limitations of computing time, only three angles (30$\degr$, 45$\degr$ and 60$\degr$) are chosen. Take $\hat{D}$ (60$\degr$, 0$\degr$) as an example, Figure \ref{F8} shows the schematic diagram of the RF method with different $\theta_\mathrm{max}$. Smaller $\theta_\mathrm{max}$ will reduce the overlapping data between sub-samples generated by adjacent random directions. The corresponding all-sky distributions will present more details on the current state of the Universe. For the Pantheon+ sample, lowering $\theta_\mathrm{max}$ can be expected to produce some poorly fitting results. Therefore, we give a constrained culling of possibly poor constraints. Here, we set a loose condition of $0 < \Omega_{m} < 1.00$. The final results are shown in Fig. \ref{F9}. For the screening angles 30$\degr$, 45$\degr,$ and 60$\degr$, the proportions of the wrong fitting results are 30.34\%, 6.47\%, and 0.70\%, respectively. From Fig. \ref{F9}, we find that \tm = 60$\degr$ seems to be the limit that the Pantheon+ sample can bear. At this time, nearly 100\% (99.30\%) of the random directions ($\hat{D}$) can be reliably constrained.

\begin{figure}[h]
        \centering
        \includegraphics[width=0.45\textwidth]{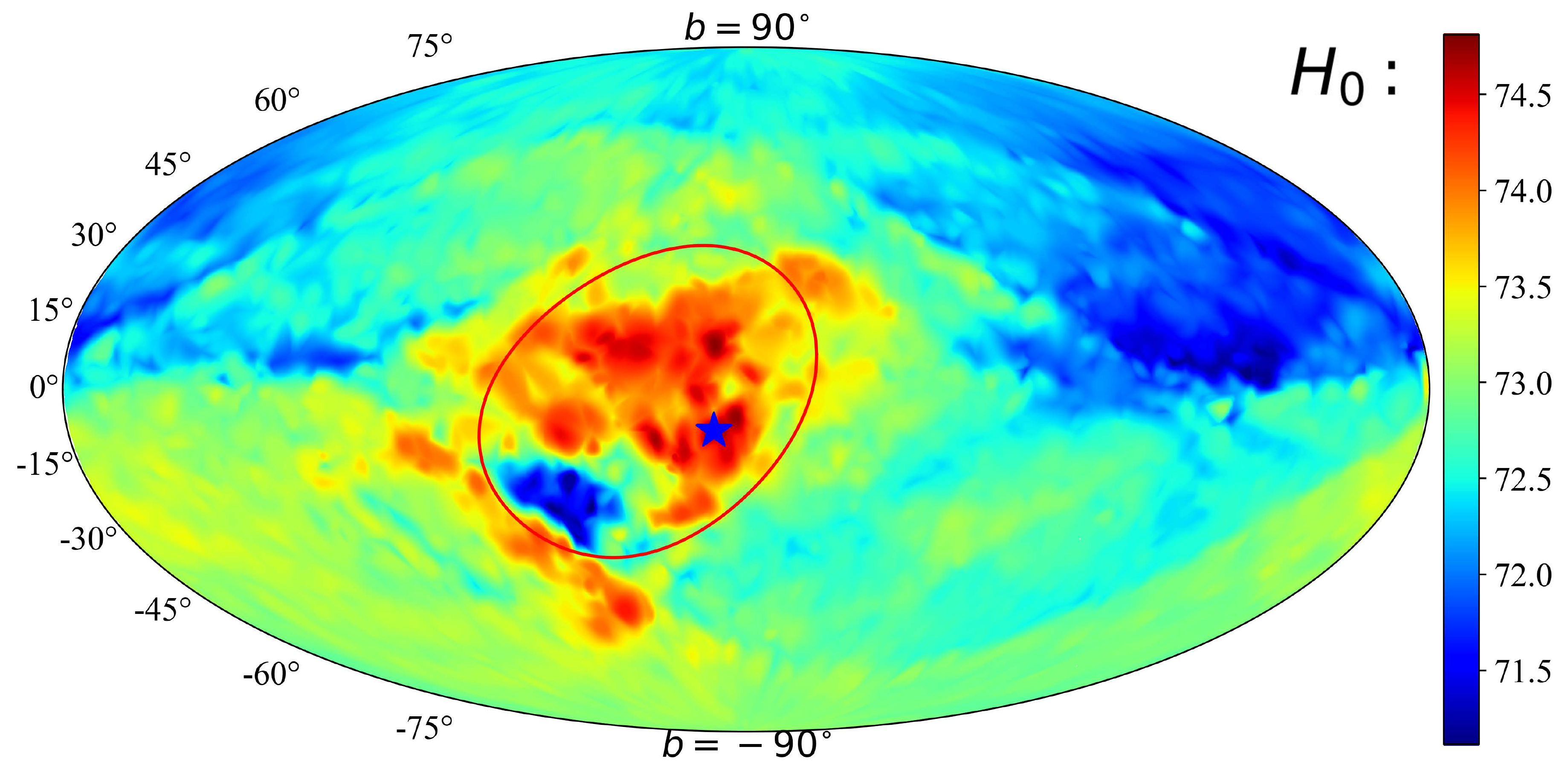}
        \caption{All-sky distribution of $H_{0}$ utilizing Pantheon+ sample combined with 60$^{\circ}$ RF method. The direction and 1$\sigma$ area are parameterized as (${351.4^{\circ}}_{-64.2}^{+28.0}$, ${-8.8^{\circ}}_{-27.8}^{+40.3}$). The statistical discrepancy $D_\mathrm{max}$ is 4.39$\sigma$.}
        \label{F10}       
\end{figure}

\begin{figure}[htp]
        \centering
        \includegraphics[width=0.4\textwidth]{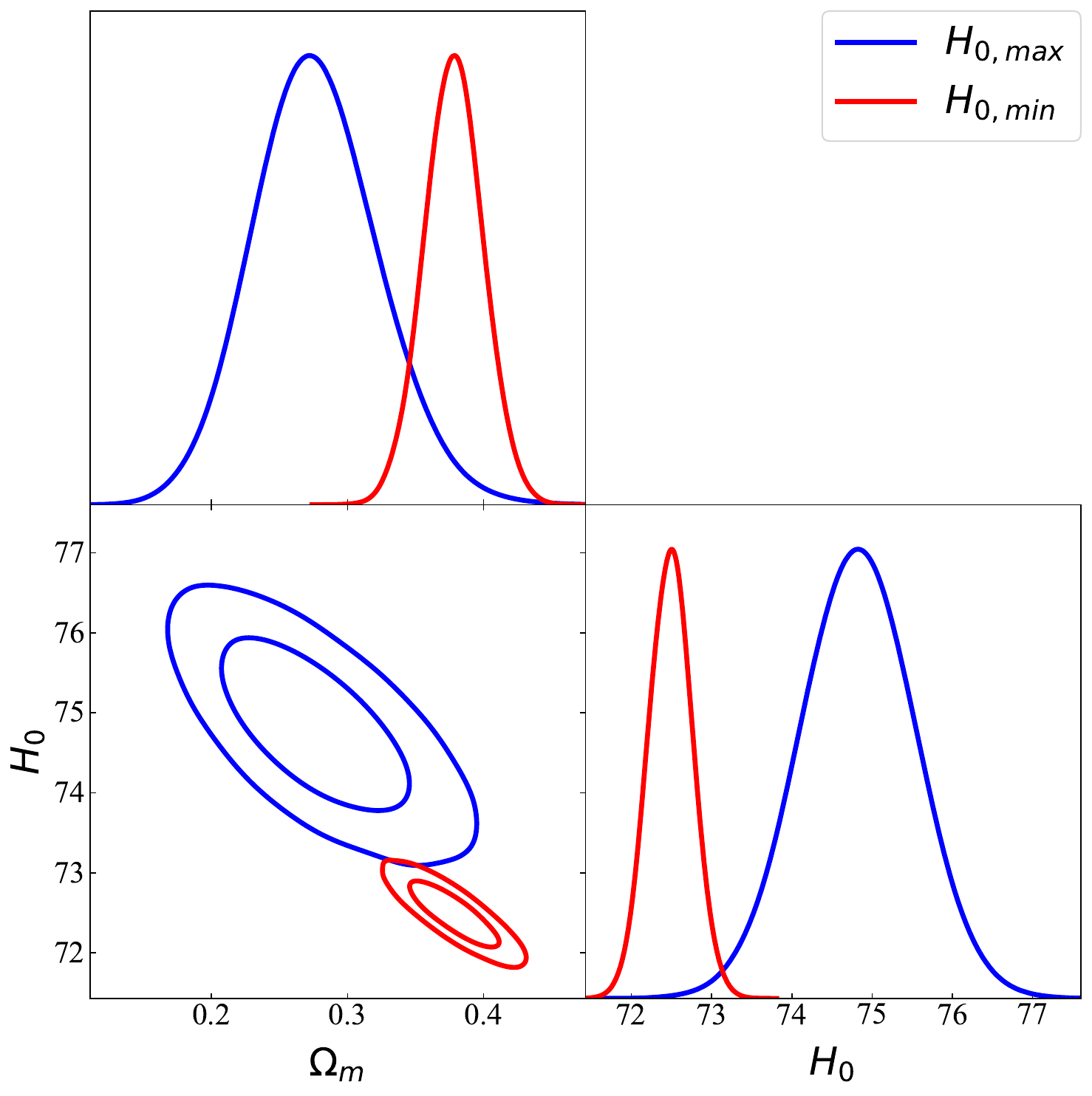}
        \caption{Confidence contours ($1\sigma$ and $2\sigma$) and marginalized likelihood distributions for parameter space ($\Omega_{m}$ and $H_{0}$) in the spatially flat $\rm \Lambda$CDM model from the SNe Ia subsamples, which corresponds to $H_{0, \mathrm{min}}$ (red line) and $H_{0, \mathrm{max}}$ (blue line). The best fitting results of preferred direction are $H_{0, \mathrm{max}}$ = 74.81$^{+0.71}_{-0.71}$ km s$^{-1}$ Mpc$^{-1}$ and $\Omega_{m}$ = 0.28$^{+0.05}_{-0.04}$ (blue line). The best fitting results of opposite directions are $H_{0, \mathrm{min}}$ = 72.49$^{+0.27}_{-0.27}$ km s$^{-1}$ Mpc$^{-1}$ and $\Omega_{m}$ = 0.38$^{+0.02}_{-0.02}$ (red line).}
        \label{F60}       
\end{figure}

\begin{figure}[h]
        \centering
        \includegraphics[width=0.24\textwidth]{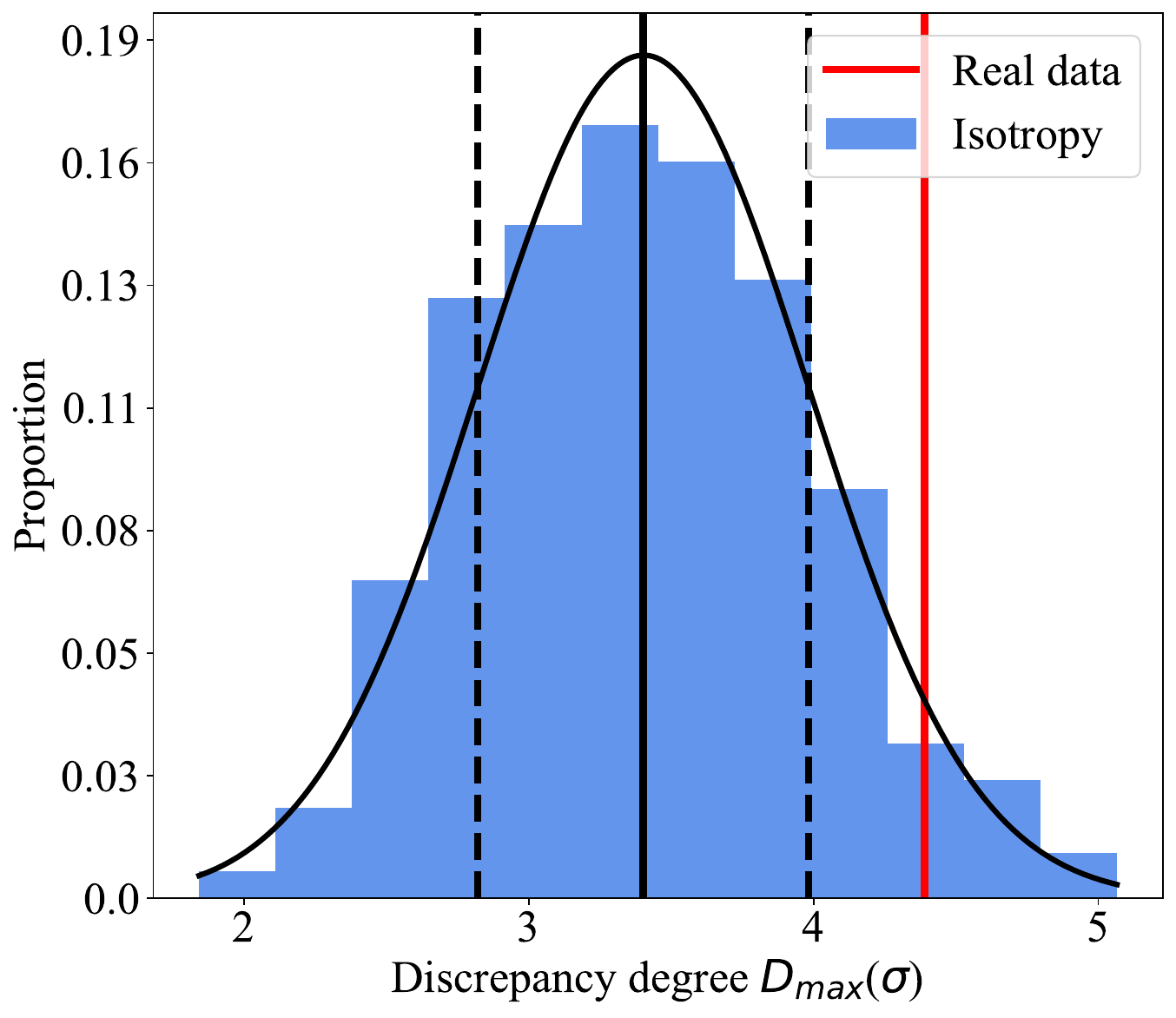}
        \includegraphics[width=0.24\textwidth]{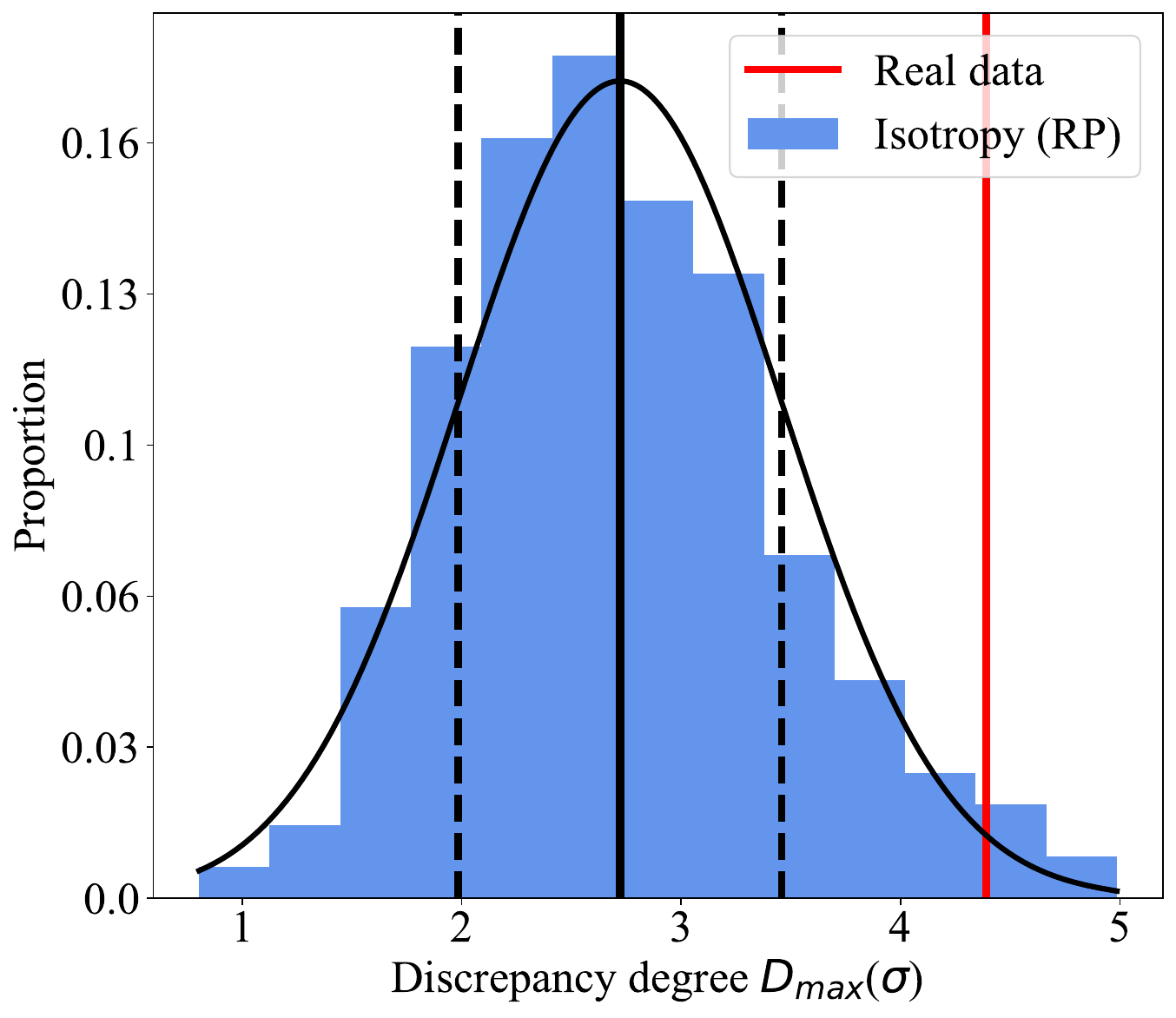}
        \caption{Distribution of discrepancy degree $D_\mathrm{max}$ obtained from $H_{0}$ distribution in 500 simulated isotropic datasets. The left panel and right panel show the results of isotropy and isotropy RP analyses, respectively. The black curve is the best fit to the Gaussian function. The solid black and vertical dashed lines are commensurate with the mean and the standard deviation, respectively. The red lines represent the discrepancy degree from the real data. The statistical significances of the real data are 1.70$\sigma$ for the isotropy analysis and 2.27$\sigma$ for the isotropy RP analysis.}
        \label{F11}       
\end{figure}

In Fig. \ref{Ommax}, we give the best fits for the hemispheres where $\Omega_\mathrm{m, min}$ and $\Omega_\mathrm{m, max}$ are located. We find that taking into account narrowing the screening angle \tm significantly increases the 1$\sigma$ error of $\Omega_{m}$ constraints. The reduced chi-square ($\chi^{2}_{r}$) corresponding to $\Omega_\mathrm{m, min}$ is 0.69, which is much smaller than 1.00. Therefore, we only show the analysis results of $H_{0}$. The all-sky distribution, cosmological constraints and isotropic statistical results are displayed in Figs. \ref{F10}, \ref{F60}, and \ref{F11}, respectively. As shown in Fig. \ref{F10}, the preferred direction of cosmic anisotropy given by the $H_{0}$ distribution is (${351.4^{\circ}}_{-64.2}^{+28.0}$, ${-8.8^{\circ}}_{-27.8}^{+40.3}$), and $D_\mathrm{max}$ is 4.39$\sigma$. The corresponding constraints of a preferred direction are $H_{0, \mathrm{max}}$ = 74.81$^{+0.71}_{-0.71}$ km s$^{-1}$ Mpc$^{-1}$ and $\Omega_{m}$ = 0.28$^{+0.05}_{-0.04}$. The statistical isotropy results show that the statistical confidences of the real data are 1.70$\sigma$ for the isotropy analysis and 2.27$\sigma$ for the isotropy RP analysis.

All results in this section are summarized in Table \ref{T1}. Additionally, we provide the reduced chi-square $\chi^{2}_{r}$ for the anisotropic direction.


\begin{table*}\footnotesize 
        \caption{ Detailed information of analysis results from the all-sky distribution of cosmological parameters. \label{T1}}
        \centering
        \begin{spacing}{1.2}
                \begin{tabular}{ccccccccc} 
                        \hline\hline
                        Sample & $\theta_\mathrm{max}$$^{a}$   & $\Omega_{m,min}$ & $H_{0}$ & $D_\mathrm{max}$$^{b}$ & $\alpha$$^{c}$ & $\beta$$^{d}$ & Direction & $\mathrm{\chi^{2}_{r}}$ \\
                        &   &  & (km s$^{-1}$ Mpc$^{-1}$) & ($\sigma$) & ($\sigma$) & ($\sigma$) & (l, b) & \\
                        \hline
                        Pantheon+ & 90$\degr$  & 0.29$^{+0.03}_{-0.02}$ & 74.11$^{+0.40}_{-0.40}$ &3.29 & 2.78 & 2.34 & $({308.4^{\circ}}_{-48.7}^{+47.6}, {-18.2^{\circ}}_{-28.8}^{+21.1})$ & 0.94 \\
                        Pantheon+ & 60$\degr$  & 0.23$^{+0.20}_{-0.17}$ & 74.05$^{+1.00}_{-1.00}$ & - & - & - & -- & 0.69$^{*}$ \\
                        lp+ & 90$\degr$  & 0.25$^{+0.05}_{-0.05}$ & 73.76$^{+0.41}_{+0.41}$  & 4.17  &  3.68 & 2.09  & (${26.4^{\circ}}_{-100.9}^{+26.3}$, ${-19.3^{\circ}}_{-17.0}^{+35.4}$) & 0.94 \\
                        \hline\hline
                        Sample & $\theta_\mathrm{max}$$^{a}$   & $H_{0,max}$ & $\Omega_{m}$ & $D_\mathrm{max}$$^{b}$&  $\alpha$$^{c}$ & $\beta$$^{d}$ &  Direction & $\mathrm{\chi^{2}_{r}}$ \\
                        &   & (km s$^{-1}$ Mpc$^{-1}$) & & ($\sigma$) &  ($\sigma$)  &  ($\sigma$) &  (l, b) & \\
                        \hline
                        Pantheon+ & 90$\degr$  & 74.26$^{+0.40}_{-0.39}$  & 0.30$^{+0.03}_{-0.03}$ & 4.48 &  3.96 & 3.15  & $({313.4^{\circ}}_{-18.2}^{+19.6}, {-16.8^{\circ}}_{-10.7}^{+11.1})$ & 0.98 \\
                        Pantheon+ & 60$\degr$  & 74.81$^{+0.71}_{-0.71}$  & 0.28$^{+0.05}_{-0.04}$ & 4.39 & 1.70 & 2.27 & (${351.4^{\circ}}_{-64.2}^{+28.0}$, ${-8.8^{\circ}}_{-27.8}^{+40.3}$) & 0.99 \\
                        lp+ & 90$\degr$   & 74.11$^{+0.47}_{-0.48}$ & 0.27$^{+0.06}_{-0.06}$ &  4.49 &  3.01 & 1.35 & (${321.9^{\circ}}_{-33.5}^{+72.5}$, ${-18.9^{\circ}}_{-11.5}^{+16.6}$) & 1.14\\
                        \hline\hline
                \end{tabular}
                \begin{itemize} 
                        \tiny
                        \item[\rm{$^{a}$}] $\theta_\mathrm{max}$ is the screening angle.
                        \item[\rm{$^{b}$}] $D_\mathrm{max}$ represents the degree of deviation from the cosmological principle; the larger the value, the higher the degree of deviation.
                        \item[\rm{$^{c}$}] $\alpha$ indicates the statistical significance of real data from the isotropy analysis.
                        \item[\rm{$^{d}$}] $\beta$ indicates the statistical significance of real data from the isotropy RP analysis.
                        \item[\rm{$^{*}$}] For the RF (60$\degr$) results of the Pantheon+ sample, the reduced chi-square ($\chi^{2}_{r}$) corresponding to $\Omega_\mathrm{m, min}$ is 0.69, which is much smaller than 1.00. Therefore, the RF (60$\degr$) results are not given.
                \end{itemize}
        \end{spacing}
\end{table*}

\section{Discussion}
All-sky distributions of cosmological parameters ($\Omega_{m}$ and $H_{0}$) from the total sample show that there is an obvious local underdensity area and a preferred direction of cosmic expansion. From Fig. \ref{F4}, we can find that there are some weird features that may result from fluctuations in the matter density, or fluctuations in the Hubble expansion \citep{Gurzadyan2023}. In the upper panel, the red and blue areas represent areas of higher and lower material density, respectively. In the lower panel, the red areas represent regions of higher Hubble expansion. The continuity of these structures suggests that the results may not be sensitive to individual SNe Ia. Combining these two panels of Fig. \ref{F4}, we find that the local underdensity in the upper panel overlaps with the higher Hubble expansion (cosmic anisotropy) in the lower panel. This may imply that local matter density might be responsible for the anisotropy of the accelerated expansion of the Universe. The corresponding 1$\sigma$ regions are quite obvious. In other words, the 1$\sigma$ range of $\Omega_{m}$ is significantly larger than that of $H_{0}$. The main reason for this phenomenon might be that these two parameters have different sensitivities on the redshift. The inhomogeneous matter-density distribution could also affect the decelaration parameter $q_{0}$, causing a larger $q_{0}$ in the local underdensity. Theoretically, the maximum $q_{0}$ direction should be consistent with the local underdensity direction. For the determination of $H_{0}$ value with local distance indicators, the observed $H_{0}$ values depend on the average matter density within the distance range covered \citep{2020A&A...633A..19B}. Combining the $\Omega_{m}$ distribution, we can find that there might be a region of low matter density, which leads to a smaller average matter density which makes the $H_{0}$ measurements higher. Therefore, the local underdensity can exacerbate the magnitude of the Hubble Tension \citep{2020A&A...633A..19B,2022MNRAS.517..576H,2023A&A...674A..45J,2023Univ....9...94H}. In other words, accounting for the local underdensity or cosmic anisotropy could alleviate the current Hubble tension.

\begin{figure}[h]
        \centering
        \includegraphics[width=0.45\textwidth]{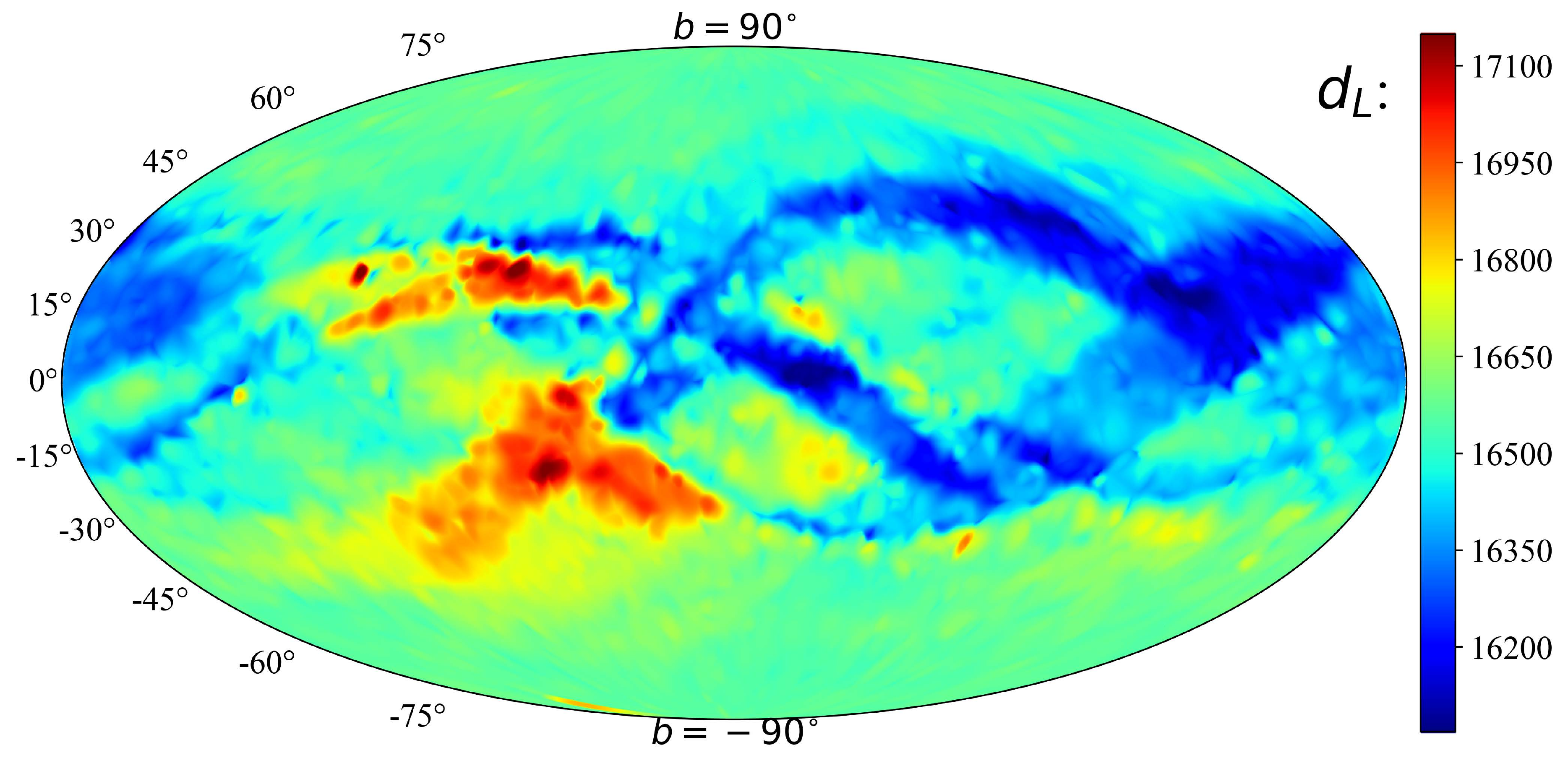}
        \caption{All-sky distribution of luminosity distance $d_{L}$ utilizing the Pantheon+ sample combined with the 90$^{\circ}$ RF method. Here, we fix $z$ = 2.26.}
        \label{DLD}       
\end{figure}

\begin{figure}[h]
        \centering
        \includegraphics[width=0.35\textwidth]{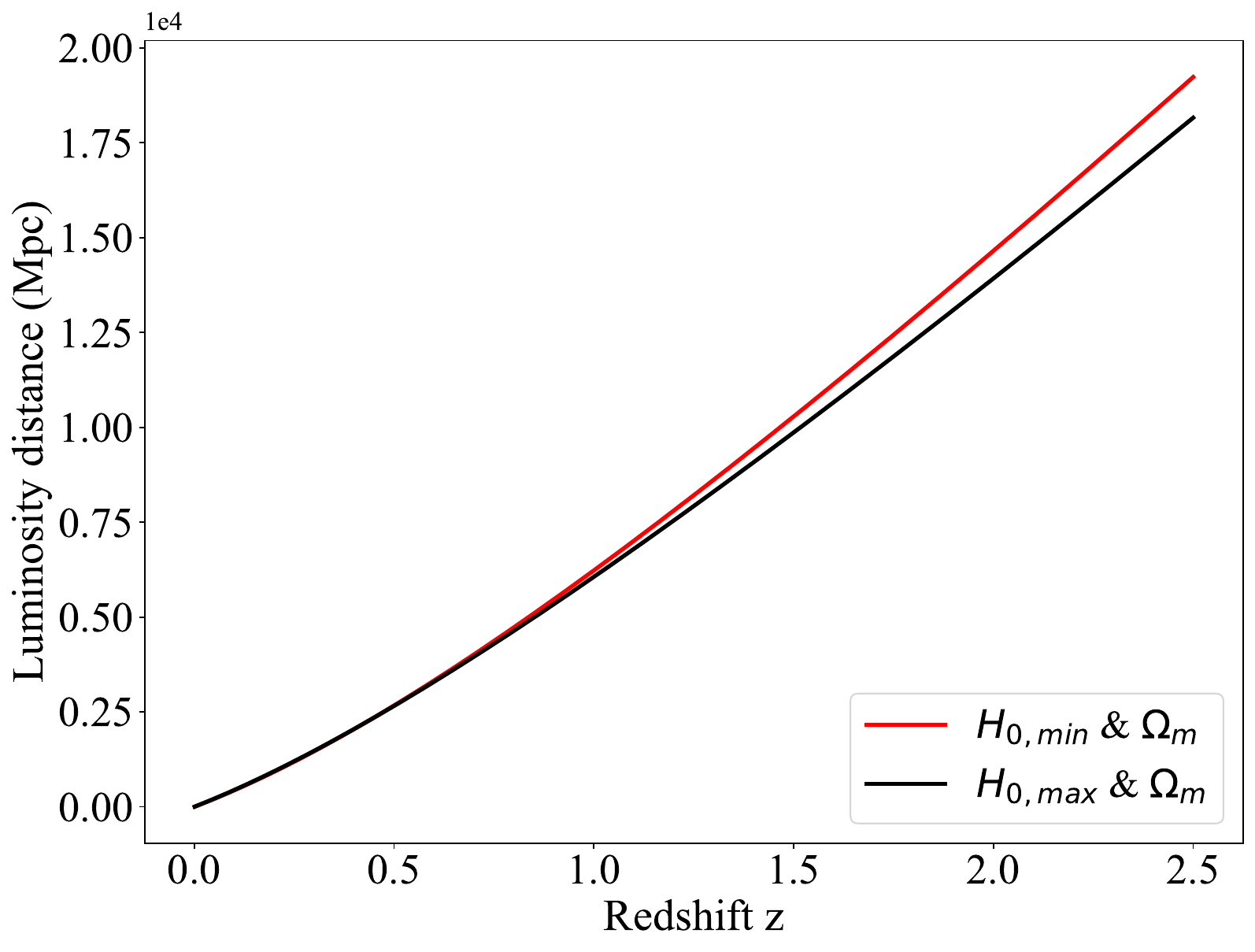}
        \caption{Relationships between $d_{L}$ and $z$ of two anisotropic hemispheres using the two sets of best fits ($H_\mathrm{0,min}$, $\Omega_{m}$  and $H_\mathrm{0,max}$, $\Omega_{m}$) in the $H_{0}$ distribution.}
        \label{DL}       
\end{figure}

From Fig. \ref{F5}, we find that the isotropy analysis shows an obvious statistical significance, especially the study of $H_{0}$, which is nearly 4$\sigma$. Afterwards, considering the spatial inhomogeneity of the real sample, we performed the isotropy RP analysis and find a slight decrease in statistical significance. This suggests that inhomogeneous spatial distribution of a real sample can increase the deviation from isotropy. In addition, combining all the statistical results, we find the statistical significations of $H_{0}$ anisotropy are more obvious than that of $\Omega_{m}$ anisotropy. From the distributions of a single parameter, we find an obvious anisotropy. But the effect of one on the other might be canceled out, making the all-sky distribution of luminosity distances isotropic. Therefore, in order to clear up this doubt, we plotted the all-sky distribution of luminosity distances when $z$ is set to 2.26 and give the relationships between $d_{L}$ and $z$ of two anisotropic hemispheres, as shown in Figs. \ref{DLD} and \ref{DL}, respectively. The results of these two figures show that the all-sky distribution of luminosity distances is also anisotropic and the $d_L-z$ relationships obtained from two anisotropic hemispheres have obvious differences. The direction of larger luminosity distance is consistent with the anisotropy direction from the Pantheon+ sample. All in all, our investigations from the Pantheon+ sample display an inhomogeneous and anisotropic universe, and the statistical results show no low confidence level. 

\begin{figure}[h]
        \centering
        \includegraphics[width=0.35\textwidth]{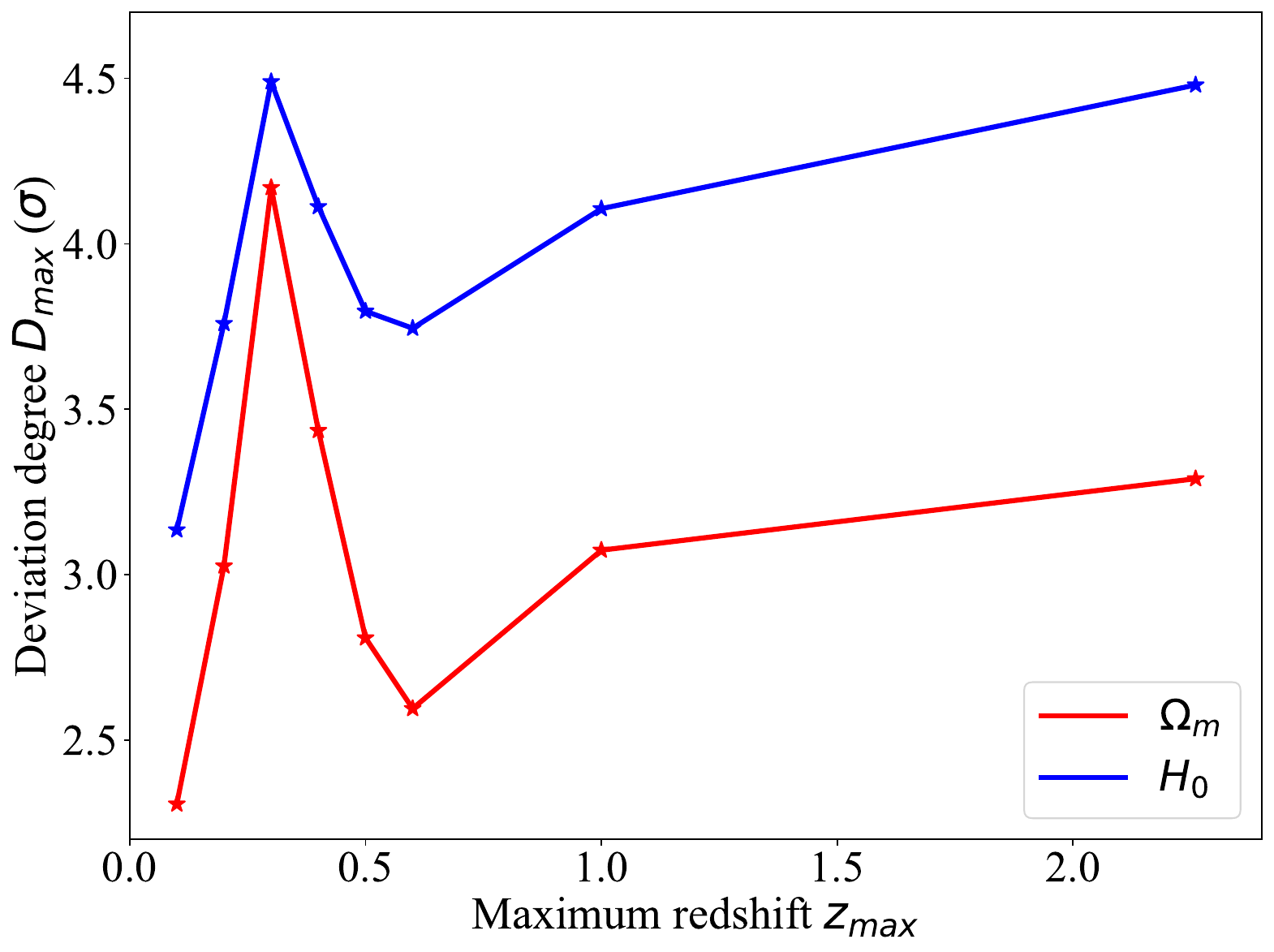}
        \caption{Relationship between discrepancy degree $D_\mathrm{max}$ and $z_\mathrm{max}$.}
        \label{Dz}       
\end{figure}

Motivated by recent research \citep{2020PhRvD.102j3525K,2021PhRvD.103j3509K,2021ApJ...912..150D,2022Galax..10...24D,2022arXiv220611447C,2022MNRAS.517..576H,2022PhRvD.106d1301O,2023A&A...674A..45J,2023arXiv230112725M} hinting that $H_{0, z}$ might evolve with redshift, we conducted reanalyses using the low-redshift (lp+) sample. If $H_{0, z}$ evolves with redshift (may be caused by the local void, and so on), then the measured $H_{0}$ of low-redshift and high-redshift SNe samples are different. In that case, if the redshift distribution is not uniform across the sky, this could create biases in the anisotropy detection, especially when the full sample is used; moreover, the redshift range is wider. The value of $H_{0}$ obtained from the higher redshift hemisphere is smaller than that obtained from the lower redshift hemisphere. Severely uneven redshift distribution might lead to an increase in the degree of anisotropy or a bias in the anisotropy detection. The reanalysis results are similar to those of the Pantheon+ sample and verify our previous findings. From the results of the lp+ sample (Fig. \ref{F6}), we find that directions of the local matter underdensity and cosmic anisotropy given by the lp+ sample are inconsistent. Comparing the results in Figs. \ref{total} and \ref{lowr}, we can find that the best fits of the two anisotropic hemispheres obtained from the lp+ sample are significantly worse than the results of the total sample. In Fig. \ref{Dz}, we also show the relationship between $D_\mathrm{max}$ and $z_\mathrm{max}$, the detailed information are display in Table \ref{TT}. From Fig. \ref{Dz}, we can find that $D_\mathrm{max}$ changes with $z_\mathrm{max}$, and has a maximum value near $z_\mathrm{max}$ = 0.30. This might imply that the structure of the Universe changes with redshift. In addition, we also find that the influence of redshift on the all-sky distribution of $\Omega_{m}$ is larger than that of $H_{0}$. Finally, from the statistical results of lp+ sample, we also find that the statistical significance obtained from the isotropy RP analysis are lower than that from the isotropy analysis. This finding confirms that inhomogeneous spatial distribution of a real sample can increase the deviation from isotropy.

\begin{table}\footnotesize
        \caption{Detailed information on the investigation of the relationship between the discrepancy degree $D_\mathrm{max}$ and $z_{max}$. \label{TT}}
        \centering
        \begin{spacing}{1.2}
                \begin{tabular}{cccc}
                        \hline\hline
                        $z_\mathrm{max}$ & Number & $D_{max}$ ($\Omega_{m}$)& $D_{max}$ ($H_{0}$)  \\
                        &  & ($\sigma$) & ($\sigma$)  \\
                        \hline
                        0.10 & 747  & 2.31 & 3.14 \\
                        0.20 & 962  & 3.03 & 3.76 \\
                        0.30 & 1218 & 4.17 & 4.49 \\
                        0.40 & 1399 & 3.43 & 4.11 \\
                        0.50 & 1497 & 2.81 & 3.80 \\
                        0.60 & 1575 & 2.60 & 3.74 \\
                        1.00 & 1677 & 3.07 & 4.11 \\
                        2.26 & 1701 & 3.29 & 4.48 \\
                        
                        \hline\hline
                \end{tabular}
                \begin{itemize} 
                        \tiny
                        \item[*] $D_\mathrm{max}$ represents the degree of deviation from the cosmological principle, and the larger the value, the higher the degree of deviation.
                \end{itemize}
        \end{spacing}
\end{table}

\begin{figure}[h]
        \centering
        \includegraphics[width=0.24\textwidth]{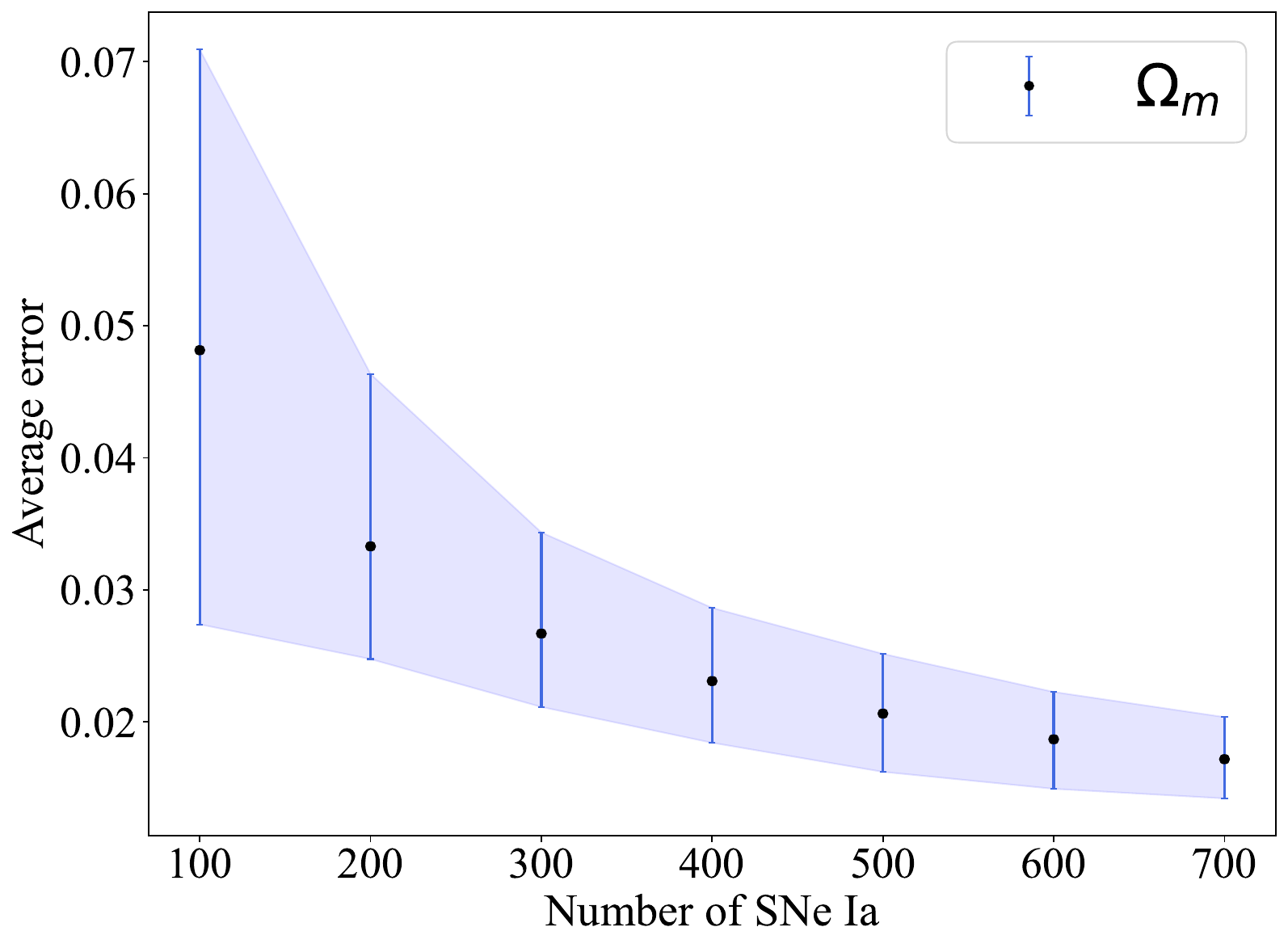}
        \includegraphics[width=0.24\textwidth]{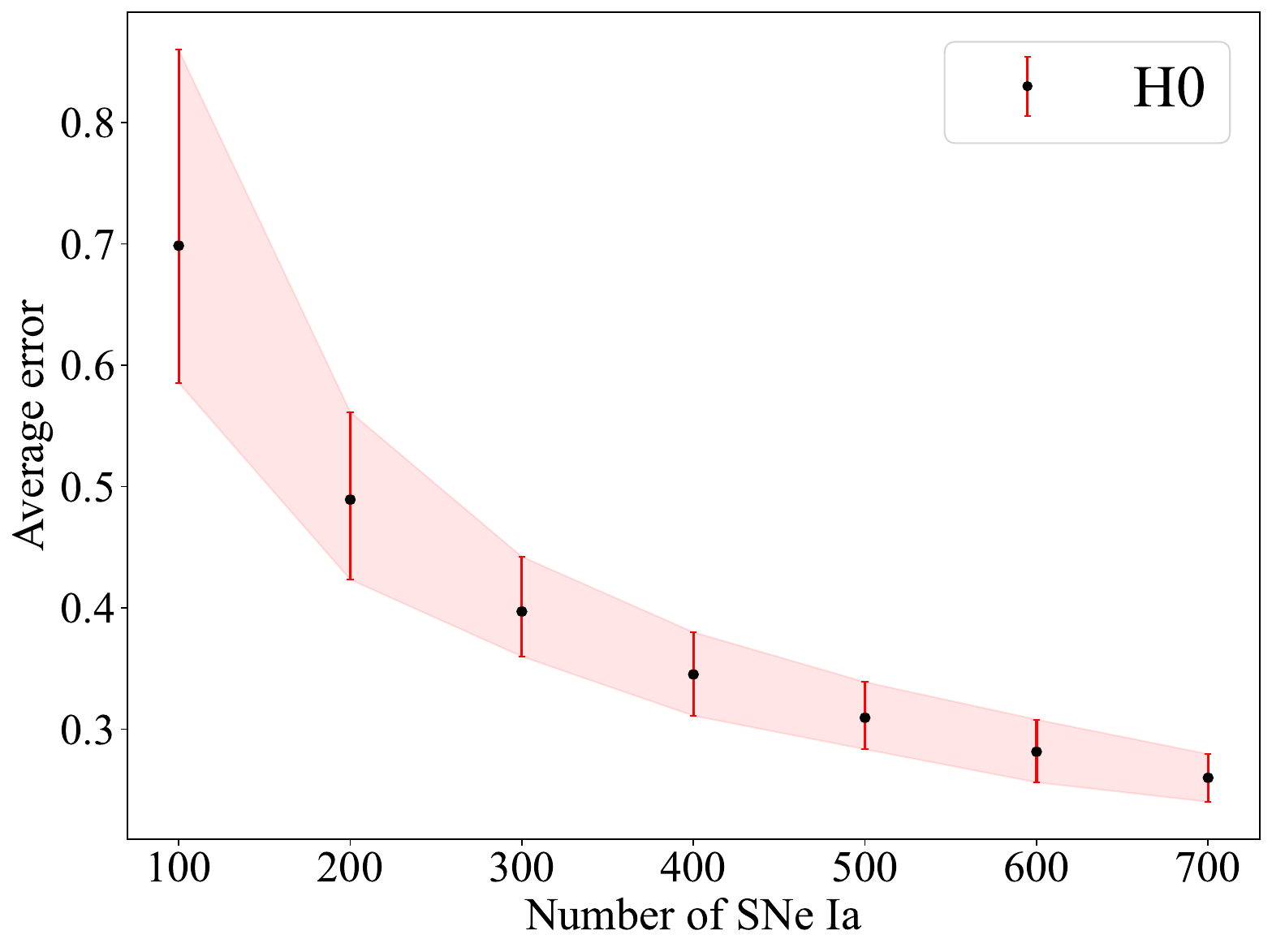}
        \includegraphics[width=0.24\textwidth]{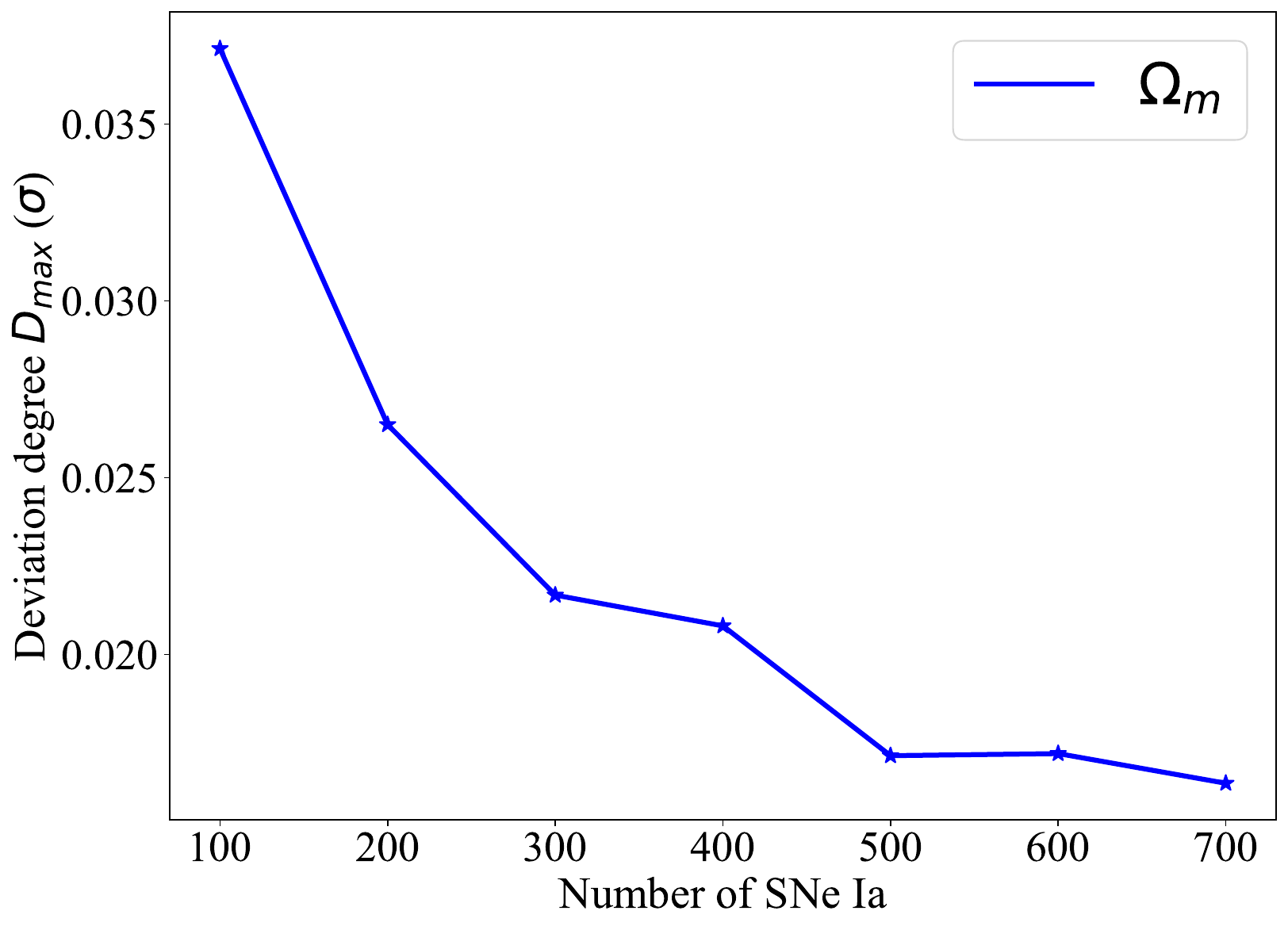}
        \includegraphics[width=0.24\textwidth]{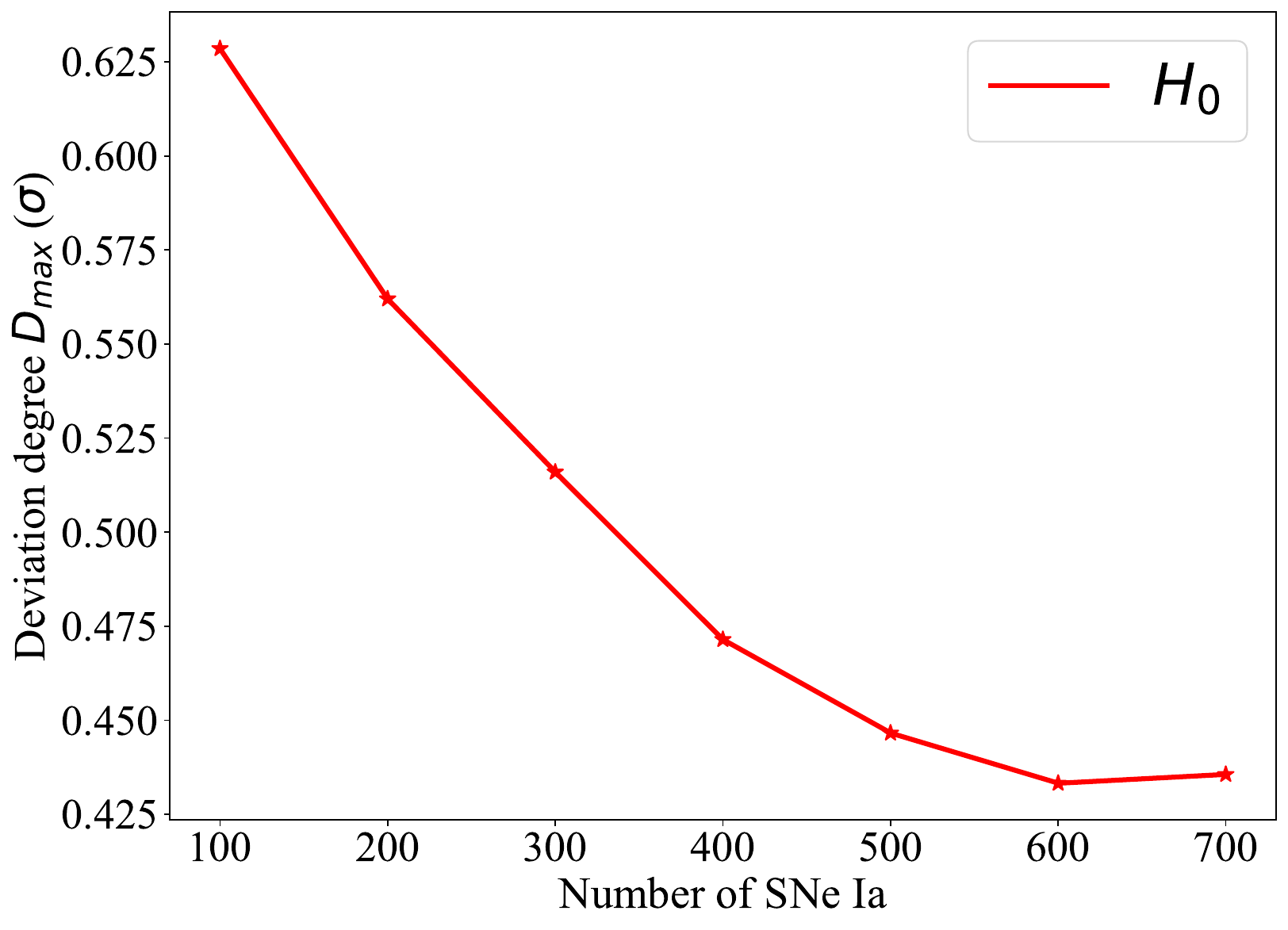}
        \caption{Relationship between average error and discrepancy degree $D_\mathrm{max}$ and SN number used. The upper panels show the relationship between the average error and the used SN number. The lower panels show the relationship between the discrepancy degree $D_\mathrm{max}$ and the SN number used. Blue and red correspond to parameters $\Omega_{m}$ and $H_{0,}$ respectively.  }
        \label{err}       
\end{figure}

Theoretically, the screening angle \tm in the RF method can be any value between 0$\degr$ and 180$\degr$. However, due to the limitations of real data in quantity, spatial part, and redshift distribution, \tm cannot be selected arbitrarily. In Sect. \ref{theta}, we combine the RF method with the latest Pantheon+ sample, hoping to find a suitable screening angle \tm. The analysis results are shown in Fig. \ref{F9}. Finally, we find that the screening angle \tm of the Pantheon+ sample is 60$\degr$. Then, based on the 60$\degr$ RF method, we restudied the Pantheon+ sample. The corresponding results are shown in Figs. \ref{F10}, \ref{F60}, and \ref{F11}. From Fig. \ref{F10}, we find that there is a small area (314.16$\degr$, -22.92$\degr$) with a lower $H_{0}$ value in the higher $H_{0}$ area, but this structure does not appear in the RF (90$\degr$) results. The lower $H_{0}$ area may be caused by high material density structures (e.g. dark matter halo) or statistical uncertainty. Narrowing the screening angle reduces the number of SNe constraining cosmological parameters, which may increase the uncertainty of the constraint. We performed a bootstrap resampling of the sample (ignoring the direction, repeated 2000 times) to study the dependence of the best-fit result error and $D_\mathrm{max}$ on the number of SNe Ia used. Here, the number of SNe ranges from 100 to 700, with a total of seven groups. From the results of Fig. \ref{err}, we can find that the average error and $D_\mathrm{max}$ increases as the used number of SNe decreases. The number of SNe used significantly affects the average error but does not significantly affect the $D_\mathrm{max}$ value. The $D_\mathrm{max}$ difference caused by the reduction in number is much smaller than the total $D_\mathrm{max}$. The preferred direction of cosmic anisotropy using the RF method with \tm = 60$\degr$ is in line with that using \tm = 90$\degr$ within a 1$\sigma$ error. However, narrowing the screening angle means that the statistical significance of the isotropy analyses are significantly reduced. Interestingly, unlike previous findings, the statistical significance of the isotropy RP analysis is actually higher than that of the isotropy one.

\begin{figure*}
        \centering
        \includegraphics[width=0.8\textwidth]{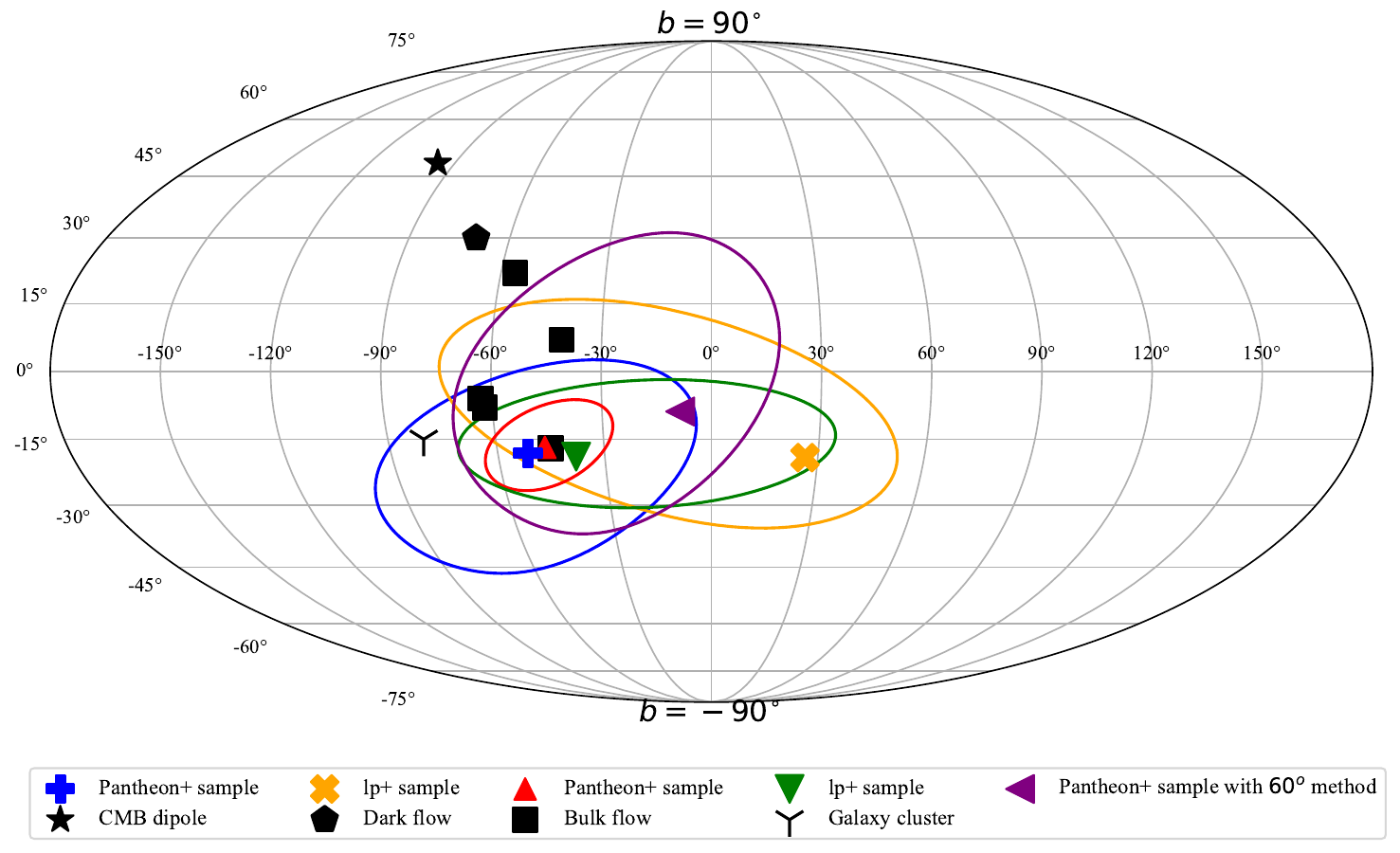}
        \caption{Distribution of preferred directions (l, b) with 1$\sigma$ range in other independent observations. We mark the directions given by $\Omega_{m}$ with plus signs at different screening angles, and the directions given by $H_{0}$ are shown with triangles at different screening angles. The color represents the important result obtained in this work; black shows the result given by other independent observations including the CMB dipole \citep{2016AA...594A...1P,2020AA...641A...1P}, dark flow \citep{2022JHEAp..34...49A}, bulk flow \citep{2012MNRAS.420..447T,2017MNRAS.468.1420F,2023MNRAS.524.1885W}, and galaxy cluster \citep{2021AA...649A.151M}. }
        \label{F12}       
\end{figure*}

Overall, we find that the all-sky distributions of cosmological parameters deviate significantly from isotropy, as shown in Figs. \ref{F4}, \ref{F6}, and \ref{F10}. All preferred directions we obtained are consistent with each other within a 1$\sigma$ range, and they are in line with previous research that traced the anisotropy of $\Omega_{m}$ and $H_{0}$ \citep{2010JCAP...12..012A,2012JCAP...02..004C,2013AA...553A..56K,2015MNRAS.446.2952C,2020A&A...643A..93H} and other dipole researches \citep{2014MNRAS.443.1680W,2014MNRAS.437.1840Y,2016MNRAS.456.1881L,2017MNRAS.468.1953P,2019MNRAS.486.1658C,2023MNRAS.525..231D}. However, they are different from those of \citet{2022PhRvD.105j3510L} and \citet{2023arXiv230402718M}, which are consistent with the results of the CMB dipole \citep{2016AA...594A...1P,2020AA...641A...1P}. In addition, comparing with other independent observations (as shown in Table \ref{T2}) including the CMB dipole \citep{2016AA...594A...1P,2020AA...641A...1P}, dark flow \citep{2022JHEAp..34...49A}, bulk flow \citep{2012MNRAS.420..447T,2017MNRAS.468.1420F,2023MNRAS.524.1885W}, and galaxy cluster \citep{2021AA...649A.151M}, it is easy to find that the directions of the larger $H_{0}$ and the smaller $\Omega_{m}$ are not consistent with the CMB dipole \citep{2016AA...594A...1P,2020AA...641A...1P}, but they coincide with the bulk flow \citep{2012MNRAS.420..447T,2017MNRAS.468.1420F,2023MNRAS.524.1885W} and the galaxy cluster \citep{2021AA...649A.151M}. To facilitate understanding, we aggregated these results with the ones we obtained, marking them on the galactic coordinate system, as shown in Fig. \ref{F12}. The effect of peculiar velocities and the bulk flow on SNe Ia cosmology has already been discussed \citep{2006PhRvD..73l3526H,2011ApJ...741...67D,2014A&A...568A..22B,2018MNRAS.477.1772R}. They can make a tiny shift in the best-fit cosmological parameters and the preferred direction locally \citep{2019A&A...631L..13C}.

\begin{table*}\footnotesize
        \caption{Preferred directions given by other independent observations. \label{T2}}
        \centering
        \begin{spacing}{1.2}
                \begin{tabular}{ccc}
                        \hline\hline
                        Cosmological obs. & Direction (l, b)  & Ref.  \\
                        \hline
                        CMB dipole & (264.00$^{\circ}$$\pm$0.03, 48.24$^{\circ}$$\pm$0.02)  & \citep{2016AA...594A...1P} \\
                        & (264.02$^{\circ}$$\pm$0.01, 48.25$^{\circ}$$\pm$0.01)  & \citep{2020AA...641A...1P} \\
                        Dark flow  &(290$^{\circ}$$\pm$20, 30$^{\circ}$$\pm$15)  & \citep{2022JHEAp..34...49A} \\
                        Bulk flow  & (319$^{\circ}$$\pm$18, 7$^{\circ}$$\pm$14)  & \citep{2012MNRAS.420..447T} \\
                        & (315$^{\circ}$$\pm$52, -17$^{\circ}$$\pm$15)  & \citep{2017MNRAS.468.1420F} \\
                        & (304$^{\circ}$$\pm$14, 22$^{\circ}$$\pm$11)  & \citep{2017MNRAS.468.1420F} \\
                        & (297$^{\circ}$$\pm$4, -6$^{\circ}$$\pm$3)  & \citep{2023MNRAS.524.1885W} \\
                        & (298$^{\circ}$$\pm$5, -8$^{\circ}$$\pm$4)  & \citep{2023MNRAS.524.1885W} \\                 
                        Galaxy cluster & (280$^{\circ}$$\pm$35, -15$^{\circ}$$\pm$20)  & \citep{2021AA...649A.151M} \\
                        Pantheon+ ($H_{0}$, 90$\degr$) & $({313.4^{\circ}}_{-18.2}^{+19.6}, {-16.8^{\circ}}_{-10.7}^{+11.1})$  & this paper \\
                        Pantheon+ ($H_{0}$, 60$\degr$) & (${351.4^{\circ}}_{-64.2}^{+28.0}$, ${-8.8^{\circ}}_{-27.8}^{+40.3}$)  & this paper \\
                        lp+ ($H_{0}$, 90$\degr$) & (${321.9^{\circ}}_{-33.5}^{+72.5}$, ${-18.9^{\circ}}_{-11.5}^{+16.6}$)  & this paper \\
                        Pantheon+ ($\Omega_{m}$, 90$\degr$) & (${308.4^{\circ}}_{-48.7}^{+47.6}, {-18.2^{\circ}}_{-28.8}^{+21.1}$)  & this paper \\
                        lp+ ($\Omega_{m}$, 90$\degr$) & (${26.4^{\circ}}_{-100.9}^{+26.3}$, ${-19.3^{\circ}}_{-17.0}^{+35.4}$)  & this paper \\

                        \hline\hline
                \end{tabular}
        \end{spacing}
\end{table*}

\section{Conclusions and perspectives}
In this paper, we propose the RF method for the first time and combine this method with the Pantheon+ sample to test the cosmological principle. From the matter density and the Hubble expansion distributions mapped using the RF method, we find that the all-sky distributions of cosmological parameters deviate significantly from isotropy. The corresponding distribution of luminosity distance also deviates from isotropy; that is, the $d_{L}$-$z$ relation actually diverges from region to region. Results of statistical isotropy analyses (isotropy and isotropy RP) show relatively high confidence levels: 2.78$\sigma$ (isotropy) and 2.34$\sigma$ (isotropy RP) for the local matter underdensity, 3.96$\sigma$ (isotropy) and 3.15$\sigma$ (isotropy RP) for the cosmic anisotropy. Comparing the results of statistical isotropy analyses, we find that inhomogeneous spatial distribution of real sample can increase the deviation from isotropy. The statistical significations of $H_{0}$ anisotropy are more obvious than that of $\Omega_{m}$ anisotropy. This might hint that parameter $H_{0}$ is more sensitive to cosmic anisotropy. The similar results and findings are found from reanalyses of the lp+ sample and the lower screening angle (\tm = 60$\degr$), but with a slight decrease in statistical significance. In addition, we find that $D_\mathrm{max}$ changes with $z_\mathrm{max}$ and has a maximum value near $z_\mathrm{max}$ = 0.30. The average error and $D_\mathrm{max}$ increase as the used number of SNe decreases. Comparing with the previous researches, we find all preferred directions we obtained are in line with that were provided by \citet{2010JCAP...12..012A}, \citet{2012JCAP...02..004C}, \citet{2014MNRAS.443.1680W}, \citet{2014MNRAS.437.1840Y}, \citet{2013AA...553A..56K}, \citet{2015MNRAS.446.2952C}, \citet{2016MNRAS.456.1881L}, \citet{2019MNRAS.486.1658C}, and \citet{2020A&A...643A..93H}, but they are not consistent with those given by \citet{2022PhRvD.105j3510L} and \citet{2023arXiv230402718M}.

Until now, many SNe Ia have been observed and have been widely used for cosmological applications \citep{2018ApJ...854...46H,2021PhRvD.104l3511P,2023MNRAS.526.1482C,2022A&A...661A..71H,2022PhRvD.106f3515W,2023MNRAS.522.6024B}.\footnote{For more recent studies on the cosmological applications employing the SNe Ia sample, see \citet{2023MNRAS.521.3909B}, \citet{2023PhRvD.107j3521C}, \citet{2023arXiv230517243D}, \citet{2023PhRvD.108h3024H}, \citet{2023ChPhC..47f5104J}, \citet{2023arXiv231003509K}, \citet{2023arXiv231006028L}, \citet{2023ApJ...956..130M}, \citet{2023PDU....4001216O}, \citet{2023PDU....4001224P}, \citet{2023MNRAS.520.5110P}, \citet{2023PhRvD.108h3519S}, \citet{2023arXiv230504946V}, \citet{2023EPJC...83..859W}, and references therein.} However, there is still an inhomogeneous distribution in data, which significantly affects the testing of cosmological principle. One way to solve this problem is to add some new SNe Ia measurements, and another way is to consider other independent observations; for example, quasar \citep{2016ApJ...819..154L,2017FrASS...4...68B,2017A&A...602A..79L,2019NatAs...3..272R,2022MNRAS.516.1721C,2022MNRAS.515.3729K,2023PhLB..84538166L,2023arXiv230508179Z}, gamma-ray burst \citep[GRB;][]{2015NewAR..67....1W,2020MNRAS.496.1530S,2021MNRAS.507..730H,2022MNRAS.510.2928C,2022ApJ...935....7L,2022Univ....8..344L,2022ApJ...941...84L,2022ApJ...924...97W}, fast radio burst \citep[FRB;][]{2022MNRAS.511..662H,2022MNRAS.515L...1W,2022MNRAS.516.4862J,2022arXiv221213433Z,2023arXiv230708285G,2023ApJ...946L..49L}, Tip of the Red Giant Branch \citep[TRGB;][]{2019ApJ...882...34F,2020ApJ...891...57F,2021ApJ...919...16F}, galaxy cluster \citep{2017A&A...597A.120J,2020A&A...636A..15M,2021AA...649A.151M}, and gravitational wave \citep[GW;][]{2017Natur.551...85A,2018Natur.562..545C,2023ChPhC..47f5104J,2023ApJ...943...13W} observations, among others. At present, these observations may still have some deficiencies in quantity or quality, but this situation is expected to improve in the future. The e-ROSITA all-sky survey \citep{2012arXiv1209.3114M,2012SPIE.8443E..1RP,2013A&A...558A..89K,2020FrASS...7....8L}, Einstein Probe \citep[EP;][]{Yuan2015}, French--Chinese satellite space-based multi-band astronomical variable objects monitor \citep[SVOM;][]{Wei2016}, China Space Station Telescope (CSST) photometric survey \citep{2022MNRAS.515.5587X,2023MNRAS.519.1132M,2023SCPMA..6629511L}, and Transient High-Energy Sky and Early Universe Surveyor \citep[THESEUS;][]{Amati2018} space missions together with ground- and space-based multi-messenger facilities will provide a lot of observations and measurements, improve the observational quality, and probe the poorly explored high-redshift universe. The Australian Square Kilometer Array Pathfinder \citep[ASKAP;][]{2017ASPC..512...73C}, MeerKAT \citep{2018IAUS..337..406S}, Very Large Array \citep[VLA;][]{2018ApJS..236....8L}, and Canadian Hydrogen Intensity Mapping Experiment (CHIME)/FRB Outriggers \citep{2021AJ....161...81L} will provide a large number of positioned FRBs in the future, which will give higher precision cosmological constraints. In addition, the Advanced Laser Interferometer Gravitational-wave Observatory \citep[aLIGO;][]{2015CQGra..32g4001L} and Virgo \citep{2015CQGra..32b4001A} detectors will provide more GW events. In combination with the electromagnetic counterparts, model-independent constraints will be given on the cosmological parameters. The high-quality observations enable us to examine the cosmological principle at higher redshifts and investigate whether the Hubble tension is related to the failure of the cosmological principle. 

\noindent \emph{\emph{Notes:} \emph{As this work was about to be completed, \citet{2023PhRvD.108f3509P} used the HC method to test the cosmic isotropy of the SNe Ia absolute magnitudes from the Pantheon+ and SH0ES samples in various redshift/distance bins. They found that sharp changes of the level of anisotropy occuring at distances under 40 Mpc in the real samples. If there are enough local observations in the future, more local information about our Universe could be obtained by combining our method with the idea of \citet{2023PhRvD.108f3509P}. This will be pursued in future work.}}

\section*{Acknowledgements}
We thank the anonymous referee for constructive comments. This work was supported by the National Natural Science Foundation of China (grant No. 12273009), the China Manned Spaced Project (CMS-CSST-2021-A12), Jiangsu Funding Program for Excellent Postdoctoral Talent (20220ZB59), Project funded by China Postdoctoral Science Foundation (2022M721561), NWO, the Dutch Research Council, under Vici research programme `ARGO' with project number 639.043.815, Yunnan Youth Basic Research Projects 202001AU070013 and National Natural Science Foundation of China (grant No. 12303050).

\bibliographystyle{aa} 

\begin{appendix}

\section{Additional figures}
\begin{figure}[htp]
        \centering
        \subfigure[30$\degr$]{\label{Fig9.sub.1}\includegraphics[width=0.42\hsize]{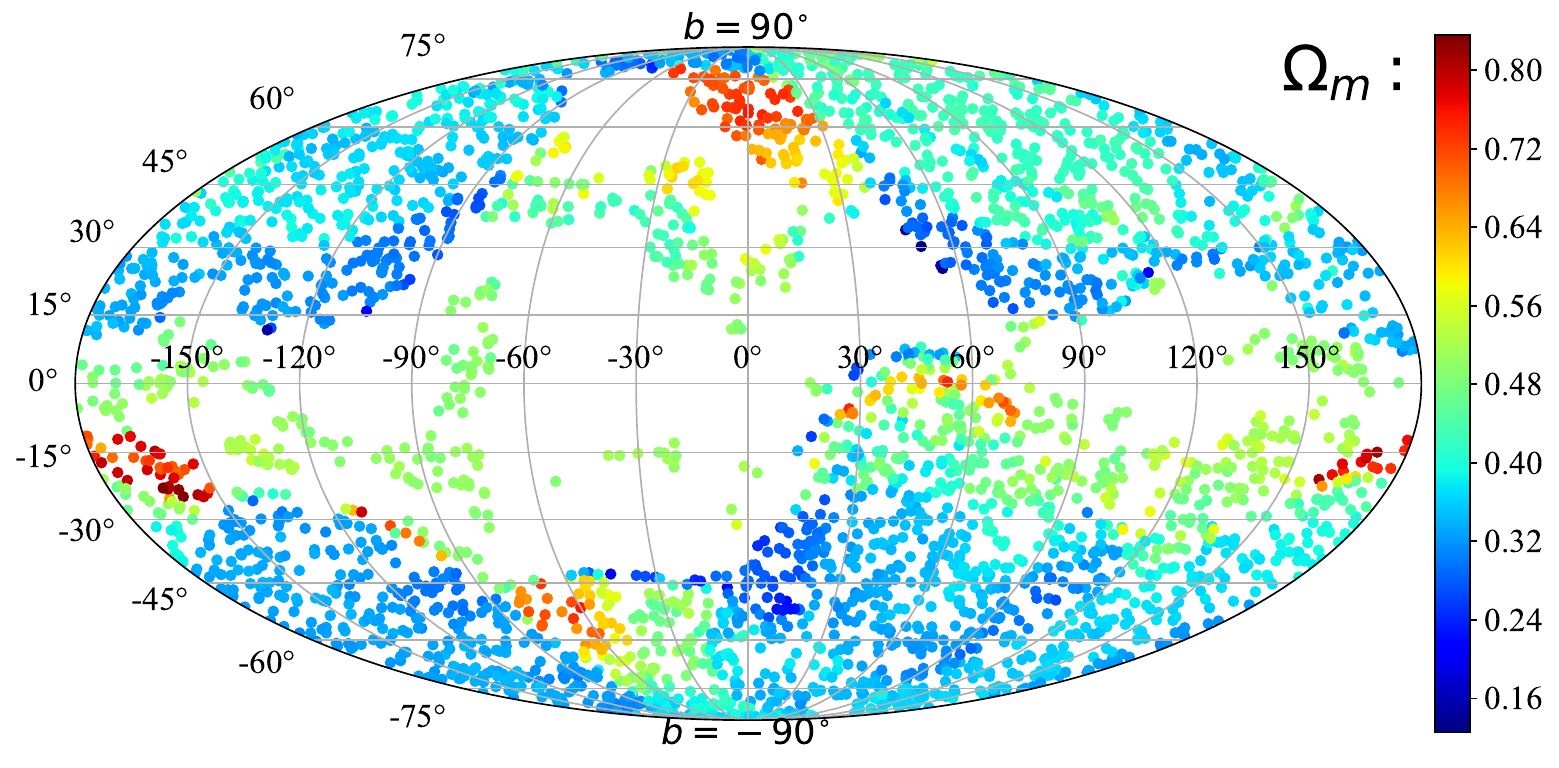}}
        \subfigure[30$\degr$]{\label{Fig9.sub.2}\includegraphics[width=0.42\hsize]{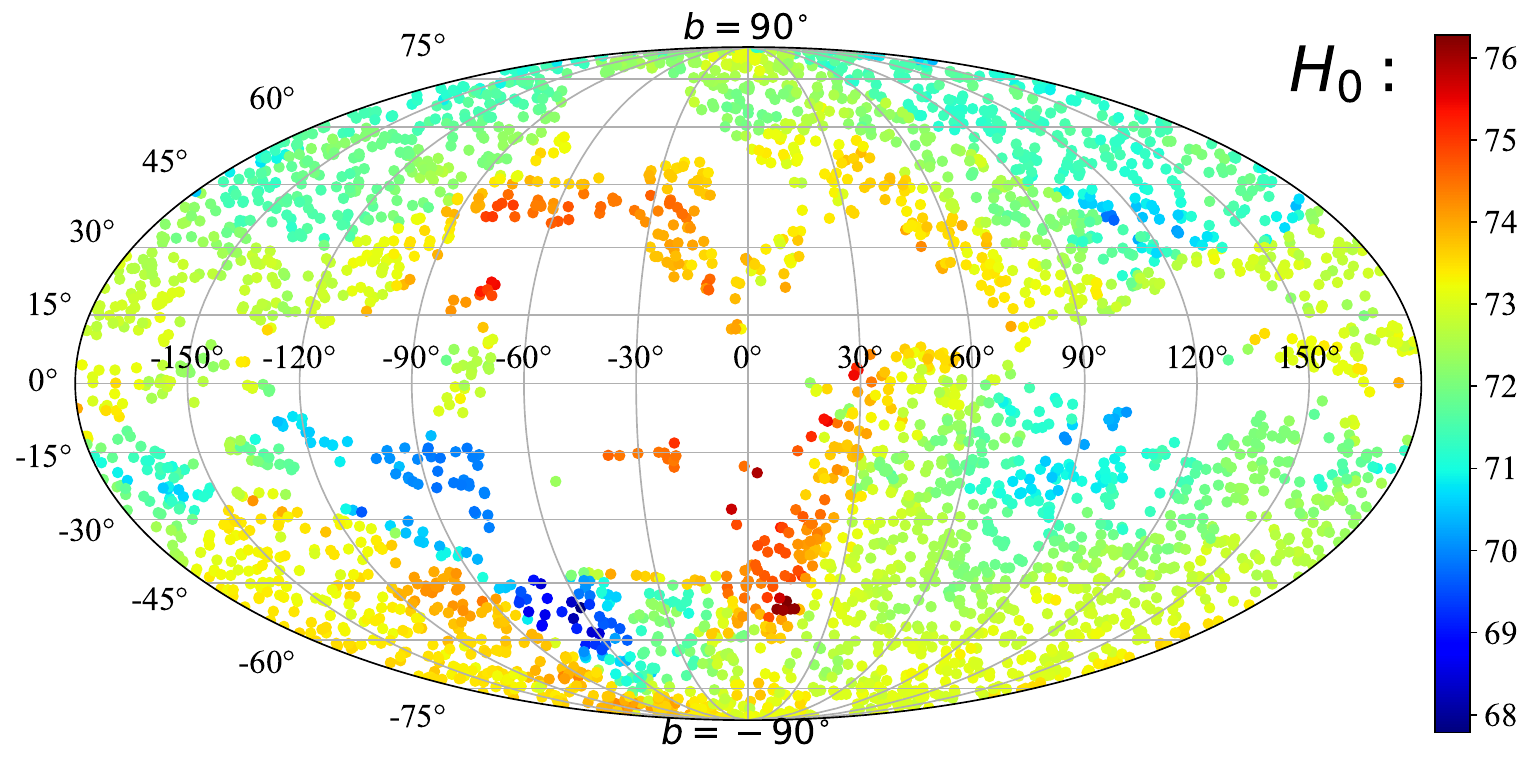}}\\
        \subfigure[45$\degr$]{\label{Fig9.sub.3}\includegraphics[width=0.42\hsize]{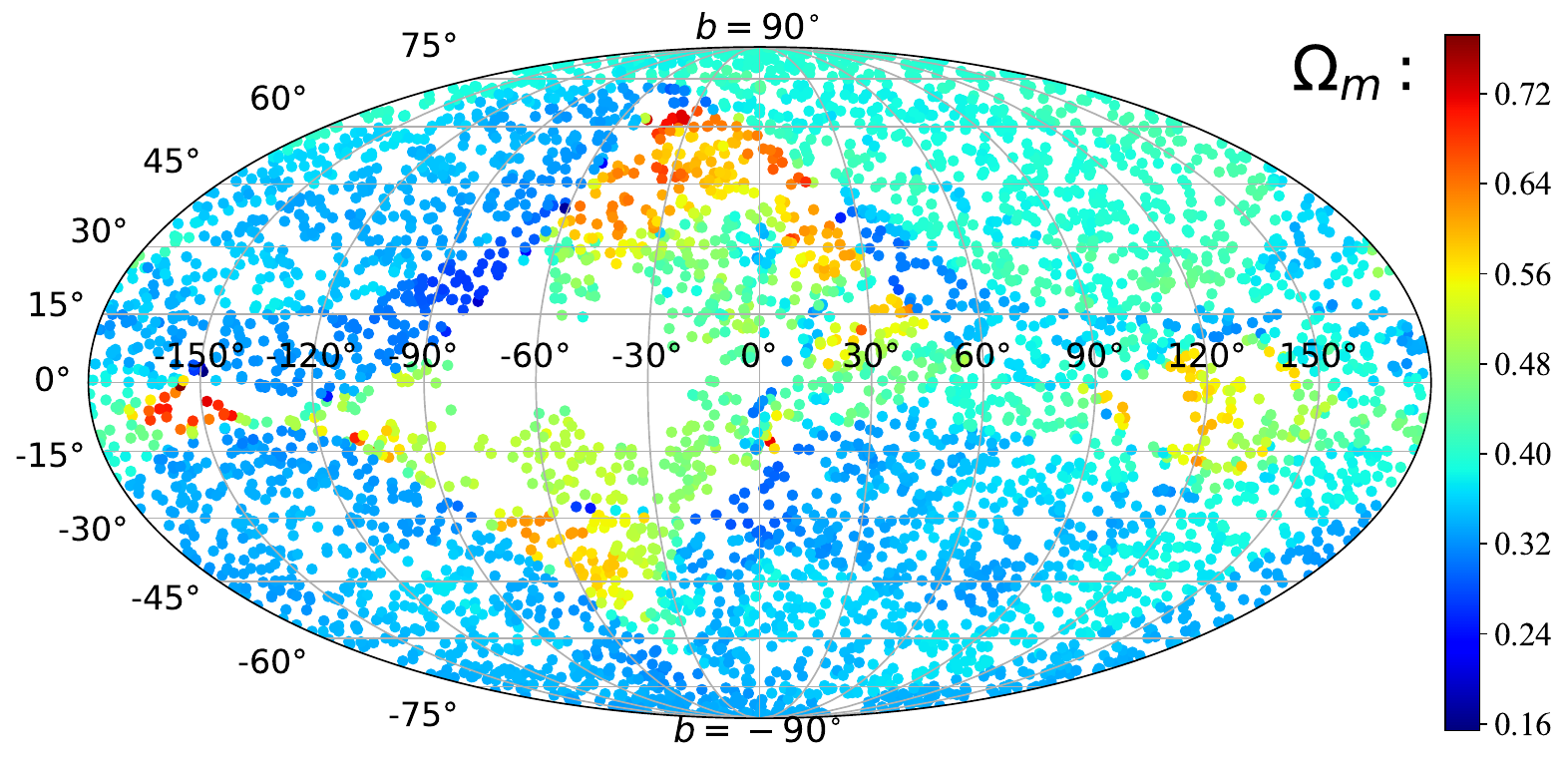}}
        \subfigure[45$\degr$]{\label{Fig9.sub.4}\includegraphics[width=0.42\hsize]{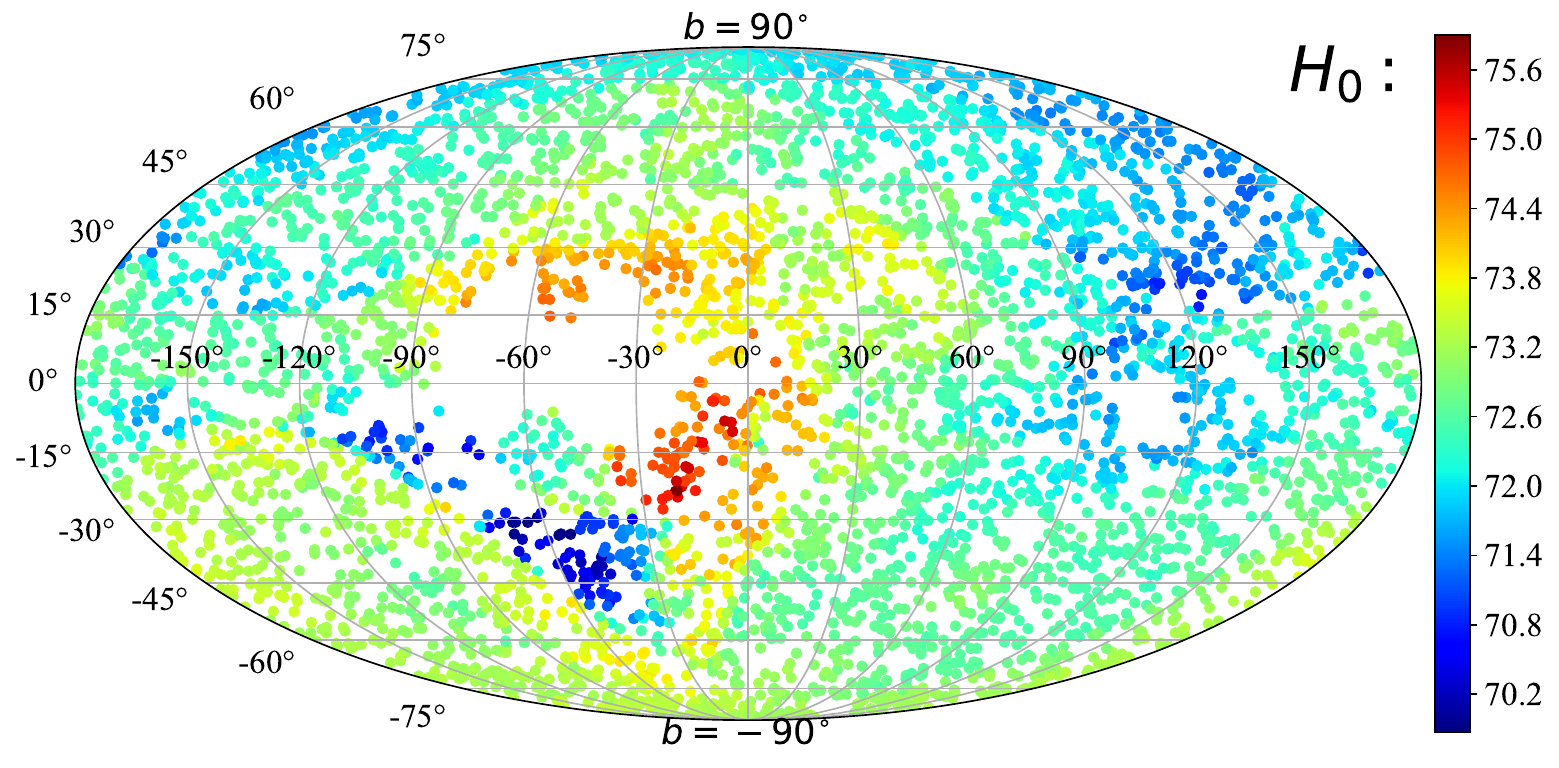}}\\
        \subfigure[60$\degr$]{\label{Fig9.sub.5}\includegraphics[width=0.42\hsize]{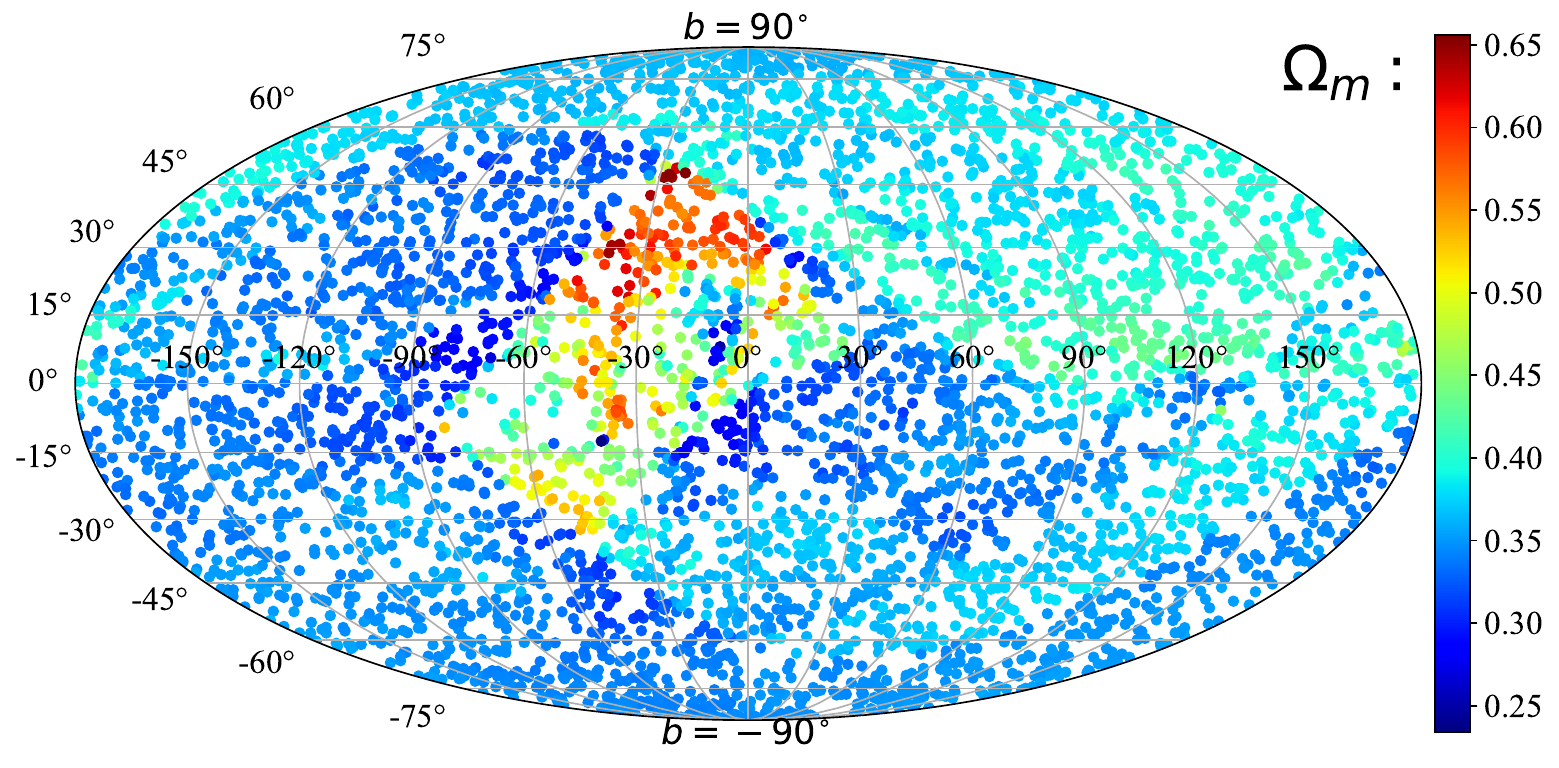}}
        \subfigure[60$\degr$]{\label{Fig9.sub.6}\includegraphics[width=0.42\hsize]{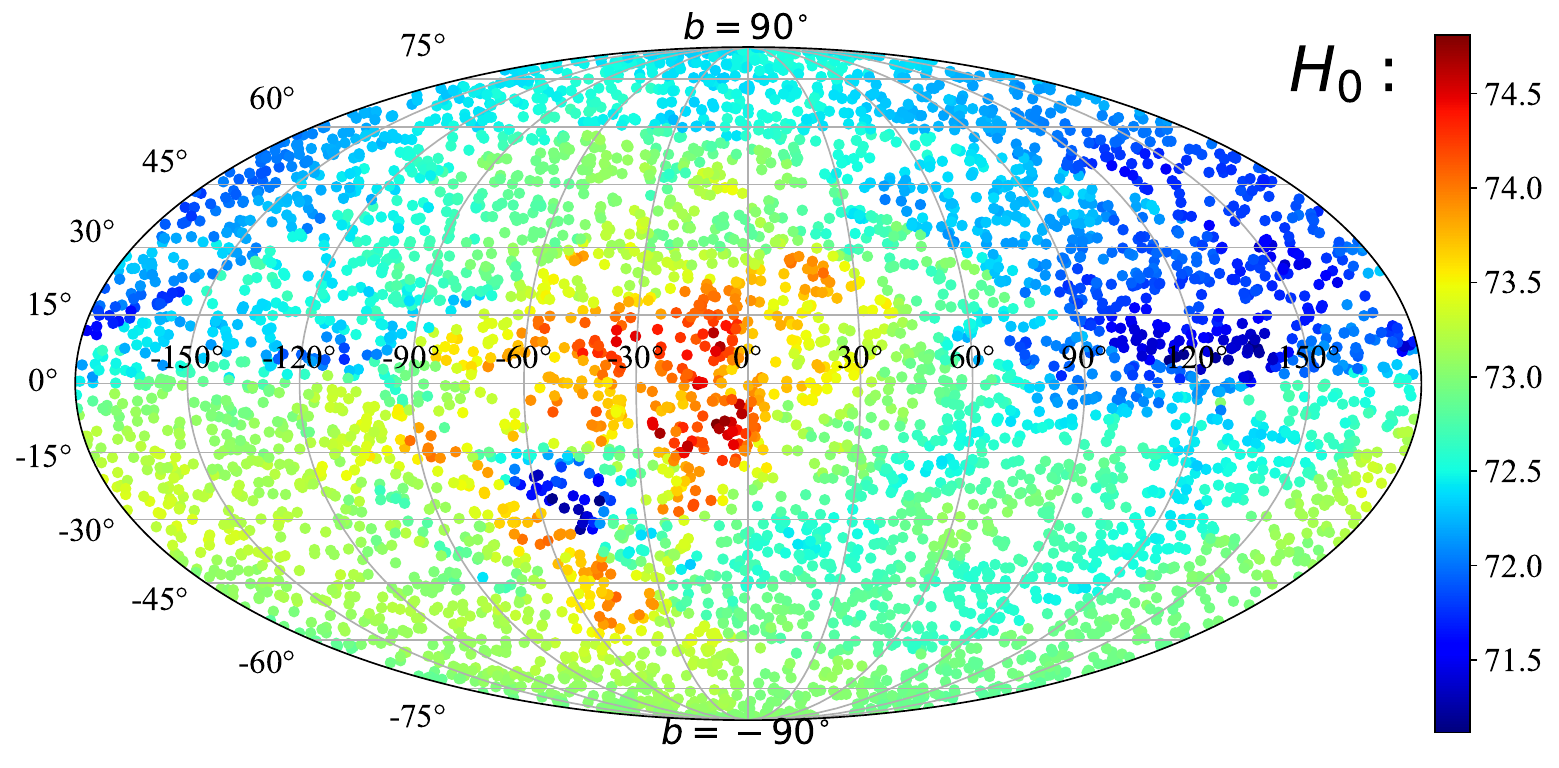}}
        \caption{All-sky distribution of cosmological parameter ($\Omega_{m}$ and $H_{0}$) utilizing the Pantheon+ sample combined with the RF method with different screening angles including 30$\degr$, 45$\degr$ , and 60$\degr$. Panels (a) and (b) show the results using the RF method with 30$\degr$. Panels (c) and (d) are the results of 45$\degr$. Panels (e) and (f) are the results of 60$\degr$. The proportions of the wrong fitting results are 30.34\%, 6.47\%, and 0.70\% for the screening angles 30$\degr$, 45$\degr$ , and 60$\degr$, respectively. }
        \label{F9}       
\end{figure}

\begin{figure}[htp]
        \centering
        \includegraphics[width=0.3\textwidth]{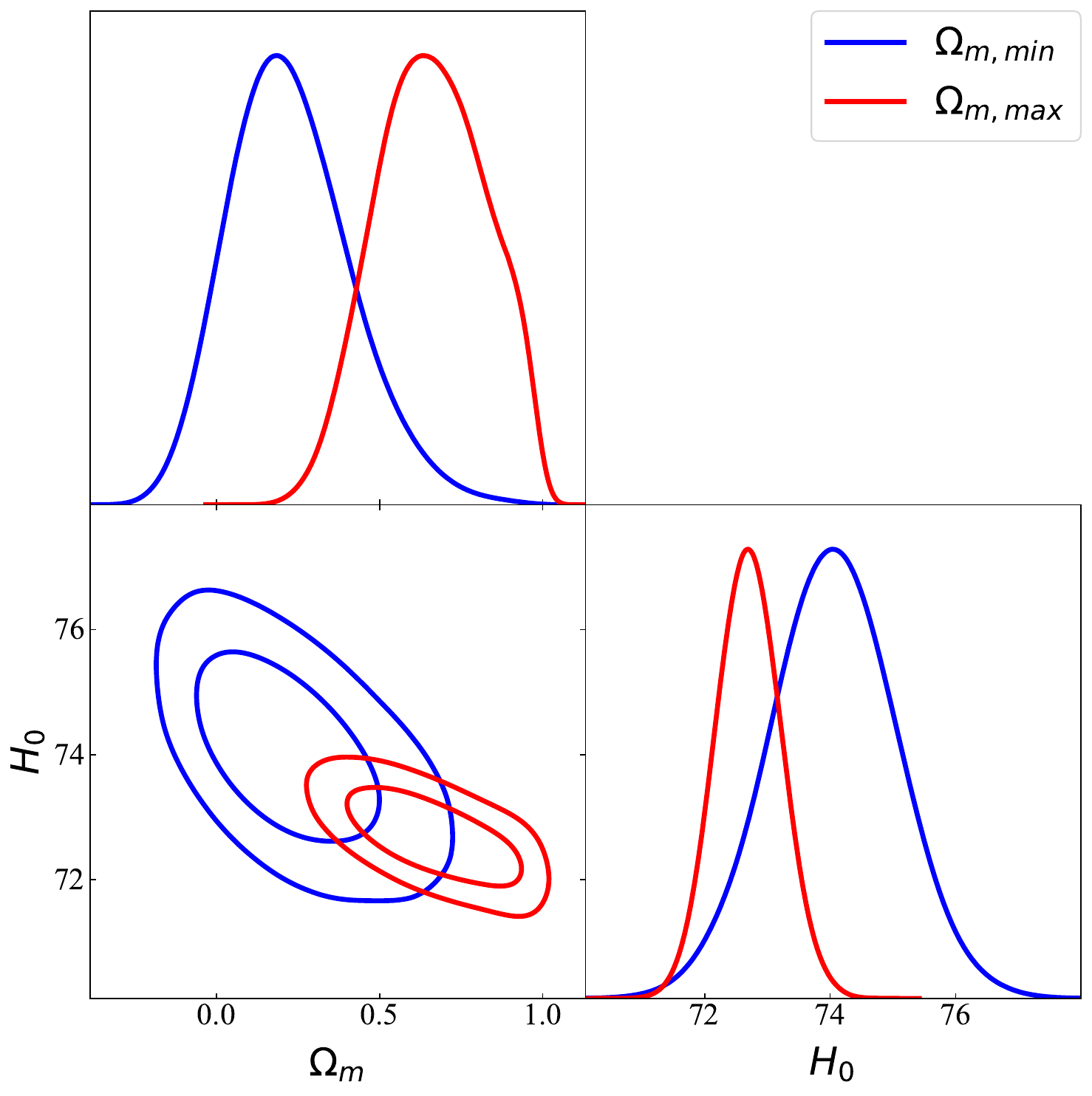}
        \caption{Confidence contours ($1\sigma$ and $2\sigma$) and marginalized likelihood distributions for parameters space ($\Omega_{m}$ and $H_{0}$) in the spatially flat $\rm \Lambda$CDM model from SNe Ia subsamples, which corresponds to $\Omega_{m, \mathrm{min}}$ and $\Omega_{m, \mathrm{max}}$. The best fits are $\Omega_{m, \mathrm{min}}$ = 0.23$^{+0.20}_{-0.17}$ and $H_{0}$ = 74.05$^{+1.00}_{-1.00}$ km s$^{-1}$ Mpc$^{-1}$ and $\Omega_{m, \mathrm{max}}$ = 0.66 $^{+0.17}_{-0.17}$ and $H_{0}$ = 72.70 $^{+0.52}_{-0.52}$ km s$^{-1}$ Mpc$^{-1}$. }
        \label{Ommax}       
\end{figure}

\end{appendix}
\label{lastpage}
\end{document}